\documentclass[11pt,a4paper]{article}
\usepackage{amsmath}
\usepackage{graphicx}
\usepackage{xcolor}
\usepackage[caption=false]{subfig}
\usepackage{mathrsfs,mathtools}
\usepackage{physics,amssymb}
\usepackage{siunitx}
\usepackage{bm}
\usepackage{float}
\usepackage{braket}
\usepackage{listings}
\DeclareMathOperator{\sinc}{sinc}
\usepackage{cases}
\usepackage{comment}
\usepackage{soul}
\usepackage{cancel}
\usepackage{cases}
\usepackage[utf8]{inputenc}
\usepackage{url}
\usepackage{float}
\usepackage{longtable}
\usepackage[normalem]{ulem}
\usepackage{xspace}
\usepackage{aas_macros}
\setstcolor{red}
\usepackage{jcappub}

\usepackage{tabularx}
   \newcolumntype{C}{>{\centering\arraybackslash}X}
   \newcolumntype{L}{>{\raggedright\arraybackslash}X}
   \newcolumntype{R}{>{\raggedleft\arraybackslash}X}
   
   \usepackage{hhline}

\hypersetup{colorlinks=true
,urlcolor=DARKBLUE
,anchorcolor=DARKBLUE
,citecolor=DARKBLUE
,filecolor=DARKBLUE
,linkcolor=DARKBLUE
,menucolor=DARKBLUE
,linktocpage=true
,pdfproducer=medialab
}

\usepackage{graphicx,epsfig,color,xcolor}

\usepackage[english]{babel}
\usepackage{amssymb,amsfonts,amsmath,physics}
\usepackage{mathtools}
\usepackage{siunitx}
\usepackage{verbatim}
\usepackage{ulem}
\usepackage{soul}
\usepackage{lipsum, babel}
\usepackage{acronym}
\usepackage{aas_macros}

\usepackage{hyperref}

\newcommand{\be}{\begin{equation}} \newcommand{\ee}{\end{equation}}
\newcommand{\bea}{\begin{eqnarray}} \newcommand{\eea}{\end{eqnarray}}

\usepackage{fancyhdr}
\usepackage{amsmath}
\usepackage{mathrsfs}
\usepackage{float}
\usepackage{xcolor}
\usepackage{graphicx, epsfig, amssymb} 
\usepackage{amsmath, amsfonts}
\usepackage{bm} 
\usepackage{tensor}
\usepackage{soul}
\usepackage{hyperref}
\usepackage{physics}

\definecolor{MONZA}{HTML}{CF000F}
\definecolor{DARKBLUE}{HTML}{00008b}
\definecolor{DARKMAGENTA}{HTML}{8b008b}
\definecolor{DARKCYAN}{HTML}{008B8B}
\definecolor{DARKORANGE}{HTML}{FF8C00}
\definecolor{OBSERVATORY}{HTML}{049372}
\definecolor{GREENBAMBOO}{HTML}{006442}
\definecolor{TURQUOISE}{HTML}{36D7B7}
\definecolor{JUNGLEGREEN}{HTML}{26C281}

\begin{document}

\title{Simulations of Ellipsoidal Primordial Black Hole Formation}



\author[a,b,c]{Albert Escriv\`a}

\author[a,d]{, Chul-Moon Yoo}

\affiliation[a]{\mbox{Division of Particle and Astrophysical Science, Graduate School of Science,}  Nagoya University. Nagoya 464-8602, Japan}
\affiliation[b]{\mbox{Division of Science, National Astronomical Observatory of Japan,} Mitaka, Tokyo 181-8588, Japan}
\affiliation[c]{Institute for Advanced Research, Nagoya University, \\
Furo-cho Chikusa-ku, Nagoya 464-8601, Japan}

\affiliation[d]{\mbox{Kobayashi Maskawa Institute,} 
Nagoya University, Nagoya 464-8602, Japan}

\emailAdd{escriva.manas.alberto.k0@f.mail.nagoya-u.ac.jp}
\emailAdd{yoo.chulmoon.k6@f.mail.nagoya-u.ac.jp}

\date{\today}
\abstract{We perform $3+1$ relativistic numerical simulations to study primordial black hole (PBH) formation from the collapse of adiabatic super-horizon non-spherical 
perturbations generated from curvature fluctuations obeying random Gaussian statistics with a monochromatic power spectrum. The matter field is assumed to be a perfect fluid of an equation of state $w:=P/\rho={\rm const.}$ with $P$ and $\rho$ being the pressure and the energy density, respectively. The initial spatial profile of the curvature perturbation is modeled with the amplitude $\mu$ and non-spherical parameters $e$ (ellipticity) and $p$ (prolateness) according to peak theory. We focus on the dynamics and the threshold for PBH formation in terms of the non-spherical parameters $e$ and $p$. We find that the critical values ($e_c, p_c$) with a fixed value of $\mu$ closely follow a superellipse curve. With $p=0$, for the range of amplitudes considered, we find that the critical ellipticity for non-spherical collapse follows a decaying power law as a function of $(\mu-\mu_{\rm c,sp})$ with $\mu_{\rm c,sp}$ being the threshold for the spherical case. Our results also indicate that, for both cases of $w = 1/3$ and $w = 1/10$, small deviations from sphericity can avoid collapsing to a black hole when the amplitude is near its critical threshold. Finally we discuss the significance of the ellipticity on the rate of the PBH production. 
}

\maketitle
\flushbottom

\acresetall

\acrodef{GW}{gravitational wave}
\acrodef{PT}{phase transition}
\acrodef{SC}{smooth crossover}
\acrodef{SM}{Standard Model}
\acrodef{QCD}{Quantum Chromodynamics}
\acrodef{EW}{electroweak}
\acrodef{CMB}{cosmic microwave background}
\acrodef{PBH}{primordial black hole}
\acrodef{DM}{Dark Matter}
\acrodef{FLRW}{Friedmann‐-Lema\^itre--Robertson--Walker}
\acrodef{MS}{Misner--Sharp}

\section{Introduction}

Primordial black holes (PBHs), if they exist, are black holes that could have been formed in the early Universe \cite{Zeldovich:1967lct,Hawking:1971ei,1974MNRAS.168..399C, 1975ApJ...201....1C,1979A&A....80..104N} (see \cite{Sasaki:2018dmp,Green:2020jor,Carr:2020gox,Carr:2020xqk,Escriva:2022duf,Yoo:2022mzl} for reviews covering different perspectives). A PBH results from the collapse of a sufficiently large over-density in the early Universe, much before the moment of matter-radiation equality, and therefore, their existence can be evidence of the existence of primordial inhomogeneities. Interestingly, PBHs may constitute a significant fraction of the dark matter \cite{Chapline:1975ojl} (especially in the so-called asteroid mass range) and explain different cosmic conundra \cite{Carr:2023tpt}. Currently, PBHs have not been detected so far, but future gravitational wave observations may establish their existence (in particular if their mass is lower than a solar mass)\cite{Sasaki:2016jop,2016PhRvX...6d1015A, 2021arXiv211103606T} and quantify their role in the dark matter.

Various mechanisms can lead to the production of PBHs (see \cite{Escriva:2022duf} for a comprehensive list). Among these, one of the most widely studied and frequently considered is the formation of PBHs from the collapse of large curvature fluctuations generated during inflation. These fluctuations, upon re-entering the cosmological horizon at sufficiently late times, can collapse to form PBHs if they are over a certain threshold. From now on, we will focus on this scenario through this work.

To accurately estimate the abundance of PBHs in our Universe, it is essential to 
clarify the initial conditions that lead to their formation. Particularly, determinating the threshold is essential for the abundance estimation as it 
is exponentially sensitive to 
the threshold
\cite{1975ApJ...201....1C}. 
The threshold estimation
typically requires relativistic numerical simulations. 
Most studies on simulation of PBH formation have focused on 
spherically symmetric systems 
(see \cite{Escriva:2021aeh} for a comprehensive review and references therein) for which we have nowadays a relatively solid understanding. 
In contrast, our understanding of PBH formation in non-spherical scenarios currently remains limited, with only a few studies numerically addressing this issue \cite{Yoo:2020lmg,deJong:2021bbo,deJong:2023gsx,Yoo:2024lhp} (see also \cite{Celestino:2018ptx,Gundlach:2016jzm,Baumgarte:2016xjw,Gundlach:2017tqq} for non-spherical collapse of perfect fluids in asymptotically flat spacetimes).

For PBHs formed during the radiation-dominated era of the Universe, the assumption of spherical symmetry is well justified within the framework of peak theory \cite{peak_theory}. In this context, large peaks in curvature fluctuations which are so rare that they are not overproduced beyond the dark matter density 
tend to be nearly spherical. However, even for such large peaks, the most likely configuration is not perfectly spherical \cite{peak_theory}. Instead, small deviations from sphericity are typical, leading to an ellipsoidal shape characterized by parameters known as ``ellipticity" ($e$) and ``prolateness" ($p$) following a probability distribution. 
Regarding the PBH formation with the initial amplitude slightly above the threshold for which critical behavior~\cite{choptuik} is relevant, 
it is not entirely clear how small deviations from sphericity impact the critical behavior of the collapse. Therefore, it is crucial to carefully examine how even a small deviation from sphericity might influence gravitational collapse and the corresponding statistical estimates of the PBH mass function and abundance. This analysis is necessary to contrast with the typical assumptions in the literature, where spherical symmetry is often assumed. For this analysis, relativistic $3+1$ simulations are essential.

Non-spherical PBH formation during a radiation-dominated Universe was studied in \cite{Yoo:2020lmg} through $3+1$ numerical simulations, considering a case with spheroidal ($p=\pm e$) initial curvature fluctuation.  
The results showed that, for a typical fluctuation amplitude and a Gaussian 
spatial profile, the threshold for non-spherical collapse is not significantly modified compared with the spherical case. In this work, we aim to consider a different curvature profile sourced by a monochromatic power spectrum following peak theory and determine the threshold values for $e$ and $p$. 
We further explore how these thresholds vary with different fluctuation amplitudes. 
Clarifying the PBH formation condition including the parameters $e$ and $p$ is crucial for statistically estimating the impact of non-spherical effects on the PBH mass function. This can also be a starting point for the generalization of analytical threshold estimations for non-spherical collapse, taking into account the
values of $e$ and $p$ as the parameters characterizing the initial profile, 
as done in \cite{universal1,Escriva:2020tak} in the case of spherical symmetry.

In the case of a soft equation of state, the reduction in pressure gradients may lead the non-sphericities to modify the gravitational collapse substantially, being the case of a dust-dominated epoch a clear example where non-spherical effects are crucial \cite{Khlopov:1980mg,Harada:2015ewt,Harada:2017fjm,deJong:2023gsx} (see also \cite{Khlopov:1985fch,Hidalgo:2017dfp,Carr:2017edp,Carrion:2021yeh,Padilla:2021zgm} where a dust epoch is modulated with scalar fields). 
It is well known, for instance, that for spherically symmetric configurations, the threshold is significantly affected by the equation of state parameter \cite{Musco:2012au,Escriva:2020tak}. The threshold of PBH formation substantially decreases due to the reduction of pressure gradients for a smaller value of $w:=P/\rho$, where $P$ and $\rho$ are the pressure and energy density, respectively.

Non-spherical effects may be relevant for some physically well-motivated scenarios where soft equations of state are considered and predictions of the mass function of PBHs may be contrasted with the gravitational wave spectrum 
of induced gravitational waves (see \cite{Domenech:2021ztg} for a review) and/or with merger rates of PBH binaries~\cite{Nakamura:1997sm,Ali-Haimoud:2017rtz}. 
The softening
can be due to a still unknown phase transition or to the standard thermal history of the Universe~\cite{Allahverdi:2020bys} when particles become non-relativistic. 
For instance, the softening of the equation of state due to 
the QCD crossover about $ \sim 10^{-5} {\rm s}$ after the Big Bang~\cite{Schmid:1998mx,Laine:2006cp,Borsanyi:2016ksw}
might cause a significant number of PBHs
in the solar mass range 
\cite{Jedamzik:1996mr,Widerin:1998my,Sobrinho:2016fay,Byrnes:2018clq,Carr:2019kxo,Jedamzik:2020ypm,Clesse:2020ghq,Juan:2022mir,Franciolini:2022tfm,Escriva:2022bwe}. 
Another possibility is from crossovers beyond standard model theories that may induce a significant softening of the equation of state \cite{Escriva:2022yaf,Escriva:2023nzn}. In addition, several scenarios with $w \neq 1/3$ have also been considered in the context of the pulsar timing array (PTA) signal \cite{NANOGrav:2023hvm,EPTA:2023xxk} and its counterpart with PBHs \cite{Domenech:2020ers,Liu:2023hpw,Liu:2023pau,Harigaya:2023pmw,Choudhury:2023fjs,Domenech:2024rks,Choudhury:2024one}.

So far, spherical symmetry is assumed to determine the threshold for PBH formation and to statistically compute the PBH mass function, which is then used to place observational constraints on the PBH scenario. 
An open question in this context is whether deviations from spherically symmetric fluctuations can significantly alter the threshold for black hole formation, particularly for a soft equation of state. 
For stiff equations of state ($w > 1/3$), it is generally expected that stronger pressure gradients would make deviations from sphericity less impactful than in the radiation case ($w = 1/3$), although a numerical confirmation would be desirable. 
If deviations from spherical symmetry do have a 
more 
significant effect 
for
softer equations of state than radiation, 
careful simulations 
are required 
for various non-spherical configurations to ensure accurate predictions.

Our investigation aims to study the dependence of the dynamics and the threshold of 
the initial amplitude for black hole formation 
on
the non-spherical parameters $e$ and $p$, in both radiation-dominated Universe ($w = 1/3$) and for a softer equation of state ($w = 1/10$). 
This will help us determine whether non-spherical effects should be considered when calculating the PBH formation threshold or if estimates assuming spherical symmetry are sufficiently accurate 
in realistic scenarios. 
For that purpose, we will employ relativistic $3+1$ numerical simulations to track the gravitational collapse of the super-horizon curvature fluctuations. 
We will consider 
the curvature fluctuation with
a monochromatic power spectrum, 
whose results are expected to apply 
to a sharp-peaked spectrum as well. 
Throughout this work, we will use geometrized units $c = G =1$.

\section{Deviation from sphericity of the initial curvature perturbation in peak theory}
\label{sec:peak_theory_nonspherical}

In this section, 
we provide some details about the peak theory 
to obtain the typical profile
of the curvature perturbation obeying a random Gaussian statistics, 
including a deviation from sphericity.
We 
briefly review and follow \cite{peak_theory}. 

Since the initial curvature perturbation is supposed to be a random Gaussian field, it
is totally characterized by the power spectrum $P_{\zeta}(k)$, defined by
\begin{equation}
    \langle \zeta(\mathbf{k}) \, \zeta(\mathbf{k'})   \rangle =   \frac{2 \pi^2}{k^3} \mathcal{P}_{\zeta} (k) (2 \pi)^3 \delta^{(3)}(\mathbf{k}+\mathbf{k'}),
\end{equation}
where $\mathbf{k}$ denotes the wavenumber vector and $k$ is its modulus. 
The variance $\sigma^2_0$ 
is given by 
\begin{equation}
\langle \zeta^2 \rangle \equiv \sigma_{0}^{2}= \int \frac{dk}{k} P_\zeta(k).
\end{equation} 
We also define the $n$-th gradient moments of the power spectrum as follows: 
\begin{equation}
\sigma_n^2 = \int k^{2n} P_\zeta(k) d\ln k.
\label{sigman}
\end{equation}
We can introduce the normalized two-point correlation function of $\zeta(\vec r)$ as
\begin{equation}
    \psi(r)\equiv \dfrac{1}{\sigma_{0}^{2}}\langle \zeta(r)\zeta(0)\rangle=\dfrac{1}{\sigma_{0}^2}\int P_\zeta(k) \sinc{(kr)} \,\dfrac{dk}{k}. 
\end{equation}

We denote
the normalized height of the peak of $\zeta$ as 
$\nu = \zeta(\vec r=0)/\sigma_0$
, and 
introduce the correlator $\gamma \equiv \sigma^2_1 /(\sigma_0 \sigma_2)$, which depends on the power spectrum $\mathcal{P}_{\zeta}$. 
Let's consider the immediate neighbour of $\zeta$ around the peak value at $\vec{r}=0$ with a Taylor expansion 
up through the
second order
as
\begin{equation}
    \zeta(\vec r) \approx \zeta(\vec{r}=0)-\sum_{l} \lambda_l \frac{r^2_l}{2},
    \label{eq:label_taylor_zeta}
\end{equation}
where $\lambda_l$ are the eigenvalues of $-\partial_i \partial_j \zeta$ and we have considered that the axes 
are oriented along its principal direction\footnote{Even if this is not the case, so that 
$ \zeta(\vec r) = \zeta(\vec{r}=0)+\sum_{i} \partial_i \partial_j \zeta \frac{r_i r_j} {2}$, 
the symmetric coefficient matrix of the second order can always be diagonalized through a rotation transformation.} with $r_l$ being the Cartesian coordinates. 
We can also arrange $\lambda_l$, so that $\lambda_1\geq\lambda_2\geq\lambda_3$ 
by considering the rotation of the axes without loss of generality. 
Around the peak, the contours of $\zeta(\vec r)$ are described by the ellipsoids 
given by $\zeta(\vec r=0)-\zeta(\vec r)\approx \sum_{l} r^2_l/(1/\lambda_l)={\rm const.}$ with the radius along each axis being proportional to 
$1/\sqrt{\lambda_l}$. 
The shape of the ellipsoid is characterized by 
\begin{equation}
e = \frac{\lambda_1-\lambda_3}{2 \sum_i \lambda_i}, \, \,\,\,\,\,\, \,\, \,\, \,\, \,\,
p = \frac{\lambda_1-2 \lambda_2 +\lambda_3}{2 \sum_i \lambda_i}. \
\label{eq:elipsoid}
\end{equation}
The parameters $e$ (``ellipticity") and $p$ (``prolateness") 
are in the range $0\leq e\leq1/2$ and $-e \leq p \leq e$. 
For instance, if 
$(\lambda_1-\lambda_2)-(\lambda_2-\lambda_3)>0\Leftrightarrow p>0$, 
the shape is oblate (pancake-like) while it is prolate (cigar-like) for $p<0$. 
It is also common to define ``spheroids" as ellipsoids with two equal eigenvalues. 


The values of $e$ and $p$ 
follow a specific probability distribution, which we will explore next. 
For instance, 
the probability density vanishes for $p=e$ and $p=-e$, 
which correspond to oblate and prolate spheroids, respectively. 
That is, approximately spheroidal configurations are improbable. 
Let's now introduce the Gaussian distributed variable 
 $\xi \equiv - \left.\nabla^{2}  \zeta\right|_{\vec r=0} / \sigma_2=\sum_i \lambda_i / \sigma_2$, 
 where we used Eq.\eqref{eq:label_taylor_zeta}. 

Solving 
Eqs.\eqref{eq:elipsoid} to obtain $\lambda_i$ in terms of $e$ and $p$, we obtain
\begin{align}
\nonumber
    \lambda_1 &= \frac{\xi \sigma_2}{3}(1+3e +p),  \\ \, 
    \label{eq:lambda_values_elipsoid}
    \lambda_2 &= \frac{\xi \sigma_2}{3}(1-2p), \, \\ \nonumber
    \lambda_3 &= \frac{\xi \sigma_2}{3}(1-3e+p). \nonumber
\end{align}
Making some algebra and using the typical trigonometric relations, we obtain\footnote{Acording to our computations, it seems there is a typo in \cite{peak_theory} at Eq.(7.4), the expansion at second order of the field should have an extra factor $3$ in the denominator.},
\begin{equation}
    \zeta(r) 
    \approx \zeta(\vec{r}=0) - \frac{\xi \,  \sigma_2}{3} \frac{r^2}{2} \left[ 1+ A(e,p)\right], 
\end{equation}
where 
\begin{equation}
\label{eq:A_function}
    A(e,p) = 3e \left[1- \sin^2 \theta (1+ \sin^2 \phi) \right]+p \left[1-3 \sin^2 \theta \cos^2 \phi \right] 
\end{equation}
with 
the convention of spherical coordinates adopted in \cite{peak_theory}: 

\begin{align}
\nonumber
    x_1 &= z = r \cos \theta,  \\ \, 
    x_2 &= x = r \sin \theta \cos \phi, \, \\ \nonumber
    x_3 &= y = r \sin \theta \sin \phi. 
\end{align}

\begin{figure}[!htbp]
\centering
\includegraphics[width=3. in]{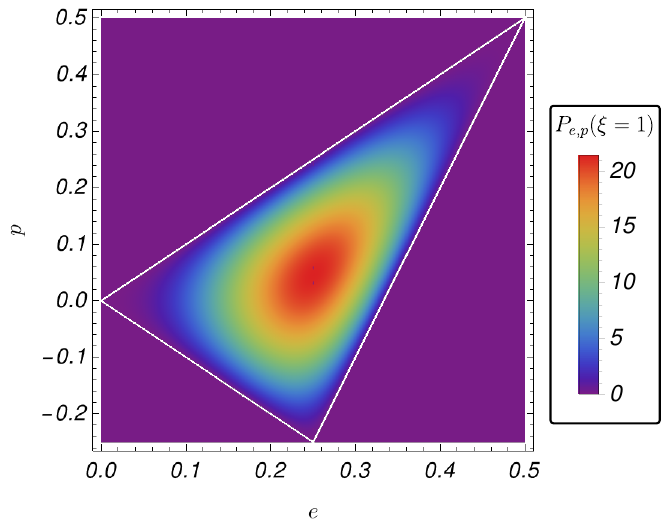}
\includegraphics[width=3. in]{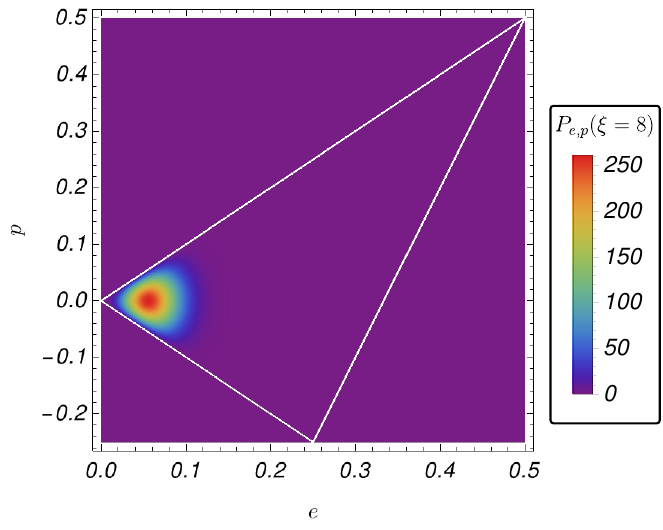}
\includegraphics[width=3.3 in]{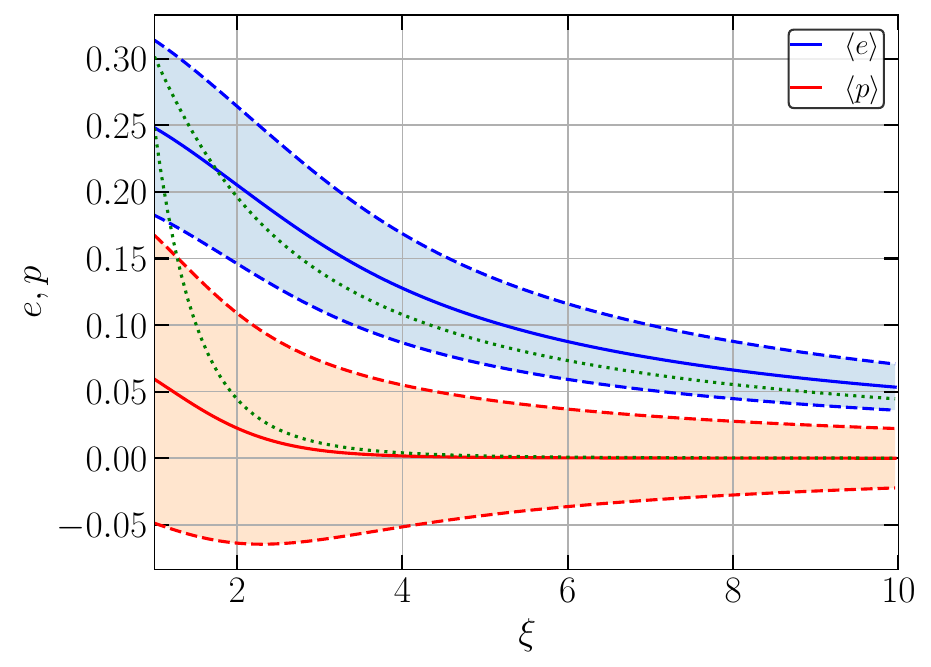}
\caption{Top panels: Provability distribution of Eq.\eqref{eq:P_ep} in $e$ and $p$ for $\xi=1$ (left panel) and $\xi=8$ (right panel).
Bottom panel: Mean values $\langle e \rangle$ and  $\langle p \rangle$ of Eq.\eqref{eq:mean_e_numerical} and \eqref{eq:mean_p_numerical} (solid line) with one standard deviation (dashed line). The green dotted lines denote the mean values $e_m$ and $p_m$ in the large $\xi$ limit given in 
Eq.\eqref{eq:mean_values_dispersione} and \eqref{eq:mean_values_dispersionp}.}
\label{fig:bbks}
\end{figure}

The parameters $e$ and $p$ follow a normalized probability distribution $\mathcal{P}_{e,p}$ for fixed $\nu$ and $\xi$, which is given by (see Eq.(7.6) in \cite{peak_theory})
\begin{align}
\label{eq:P_ep}
    \mathcal{P}_{\rm e,p}(e,p \mid \nu , \xi) &= \frac{3^{2} 5^{5/2}}{\sqrt{2\pi}} \frac{\xi^8}{f(\xi)}\exp{-\frac{5}{2} \xi^2 (3 e^2 + p^2)}W(e,p) , \\ \nonumber
    W(e,p) &= (1-2p)\left[ (1+p)^2 - (3 e )^2 \right] e (e^2-p^2) \chi(e,p),   
\end{align}
where $f(\xi)$ is given by
\begin{align}
\nonumber
    &f(\xi)=\frac{1}{2}\xi(\xi^2-3)\left(\erf\left[\frac{1}{2}\sqrt{\frac{5}{2}}\xi\right]+\erf \left[ \sqrt{\frac{5}{2}}\xi\right]\right) \\ 
\label{eq:f}
    &+\sqrt{\frac{2}{5\pi}}\left[\mathcal{C}_1(\xi)\exp{-\frac{5}{8}\xi^2}+\mathcal{C}_2(\xi)\exp{-\frac{5}{2}\xi^2}\right]
\end{align}
with 
\begin{align}
    \mathcal{C}_1(\xi) = \frac{8}{5}+\frac{31}{4}\xi^2  \ ,  \, \, \, \,  \mathcal{C}_{2}(\xi) = -\frac{8}{5}+\frac{1}{2}\xi^2,  
\end{align}
and $\chi(e,p)$ is defined as
\begin{equation}
\left\{
\begin{aligned}
\chi &= 1 , \, \, \,  \, 0 \leq e \leq 1/4 , -e \leq p \leq e, \\
\chi &= 1 , \, \, \,  \, 1/4 \leq e \leq 1/2 , -(1-3e) \leq p \leq e, \\
\chi &= 0 , \, \, \,  \, \textrm{otherwise.}
\end{aligned}
\right.
\end{equation}
This expression does not explicitly depend on $\nu$, and
the domain with a non-vanishing value is restricted to $\abs{p} \leq e$ and $e \geq 0$. Specifically, the allowed domain  $(e,p)$ is the interior of a triangle bounded by the points $(0,0)$, $(1/4,-1/4)$ and $(1/2,1/2)$. 
The mean values of $\langle e \rangle$ and  $\langle p  \rangle$ 
with 
$\mathcal{P}_{e,p}$ can be computed analytically and are shown in the Appendix \ref{sec:peak_theory_details_formulas}.
For high peaks, the mean values 
 $\langle e \rangle$ and  $\langle p  \rangle$ are small, and high peaks are more spherically symmetric than lower ones. 
 Despite that, the most likely values are not exactly zero, i.e., 
a small deviation from sphericity is likely to exist in general (see Fig.\ref{fig:bbks}).  
 Notice that the mean value of $p$ is much smaller than $e$ for large $\xi$.

We now focus on obtaining the typical profile of the curvature fluctuation $\bar{\zeta}$ characterized by 
fixed values of the parameters 
$(\nu,\xi, e, p)$. 
Euler angles are fixed, so that the coordinate axes are 
along the principle axes of $-\partial_i\partial_j\zeta$. 
Then we obtain (see (7.8) in Ref.~\cite{peak_theory})
\begin{equation}
\label{eq:typical_zeta}
\frac{\bar{\zeta}}{\sigma_0} = \frac{\nu}{1-\gamma^2}\left( \psi+R^2_s \frac{\nabla^2 \psi}{3}\right) - \frac{\xi/\gamma}{(1-\gamma^2)}\left( \gamma^2 \psi + \frac{R^2_s \nabla^2 \psi}{3} \right)+\frac{5}{2}R^2_s
\left(  \frac{\xi}{\gamma} \right) \left( \frac{\psi'}{r}-\frac{\nabla^2 \psi}{3} \right) A(e,p), 
\end{equation}
where $R_s = \sqrt{3} \sigma_1/\sigma_2$. 
Once $\nu$ and $\xi$ are fixed, 
the values of $e$ and $p$ are realized
following the probability distribution of Eq.\eqref{eq:P_ep}, 
then the peak profile is expected to be well approximated by the typical profile 
given by Eq.\eqref{eq:typical_zeta} with the given parameters of $(\nu, \xi, e, p)$. 

\begin{figure}[!htbp]
\centering
\includegraphics[width=2.6 in]{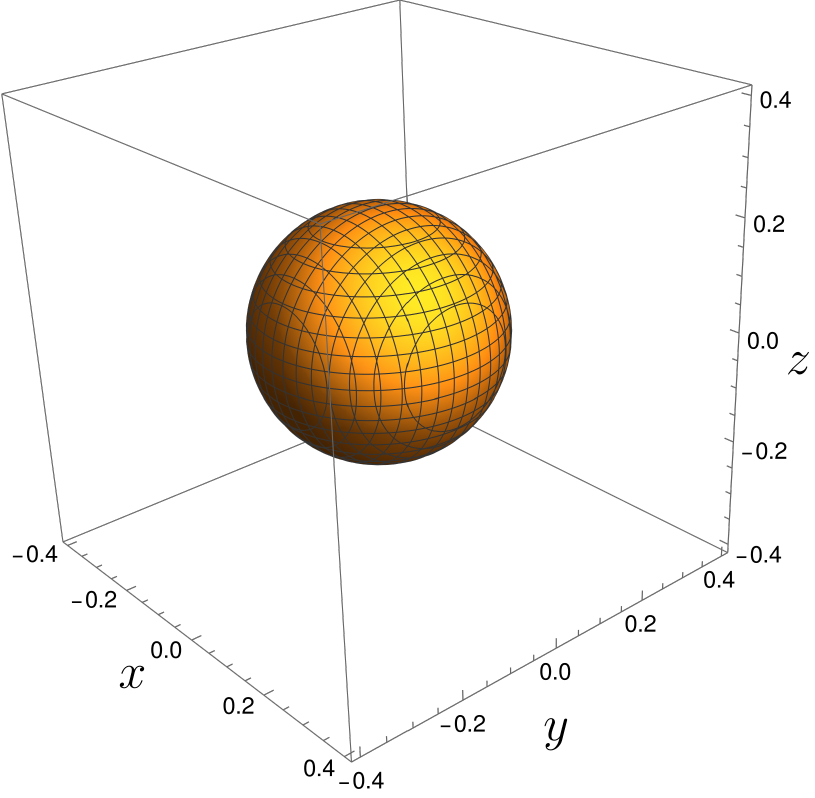}
\includegraphics[width=2.6 in]{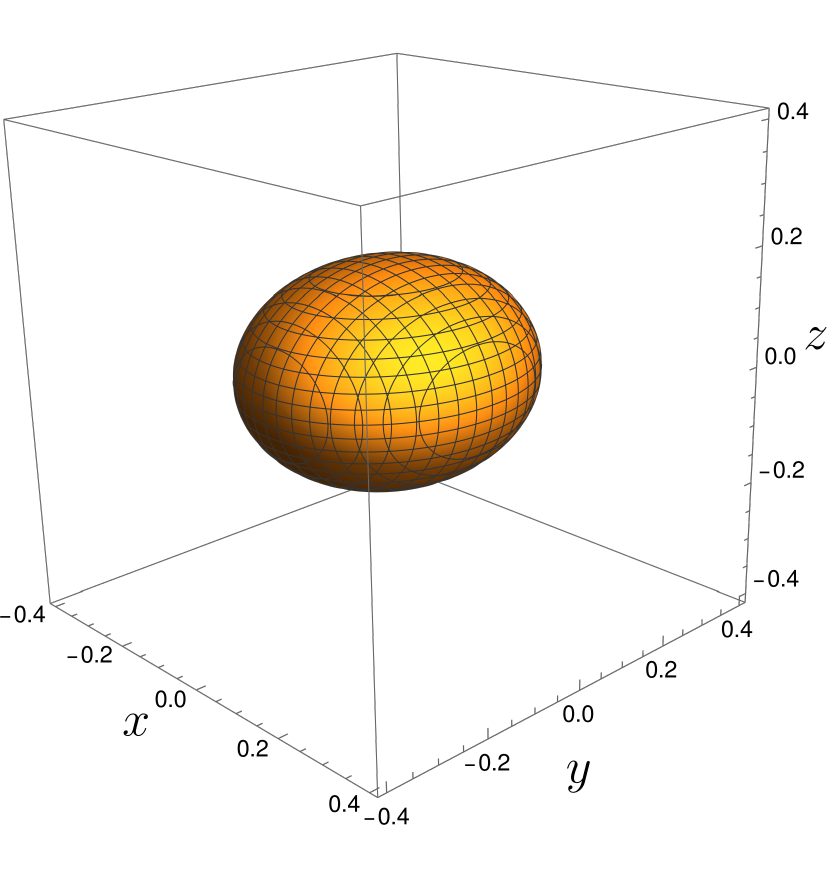}
\includegraphics[width=3. in]{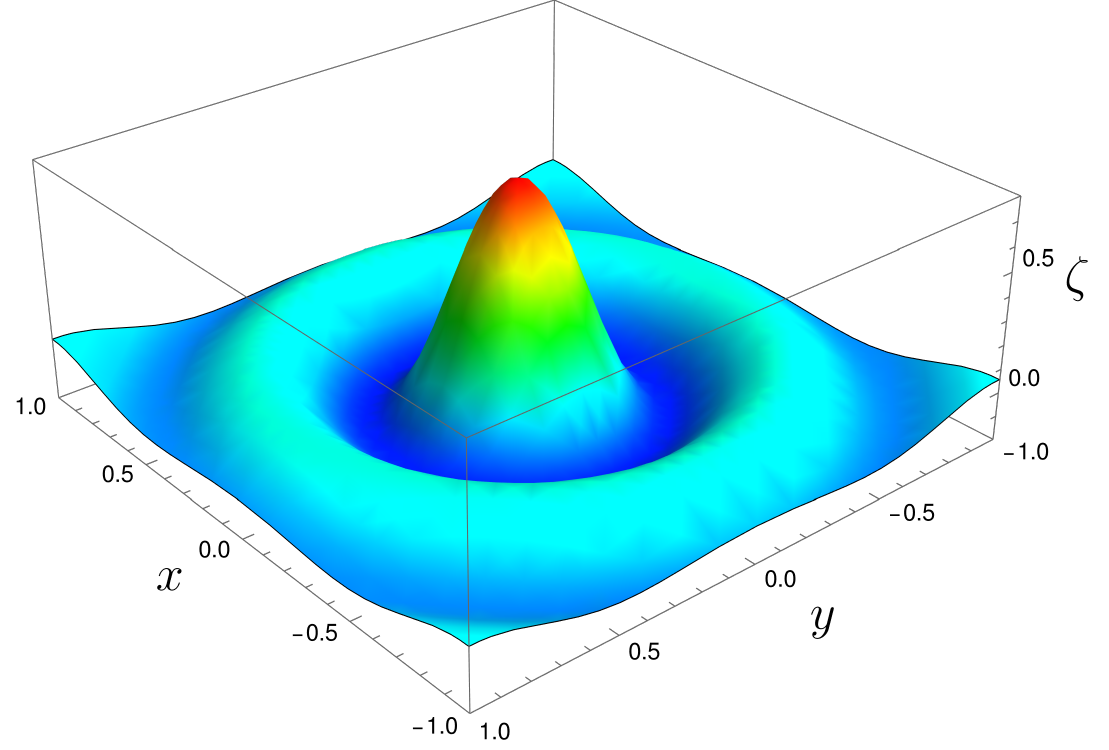}
\includegraphics[width=3. in]{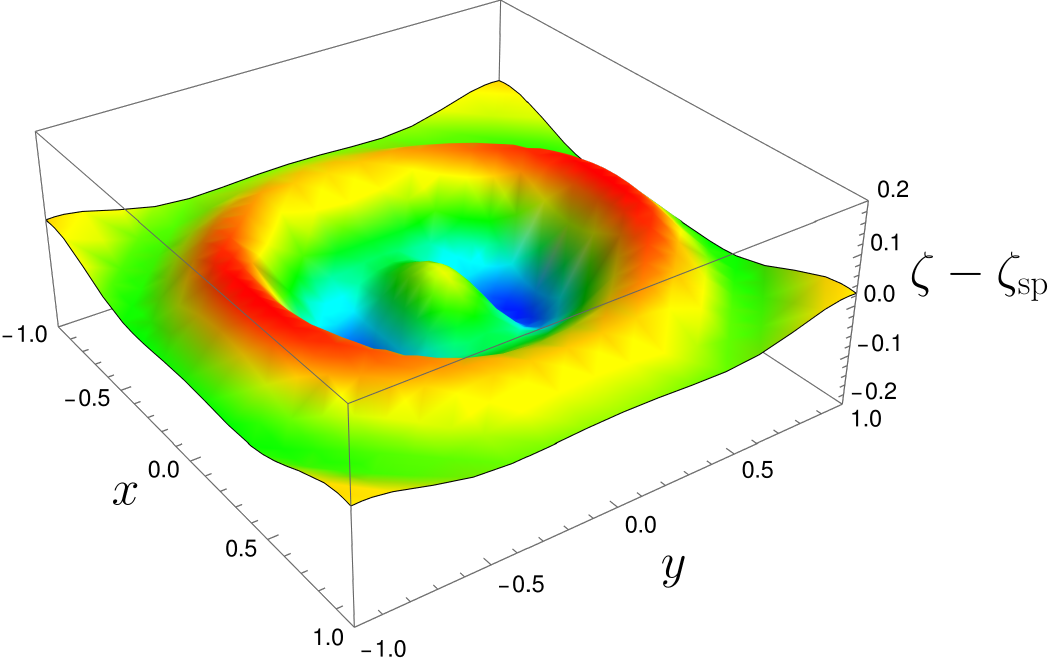}
\caption{Top panels: Contour of $\zeta$ fixing $\zeta(x,y,z) = \mu/2$ with $\mu=0.8$ for $e=p=0$ (left panel) and $e=0.1$ and $p=0.05$ (right panel). 
Bottom panels: The left panel corresponds to a $3$D plot of $\zeta$ in the plane $z=0$ for $e=p=0$, and the right panel is the difference in $\zeta$ between a spherical $\zeta_{\rm sp}$ configuration and non-spherical one with $e=0.1$ and  $p=0.05$.}
\label{fig:initial_zeta}
\end{figure}

In this work, we 
consider
a monochromatic power spectrum 
$\mathcal{P}_{\zeta} = \mathcal{A}_{\zeta} \delta(\ln(k/k_p))$ with $\mathcal{A}_{\zeta} = \sigma^2_0$, which is peaked at the scale $k_p$. 
The two-point correlation function is then given by $\psi(r) = \sinc(k_p r)$ with $\gamma=1$ and $R_s = \sqrt{3}/k_p$ since $\sigma_n = \sigma_0 k^n_p$. 
Notice that, 
in the limit of $\gamma\rightarrow 1$, the conditional probability of $\xi$ with a given value of $\nu$ is peaked precisely at $\xi=\gamma \nu=\nu$
as a Dirac delta function (see Eq.~(7.5) in Ref.~\cite{peak_theory}). 
%
Therefore we can simply set $\xi=\nu$, and obtain
%
 \begin{equation}
 \label{eq:non_spherical_zeta}
     \bar{\zeta} = \zeta_{\rm sp} + \mu \frac{5 A(e,p)}{2 k^3_p r^3} \left( 3 k_p r \cos(k_p r)+(r^2 k^2_p -3) \sin(k_p r) \right), 
 \end{equation}
where $A(e,p)$ is given by Eq.\eqref{eq:A_function} and $\zeta_{\rm sp} = \mu \sinc(k_p r)$ is the typical profile in spherical symmetry. Taking into account spherical coordinates, we can transform Eq.\eqref{eq:A_function} into the Cartesian coordinates to obtain,
 \begin{equation}
 \label{eq:A_factor}
     A(e,p) = \frac{3 e}{r^2} (z^2-y^2)+p\left[ 1-3\left(\frac{x}{r}\right)^2  \right], 
 \end{equation}
which is 
practically
more convenient 
for our numerical computation. 
Therefore, our initial condition for given values of $(\mu,e,p)$ is fixed by Eqs.\eqref{eq:non_spherical_zeta} and \eqref{eq:A_factor}. 
Although the dispersion of the profile around the typical one 
can be explicitly calculated as shown in
Eq.(7.9) of \cite{peak_theory} and in Eq.(2.10) of \cite{2020JCAP...05..022A} for the case of the monochromatic spectrum, 
we do not consider the dispersion for simplicity. 
%
%
%
%
%
%
%
%
In Fig.~\ref{fig:initial_zeta}, we show two examples of the initial curvature profile, assuming spherical symmetry with $e=p=0$ and with $e=0.1, p=0.05$, which correspond to a triaxial oblate shape. 
It should be noted that, although the spatial profile given by Eqs.\eqref{eq:non_spherical_zeta} and \eqref{eq:A_factor}
is characterized by $(\mu, e, p)$, since the probability distribution for $e$ and $p$ 
is characterized by $\xi=\nu=\mu/\sigma_0$, the typical values of $e$ and $p$ depend on the value of the standard deviation $\sigma_0$ once the value of $\mu$ is fixed.

In Appendix \ref{sec:appendix_high_peaks}, we calculate the mass function of primordial black holes under the assumption of spherical symmetry with a monochromatic power spectrum and different $w$'s. This calculation is later used to set a typical value for the fluctuation amplitude. 
We note that, 
the constraint from the total PBH fraction of dark matter $f_{\rm PBH}^{\rm tot} \leq 1$
requires 
that PBH formation is sufficiently rare, and the relevant amplitude of the peak 
is much larger than the standard deviation $\sigma_0$. 
That is, we need 
a high peak value, $\nu \gg 1$, which depends on the mass scale where $k_p$ is fixed (see Fig.\ref{fig:mass_functions_appendix}). The situation could differ for a broad spectrum, where $\gamma < 1$, but a large peak is still likely necessary.




\section{Numerical method and procedure}

We perform numerical simulations following the Baumgarte–Shapiro–Shibata–Nakamura (BSSN) formalism~\cite{PhysRevD.52.5428,Baumgarte:1998te} 
decomposing the spacetime metric into the following 3+1 form: 
\begin{equation}
ds^2 = -\alpha^2 dt^2 + \tilde \psi^4 \tilde{\gamma}_{ij}( dx^{i}+\beta^{i} dt )( dx^{j}+\beta^{j} dt ),
\label{eq:line_element}
\end{equation}
where $\alpha$ is the lapse function, 
$\beta^{i}$ is the shift vector, 
$\tilde{\gamma}_{ij}$ is the conformal metric whose determinant is given by 
that of the flat reference metric, and $\tilde\psi$ is the spatial conformal factor, which at super-horizon scales is $\tilde{\psi}=a^{1/2}\Psi = a^{1/2}\exp(-\zeta/2)$
with $a$ being the scale factor of the Friedmann-Lema\^itre-Robertson-Walker (FLRW) background. 
All these variables are functions of $t$ and $\vec x$. 
The time evolution of these variables at super-horizon scales has been analytically investigated in Ref.~\cite{Harada:2015yda}. 
The energy momentum-tensor is given by 
the following perfect fluid form:
\begin{equation}
    T_{\mu \nu} = (\rho+P) u_{\mu}u_{\nu} +P g_{\mu \nu}, 
\end{equation}
where 
$u^\mu$, $P$, and $\rho$ are the fluid 4-velocity, the pressure and the energy density, respectively. 
The equation of state is assumed to be the liner relationship $P=w \rho$ with a constant parameter $w$. 
For a later convenience, 
we define the fluid velocity relative to the Eulerian observer as 
\begin{equation}
U^\mu=u^\mu/\Gamma-n^\mu, 
\end{equation}
where $n_\mu=-\alpha\partial_\mu t$ and $\Gamma$ is the Lorentz factor defined by $\Gamma=-u^\mu n_\mu$.
We use COSMOS code written in C++~\cite{Yoo:2014boa,Okawa:2014nda}, 
which originally follows the SACRA code~\cite{Yamamoto:2008js}. The code uses the mono upstream-centered scheme for conservation laws (MUSCL) \cite{KURGANOV2000241,Shibata:2005jv} for the fluid dynamics (we refer the reader to Appendix \ref{sec:numerical_scheme} for a summary of the evolution scheme used). We basically use the same setup and parameters (like the grid spacing, the Courant–Friedrichs–Lewy number condition and others) of the code used in \cite{Yoo:2024lhp} but modify the boundary conditions to handle only one quadrant in the simulations as in \cite{Yoo:2020lmg}, taking into account the 
symmetry of our initial conditions, which makes the simulations less computationally expensive. 
First, we use the SPriBoSH code \cite{escriva_solo} to obtain efficiently the thresholds of PBH formation under the assumption of spherical symmetry (denoted by $\mu_{\rm c,sp}$) for several values of $w$. 
In this work, we consider 
$w=1/10$ and $1/3$ (soft and radiation cases), 
for what we obtain $\mu_{\rm c,sp} \approx 0.3095$ and $0.6061$, respectively (see Fig. 11 of Ref.\cite{Escriva:2022duf} for other values of $w$). 
In the COSMOS code, we obtain the consistent values of $\mu_{\rm c,sp}$ for 
the same spherical initial profile. 
This also helps us to determine the optimal number of grid points needed to achieve the desired accuracy in calculating the threshold values and make a control test.
We have found that setting $N = 60$ ($w=1/10$) and $N=100$ ($w=1/3$) grid points in each direction gives enough accuracy for our purposes. 
We check the correctness of our simulations using the averaged Hamiltonian constraint violation, for which we show detailed examples in the Appendix \ref{sec:appendix_convergence}. 

Although the number of grid points is sufficient to follow the overall dynamics of the system, it is insufficient in some cases to precisely capture the apparent horizon formation without introducing the mesh-refinement procedure adopted in Ref.~\cite{Yoo:2024lhp}. 
Instead of introducing the time-consuming mesh-refinement procedure, 
to determine whether a black hole forms or the fluid disperses in a FLRW background, we monitor the value of the lapse function at the origin. 
If the lapse function continuously decreases to a small value $\lesssim 0.1$, it suggests the eventual formation of an apparent horizon at sufficiently late times, indicating black hole formation. 
Conversely, if the lapse function exhibits a bouncing behaviour and increases 
after that, the fluid's mass excess disperses within the FLRW background, preventing black hole formation. 
A similar pattern is observed in the peak value of the energy density: a continuous increase signals black hole formation while reaching a maximum followed by a decline indicates fluid dispersion, although observing the full process of dispersion of the fluctuation requires much more computational time than what we have used. This behaviour is quite universal, appearing in both spherically symmetric and non-spherical simulations, as noted in various studies \cite{Niemeyer:1999ak,Musco:2004ak,escriva_solo,Yoo:2020lmg,2022PhRvD.105j3538Y,Uehara:2024yyp}. 
In some cases, our simulation failed 
with a significant violation of the Hamiltonian constraint 
at a late time. 
However, 
the behavior of the lapse function until the significant violation of the Hamiltonian constraint allows us
to reliably infer whether the fluctuation will ultimately collapse 
to form an apparent horizon.




In the initial profile \eqref{eq:non_spherical_zeta}, we set $k_p=10 L^{-1}$ and $H_0:=H(t_0)=50 L^{-1}$ with $L$ being the coordinate length along each axis of the numerical domain given by $x \in [0, L]$, $y \in [0, L]$ and $z \in [0, L]$. 
The other initial quantities of the BSSN formalism are computed according to \cite{Harada:2015yda, Yoo:2020lmg}. 
The initial time corresponding to $H_0$ is given by $t_0=\alpha_w / H_0$ with $\alpha_w=2/(3w+3)$, and the initial energy density of the background Universe is 
given by $\rho_b(t_0)=3 H^2_0 / (8 \pi)$. 
We also define the time of horizon crossing $t_H$ of the mode $k_p$ as 
$t_H =t_0 (H_0/k_p)^{1/(1-\alpha_w)}$.



Practically,
it is necessary to implement a window function in the curvature $\zeta$ to match the boundary condition $\Psi(r \rightarrow \infty) = 1$ at the last points of the grid. 
This is particularly relevant for the case of the $\sinc$ profile, which has small oscillations for a large radius. 
We have found that the following functional form works well for our purposes: 
\begin{equation}
    W(r) = \exp{ - \left( \frac{r-r_W}{\sigma_W} \right)^{\alpha_W}  }
    \label{eq:window_function}
\end{equation}
with values $\sigma_W = 0.09L$, $r_W \approx 0.77L$ and $\alpha_W = 4$. 
Effectively, the window function will change the functional form of Eq.\eqref{eq:non_spherical_zeta}, but only for a large radius $r>r_W$. 
As found in \cite{universal1}, the threshold of formation is mainly affected by the shape around the 
radius at which the compaction function takes the maximum value.%
\footnote{The definition of the compaction function, which is defined for spherically symmetric fluctuations goes beyond the scope of the paper. 
We refer the reader to \cite{Shibata:1999zs} for the original definition. }
This radius is much smaller than the radius  
where the window function is introduced. 
This expectation is confirmed in spherical symmetry by comparing the threshold 
with that obtained by the spherically symmetric code SPriBHoS without introducing the window function. 
The difference in the threshold values is within $0.01\%$. 
Although our case is now a simulation beyond spherical symmetry, we expect such consideration to also hold\footnote{The situation would differ for the behaviour of the PBH mass, where the shape of the profile beyond the maximum of the compaction function may have a significant effect \cite{Escriva:2021pmf,Escriva:2023qnq}.}.

To investigate the impact of non-spherical configurations on the critical threshold values, we fix the 
value of $\mu>\mu_{\rm c,sp}$, where $\mu_{\rm c,sp}$ is the threshold value in spherical symmetry. 
We consider the mass spectrum derived with spherical symmetry 
to find a typical value $\mu_{\rm t}$. 
As is shown in Appendix\ref{sec:appendix_high_peaks}, the PBH mass function 
takes the maximum at a certain value of the mass. 
Since the mass is given as a function of the initial amplitude $\mu$ for the monochromatic spectrum\cite{Kitajima:2021fpq,Yoo:2022mzl}, 
we can find the value of $\mu$ for which the mass function takes the maximum. 
We use this value as a typical value of $\mu$ throughout this paper. 
For $\mu=\mu_{\rm t}$, increasing the non-sphericity characterized by 
the two parameters $e$ and $p$, we can find a boundary between 
black hole formation and dissipation. 
That is, we can divide the $2$-dimensional parameter region of $e$ and $p$
into those two cases with the boundary curve described as 
$\mu_{\rm t}=\mu_{\rm c}(e,p)$, where $\mu_{\rm c}(e,p)$ is the function of $e$ and $p$ 
which gives the threshold for a given set of $e$ and $p$. 
The specific value of $\mu_{\rm t}$ is given by $0.6176$ for $w=1/10$ and $0.313$ for $w=1/3$.

The dynamics of the system characterized by the parameter set $(\mu,e,p)$ 
is independent of the value of the standard deviation $\sigma_0$. 
However, once the value of $\mu$ is fixed, the probability distribution of $e$ and $p$ depends on $\sigma_0$. 
Since the PBH mass function also depends on $\sigma_0$, 
we fix the value of $\sigma_0$ by imposing $f^{\rm tot}_{\rm PBH}=1$ for the spherically symmetric case. 
In Table \ref{table:parametters},  we summarize those parameters.




\begin{table} 
 \centering
\begin{tabularx}{150mm}{cc|CCCC}
\hline
$w$ & $k_p[\textrm{Mpc}^{-1}]$ & $\mu_{\rm c,sp}$ & $\mu_{\rm t}$ & $\sigma_0/10^{-2}$ & $\nu_{\rm c}$ \\ \hhline{==|====}
$1/10$& $10^{13.5}$  & $0.30948$ &  $0.313$ &  $3.524 $ & $8.781$ \\ \hline
$1/10$ & $10^{7}$ & $0.30948$ &  $0.313$ &  $3.876 $ & $7.984$ \\ \hline
$1/3$ & $10^{13.5}$ & $0.60613$ &  $0.6176$ & $6.979 $ & $8.685$ \\ \hline
$1/3$ & $10^{7}$& $0.60613$ &  $0.6176$ & $8.976 $ & $6.753$  \\ \hline
\end{tabularx}
\caption{Parameters $\mu_c$, $\mu_{t}$, $\sigma_0$, $\nu_c$  satisfying $f_{\rm PBH}^{\rm tot}\simeq1$ for both cases of $w$ with $k_p=10^{13.5}{\rm Mpc^{-1}}$ (asteroid mass range) and $k_p=10^{7}{\rm Mpc^{-1}}$ (solar mass range).  
}
\label{table:parametters}
\end{table}


\section{Numerical results}
In this section, we present the main numerical results of our work. First, we show the dynamical evolution of the gravitational collapse for a few representative cases, and later, we focus on the threshold study in terms of the non-spherical parameters $e$ and $p$.
\subsection{Dynamical evolution of the non-spherical gravitational collapse}
\label{sec:dynamics_collapse}

We analyze two specific cases to examine the dynamics of gravitational collapse. In the first case, the fluctuation exceeds the critical threshold $[\mu_{\rm t} > \mu_{\rm c}(e,p)]$, leading to black hole formation. In the second case, the fluctuation remains below the threshold $[\mu_{\rm t} < \mu_{\rm c}(e,p)]$, resulting in the dispersion of the fluctuation.  
To study the dynamical evolution of the fluctuation, we plot the ratio of $\rho/\rho_b(t)$ 
and the 4-velocity of the fluid $U^{i}$. 
We focus on the case of radiation-dominated Universe $w=1/3$ and refer the reader 
to Appendix \ref{sec:appendix_suplemental_figures} for the $w=1/10$ case. 
We found the qualitative behaviour to be the same for both $w=1/3$ and $1/10$
except that
the collapsing time is longer for $w=1/10$, 
which is consistent with the case of spherical simulations \cite{Escriva:2020tak}.

\begin{figure}[t]
\centering
\includegraphics[width=1.5 in]{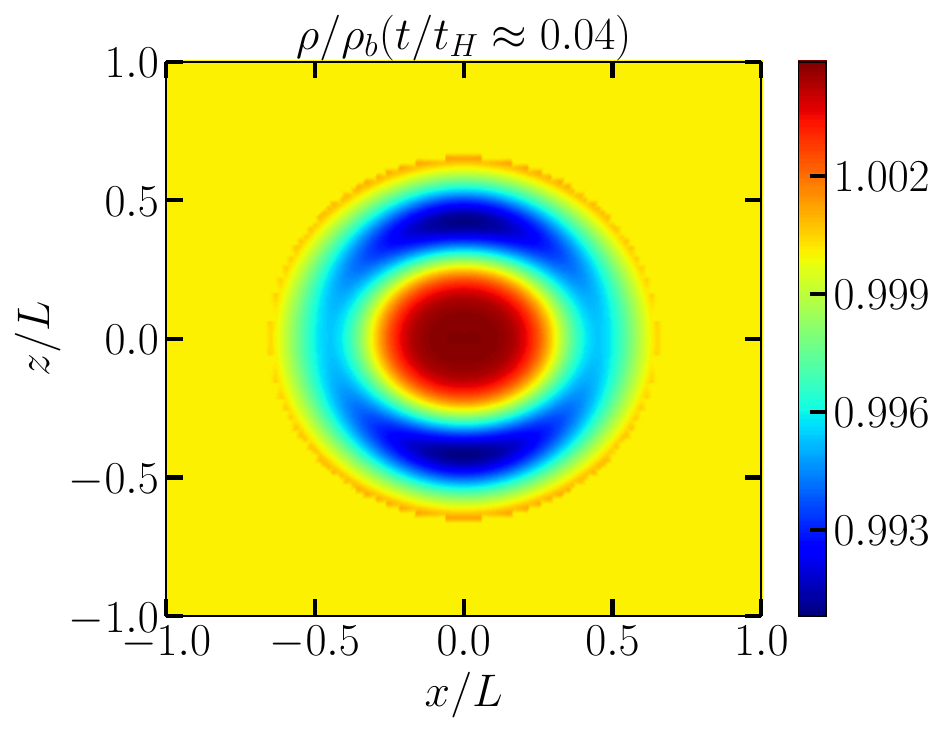}
\hspace*{-0.3cm}
\includegraphics[width=1.5 in]{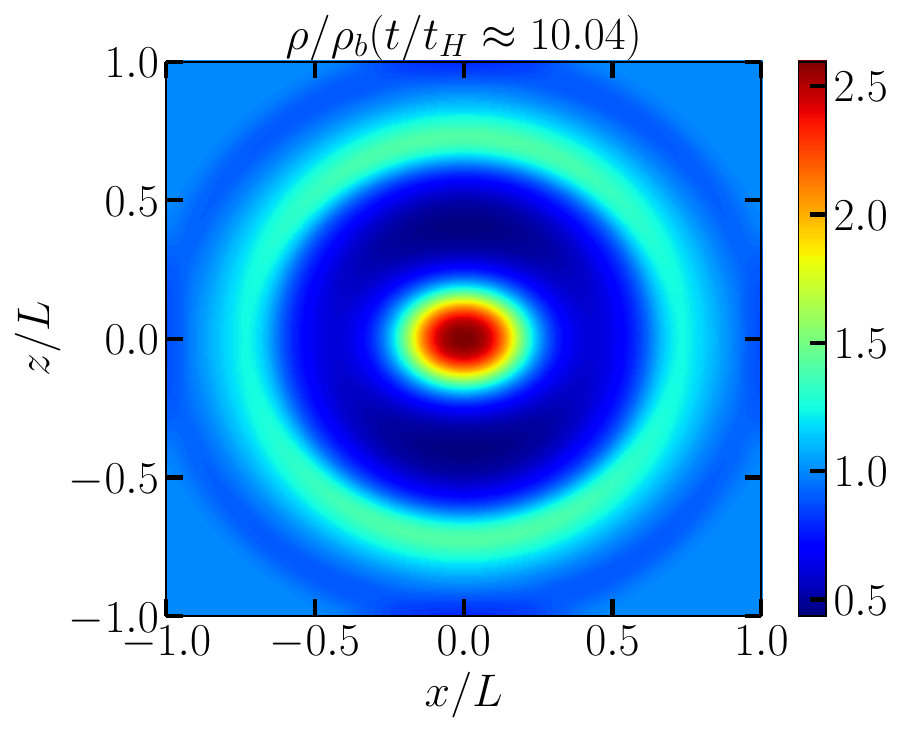}
\hspace*{-0.3cm}
\includegraphics[width=1.5 in]{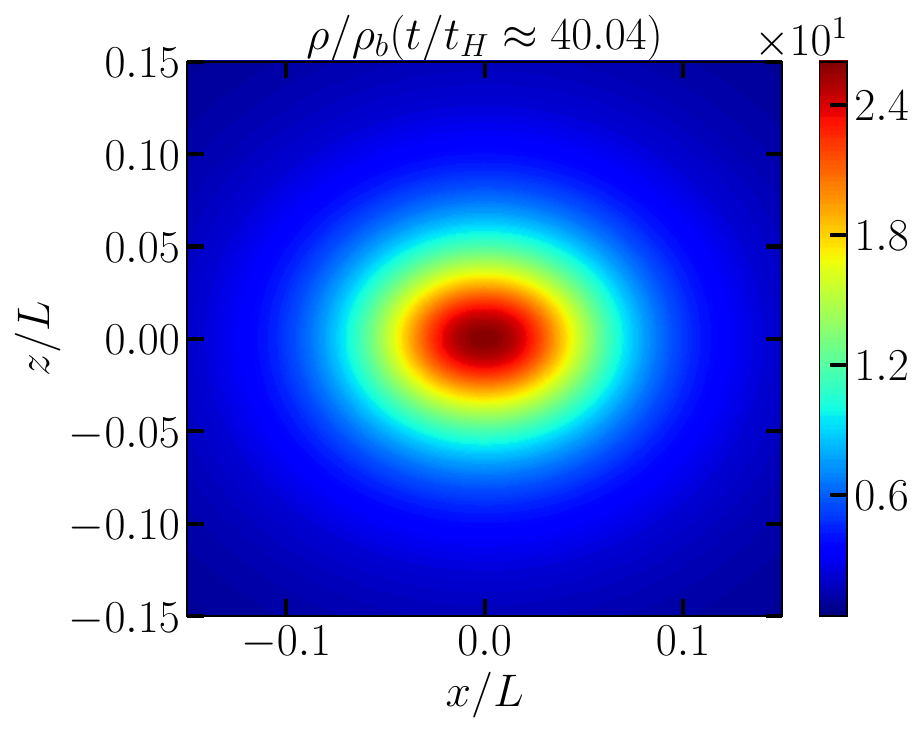}
\hspace*{-0.3cm}
\includegraphics[width=1.5 in]{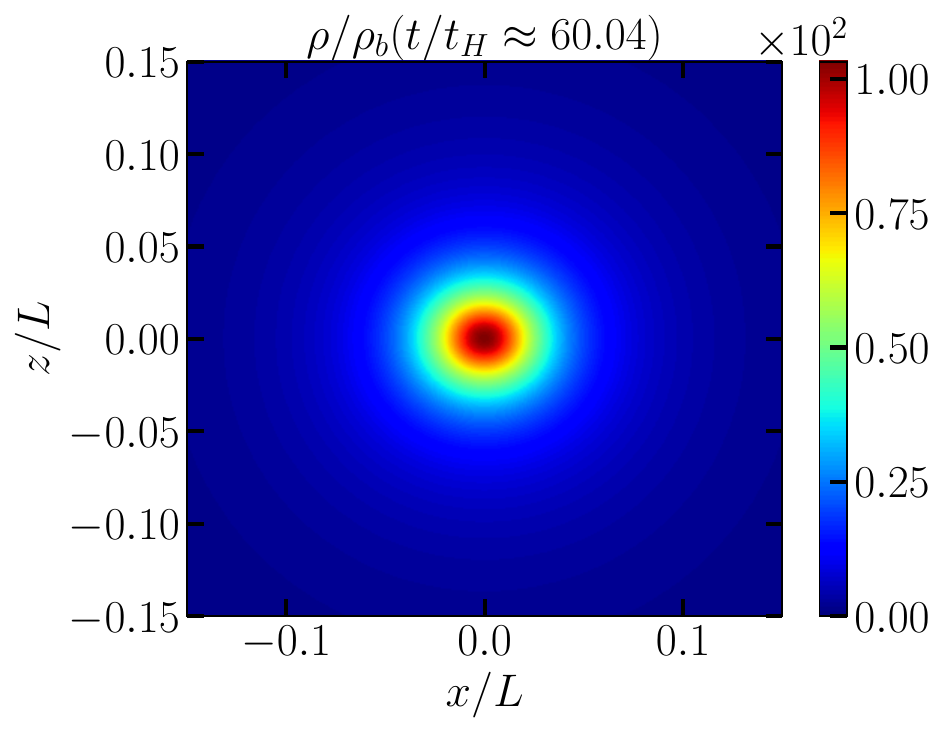}
\hspace*{-0.3cm}
\includegraphics[width=1.5 in]{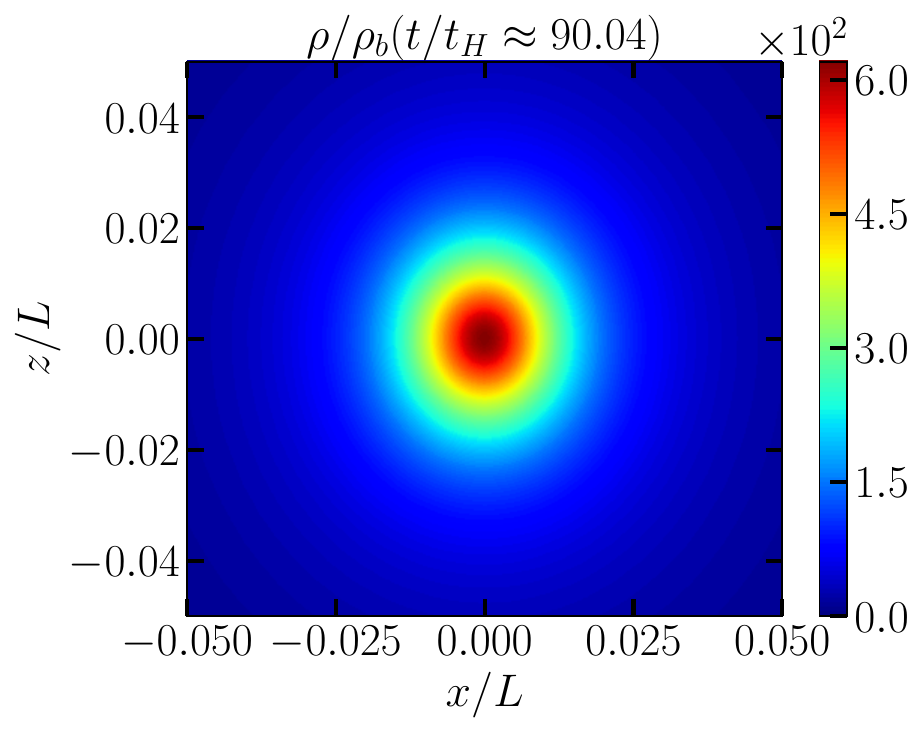}
\hspace*{-0.3cm}
\includegraphics[width=1.5 in]{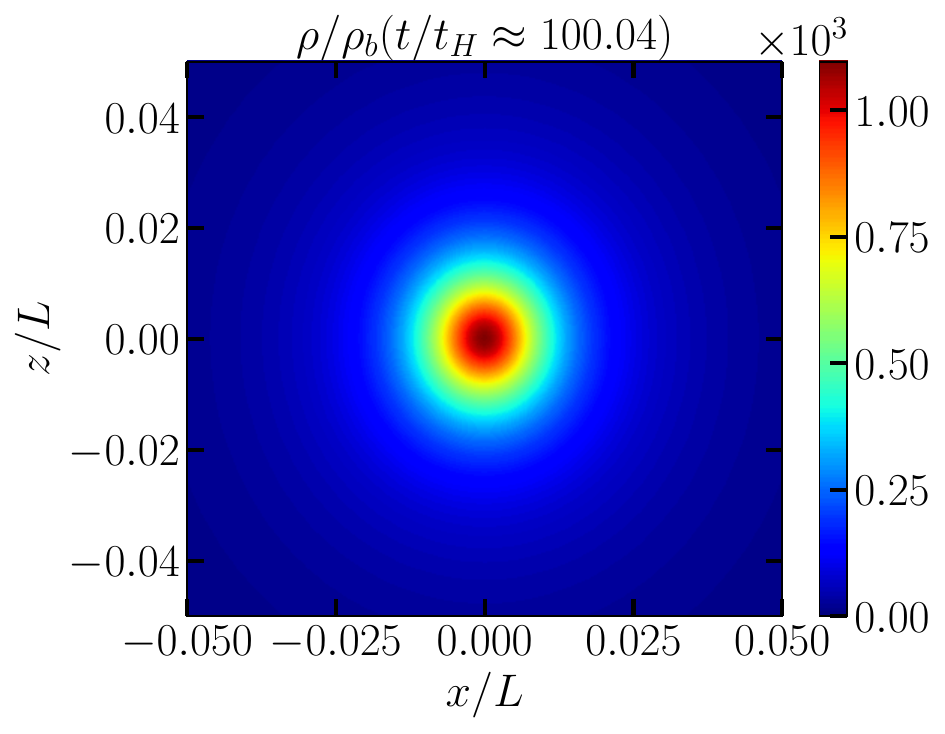}
\hspace*{-0.3cm}
\includegraphics[width=1.5 in]{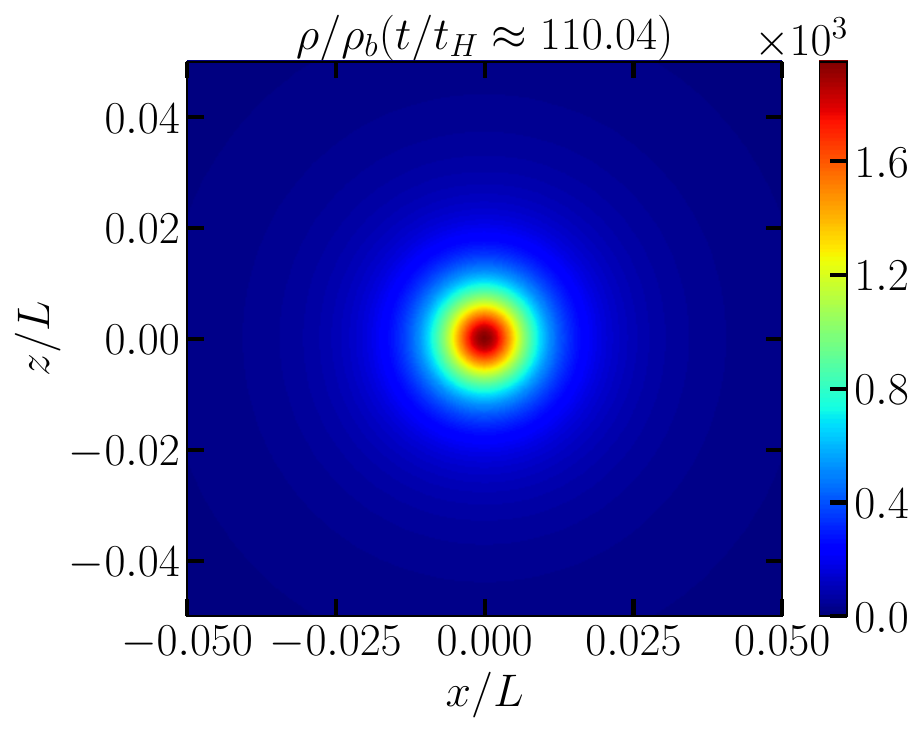}
\hspace*{-0.3cm}
\includegraphics[width=1.5 in]{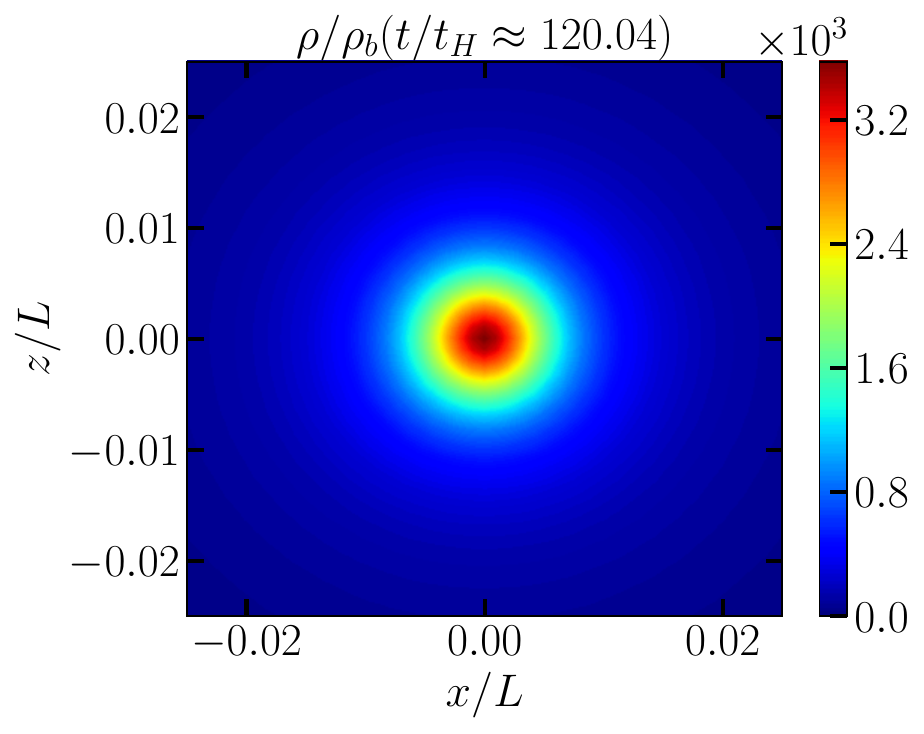}
\caption{Snapshots of the evolution of $\rho/\rho_b$ in the plane $y=0$ 
for 
$e=0.08$ and $p=0$ with $w=1/3$.}
\label{fig:energy_density_ratio_collapse_rad}
\end{figure}

Let's first consider a case where the fluctuation collapses into a black hole, 
in which
the deviation from sphericity is not sufficiently large to avoid black hole formation. In particular, we choose $e=0.08$ and $p=0$, which correspond to the eigenvalues $\lambda_1 /( \nu \sigma_2)= 1.24$ , $ \lambda_2/( \nu \sigma_2)=1$, $ \lambda_3/( \nu \sigma_2) =0.76 $ 
following Eq.~\eqref{eq:lambda_values_elipsoid}. In Fig.~\ref{fig:energy_density_ratio_collapse_rad}, we show the evolution of the energy density ratio $\rho/\rho_{b}$ for different times.
One can see 
a characteristic ellipsoidal shape that changes over time. 
Initially, the shape of the energy density is shorter in the $z$-direction than $x$ and $y$, since the length of fluctuation size in the 
$i$-axis goes like $\sim 1/\sqrt{\lambda_i}$, 
being $\lambda_1$ the largest eigenvalue (it should be noted that we follow
the convention adopted in \cite{peak_theory}, $(1,2,3) \equiv (z,x,y)$ axes). 

\begin{figure}[!htbp]
\centering
\includegraphics[width=1.5 in]{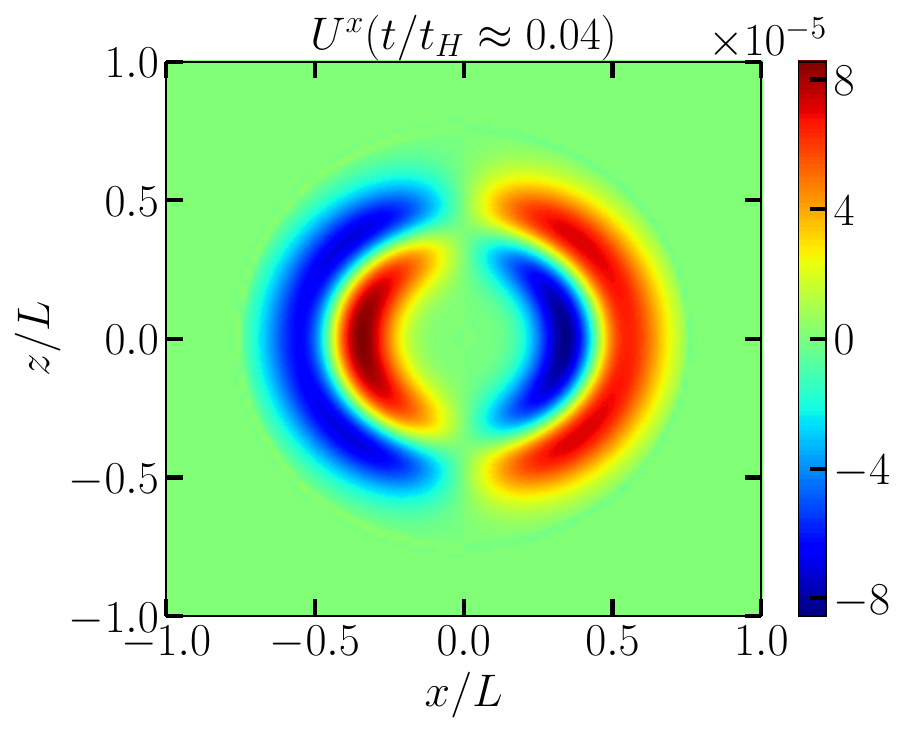}
\hspace*{-0.3cm}
\includegraphics[width=1.5 in]{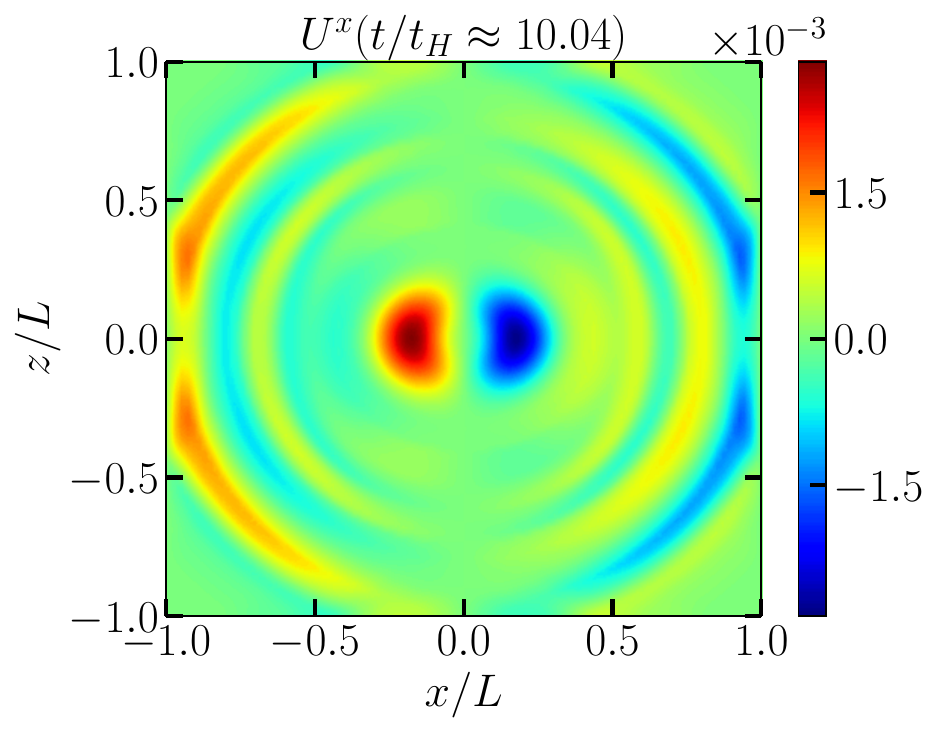}
\hspace*{-0.3cm}
\includegraphics[width=1.5 in]{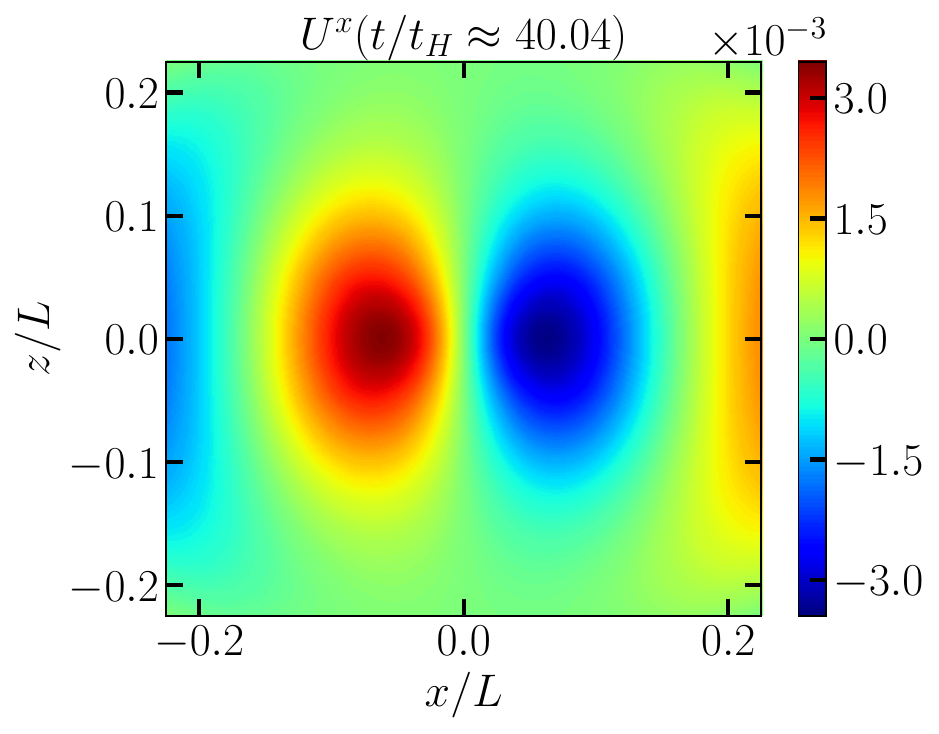}
\hspace*{-0.3cm}
\includegraphics[width=1.5 in]{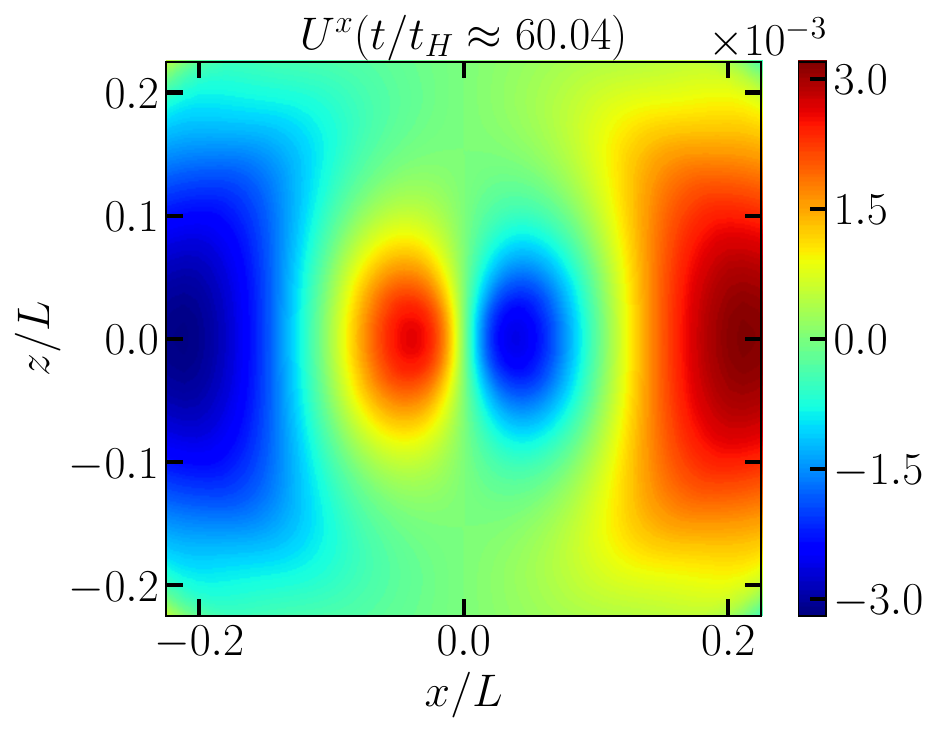}
\hspace*{-0.3cm}
\includegraphics[width=1.5 in]{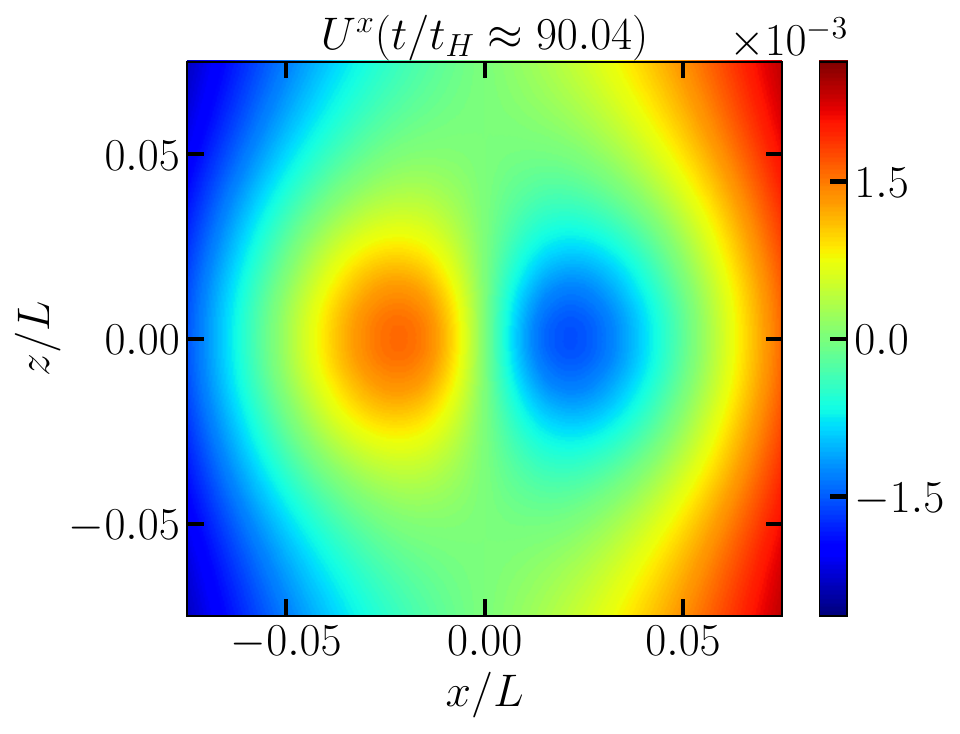}
\hspace*{-0.3cm}
\includegraphics[width=1.5 in]{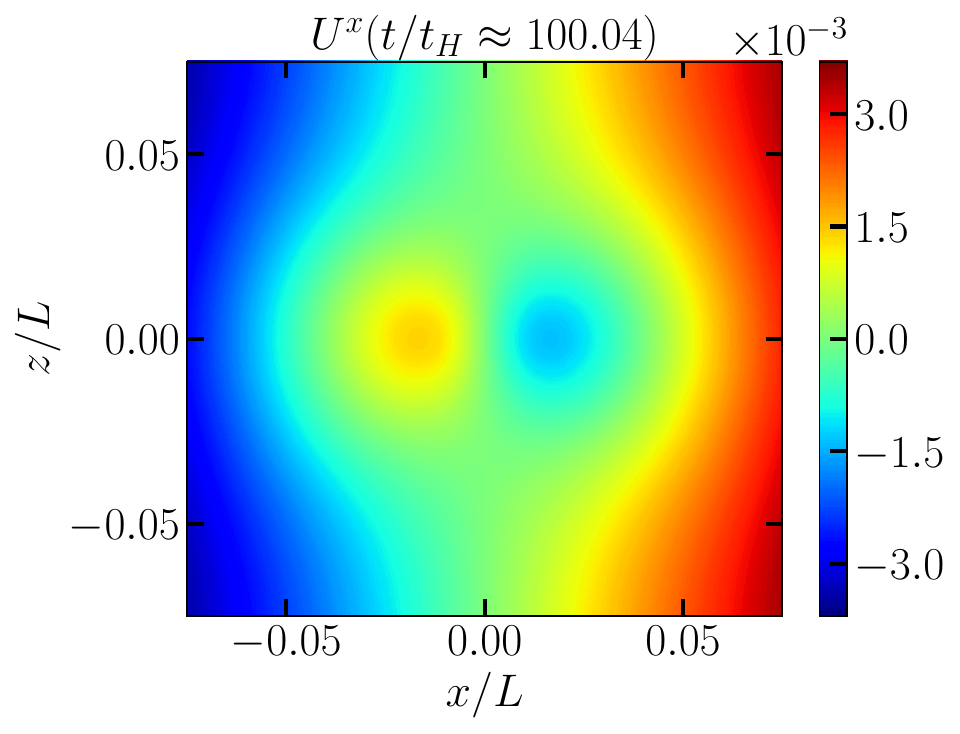}
\hspace*{-0.3cm}
\includegraphics[width=1.5 in]{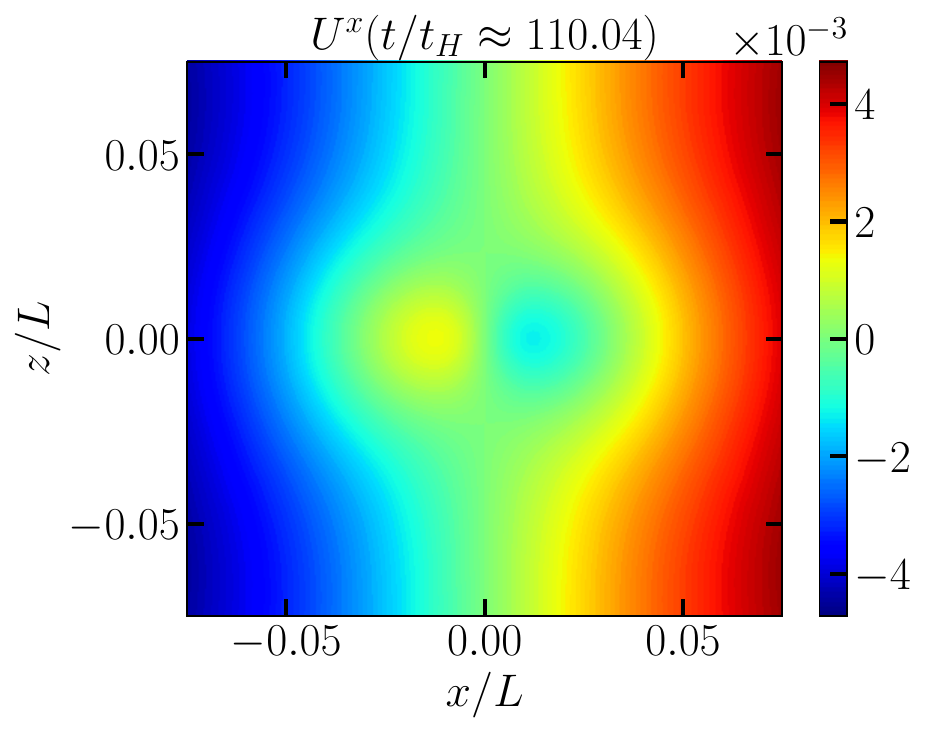}
\hspace*{-0.3cm}
\includegraphics[width=1.5 in]{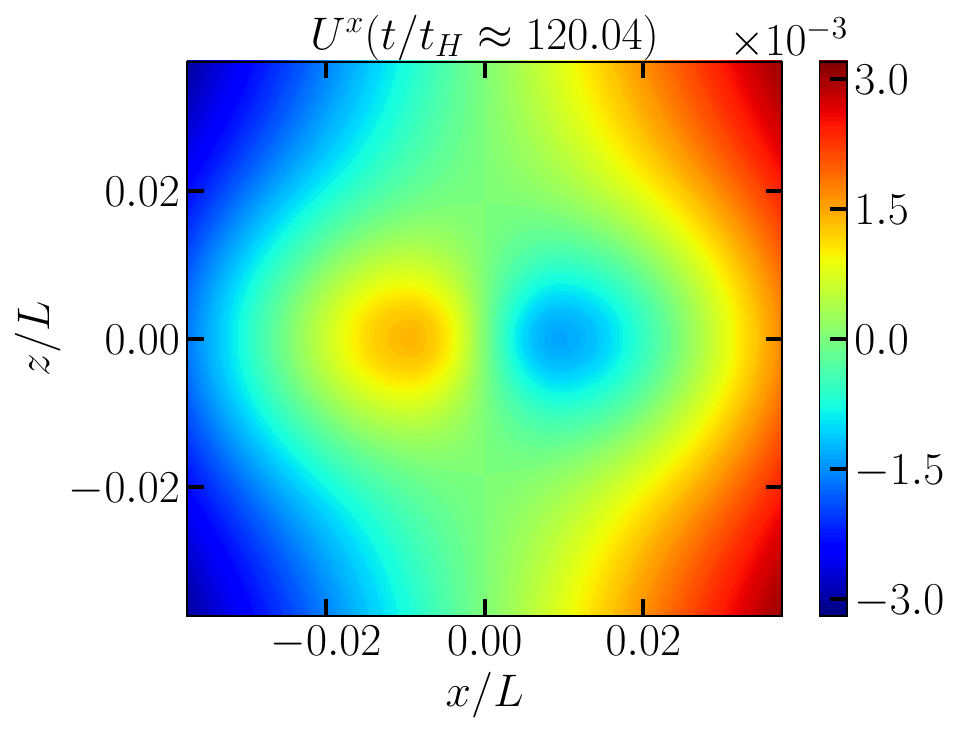}
\caption{Snapshots of the evolution of the fluid velocity $U^{x}$ in the plane $y=0$ 
for
$e=0.08$ and  $p=0$ with $w=1/3$.}
\label{fig:vel_collapse_rad_x}
\end{figure}

\begin{figure}[t]
\centering
\includegraphics[width=1.5 in]{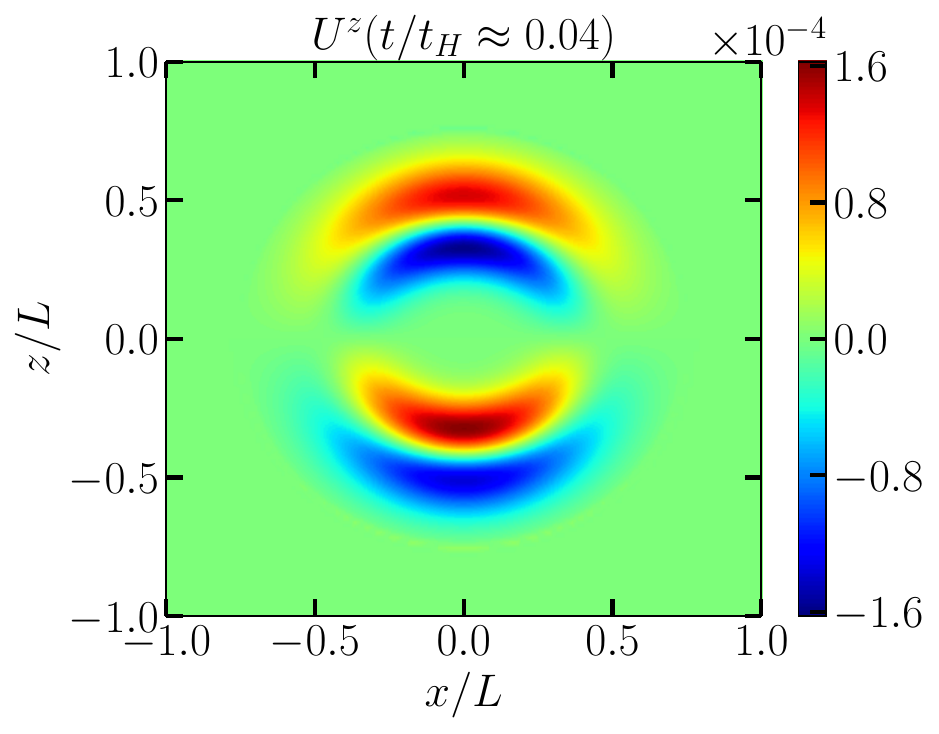}
\hspace*{-0.3cm}
\includegraphics[width=1.5 in]{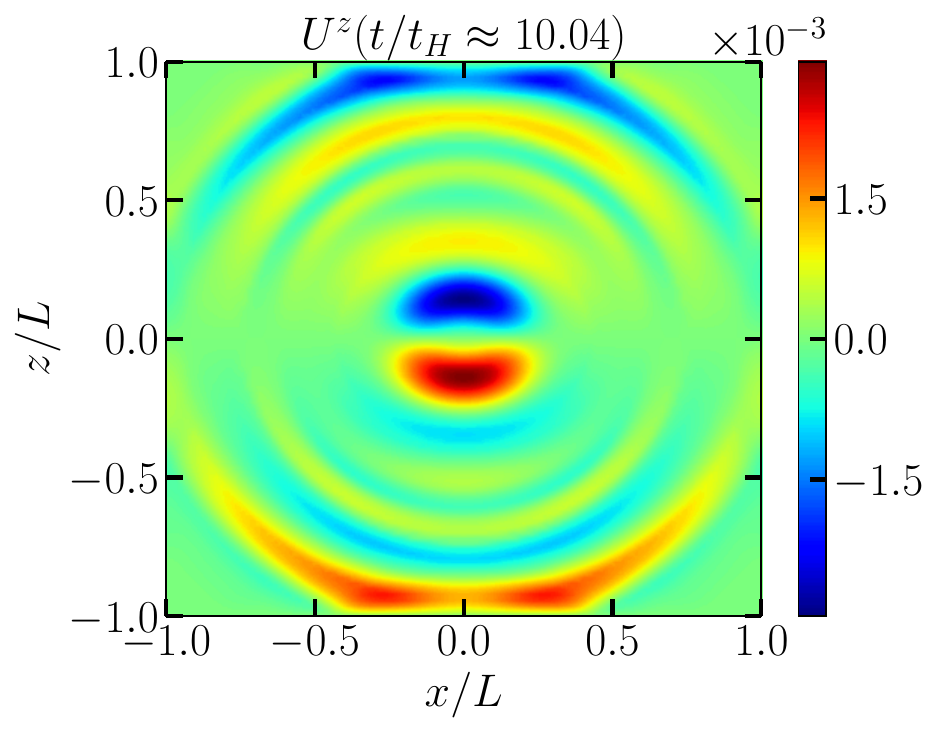}
\hspace*{-0.3cm}
\includegraphics[width=1.5 in]{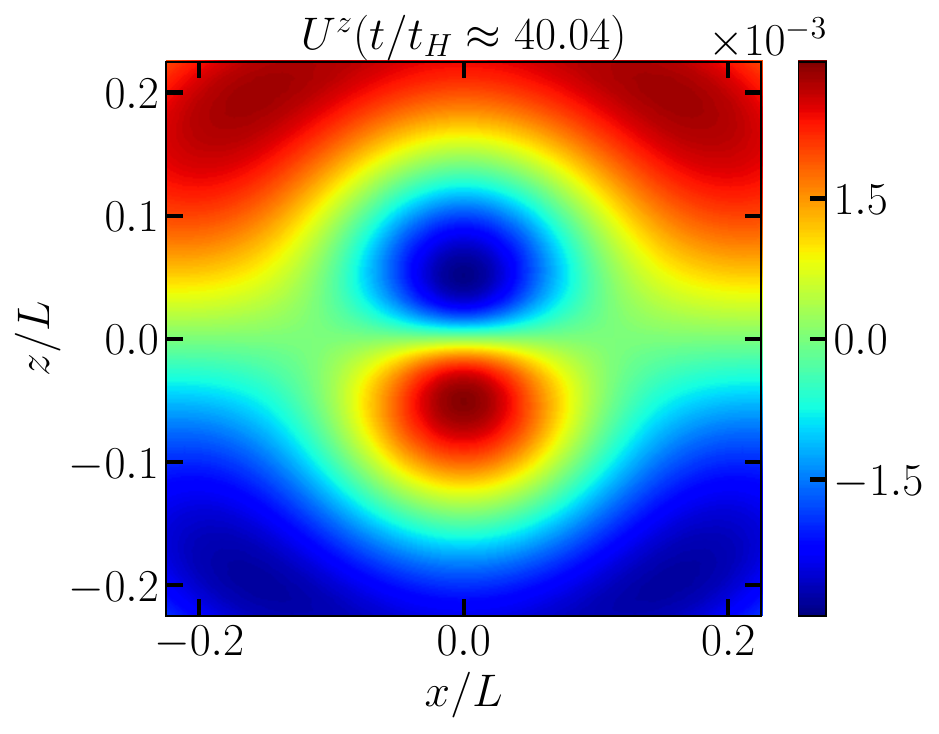}
\hspace*{-0.3cm}
\includegraphics[width=1.5 in]{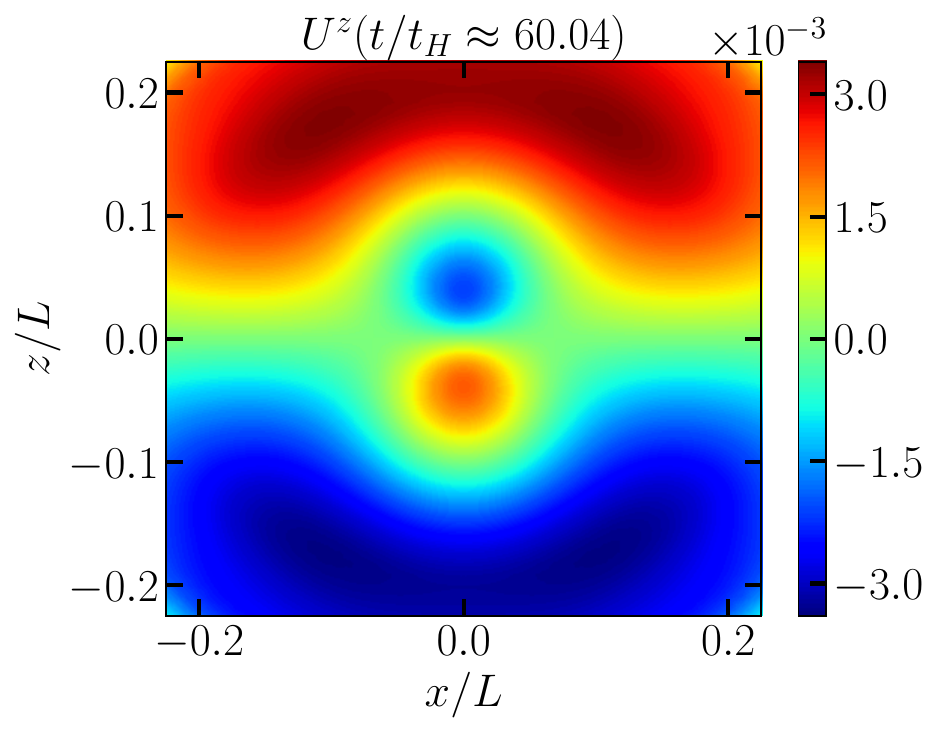}
\hspace*{-0.3cm}
\includegraphics[width=1.5 in]{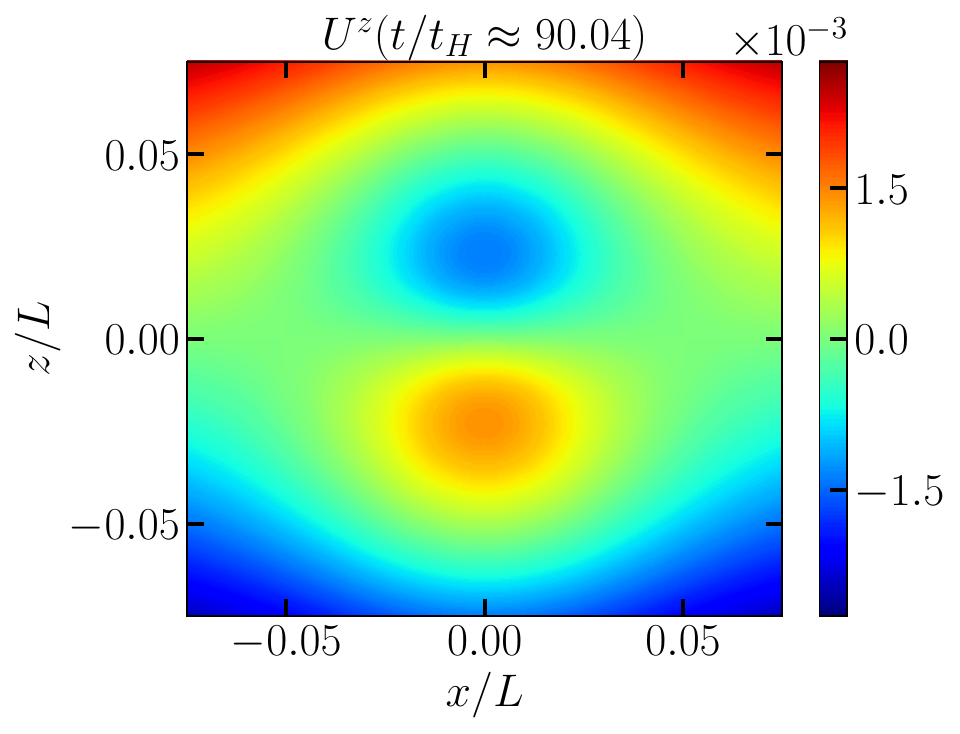}
\hspace*{-0.3cm}
\includegraphics[width=1.5 in]{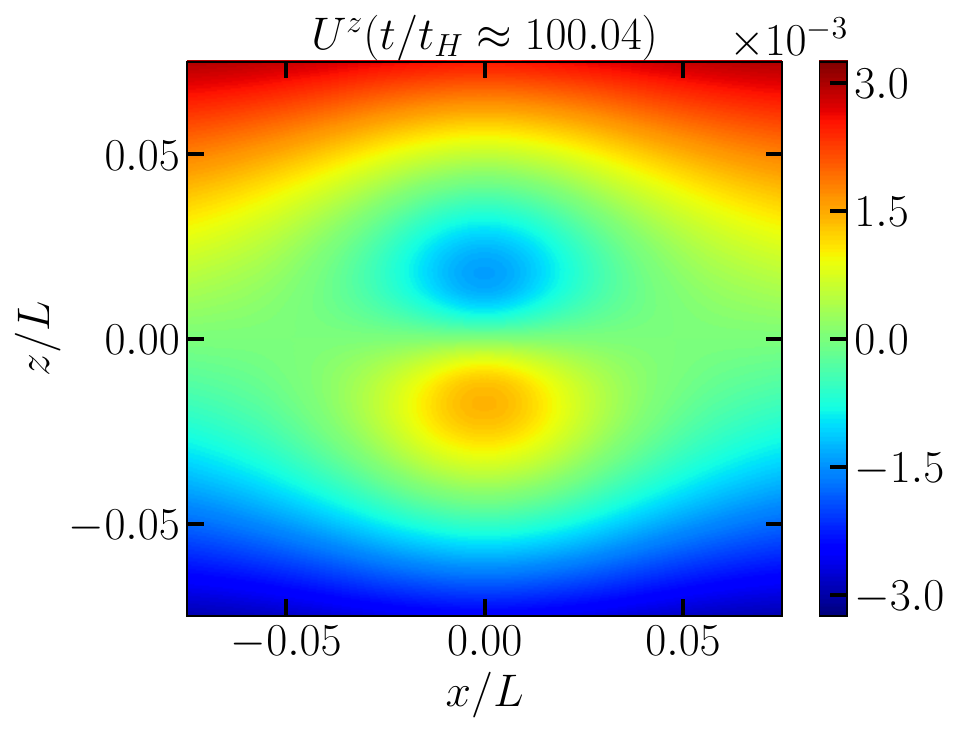}
\hspace*{-0.3cm}
\includegraphics[width=1.5 in]{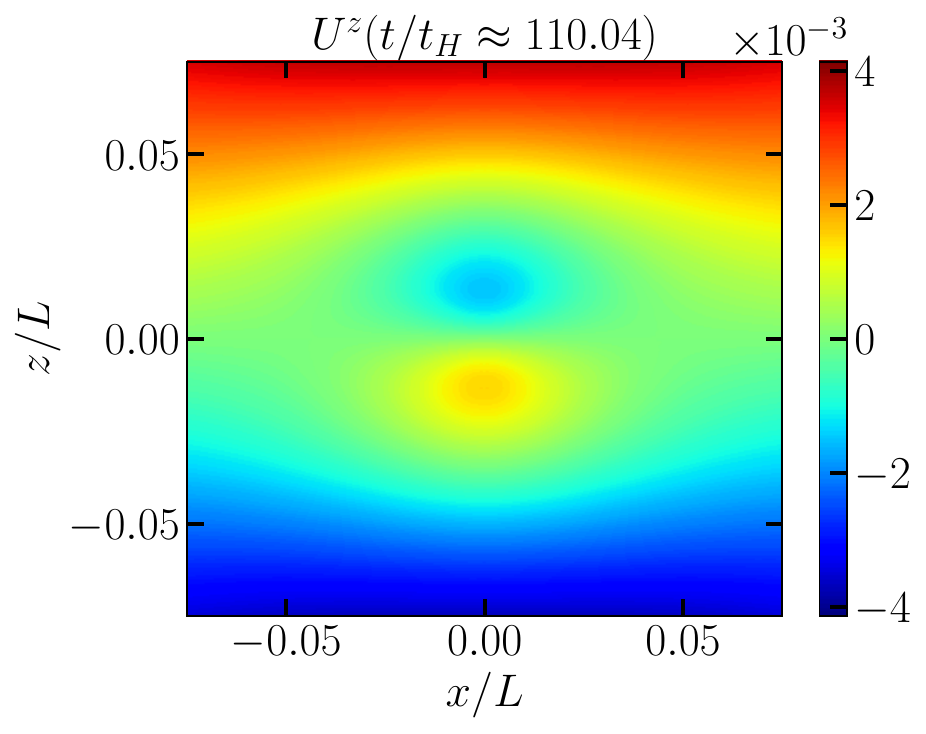}
\hspace*{-0.3cm}
\includegraphics[width=1.5 in]{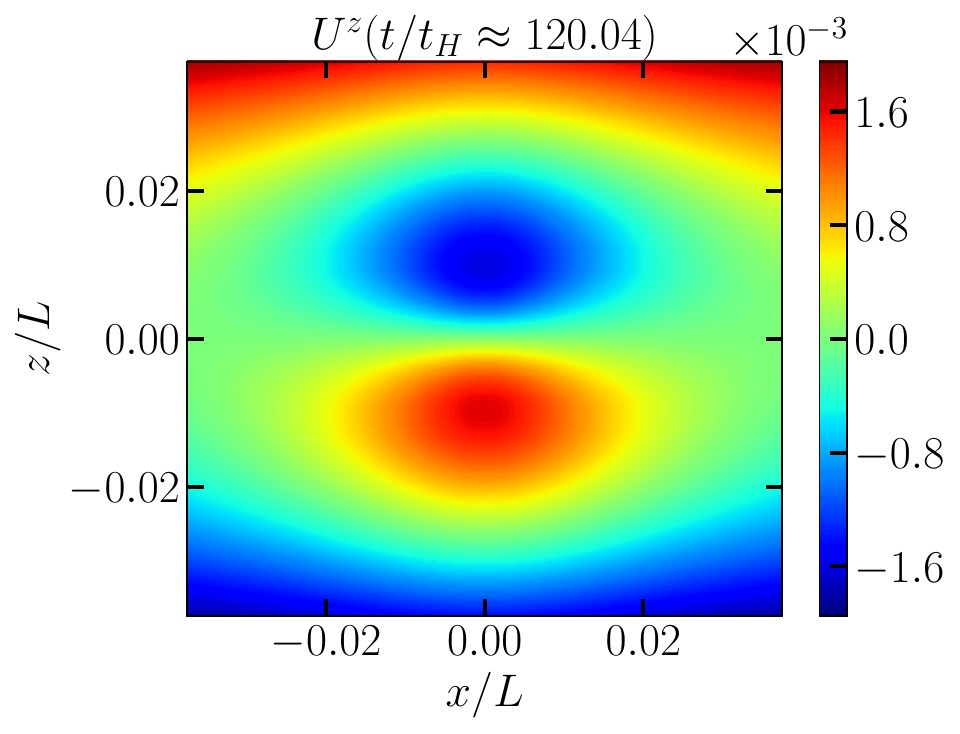}
\caption{Snapshots of the evolution of the fluid velocity $U^{z}$ in the plane $y=0$ 
for
$e=0.08$ and $p=0$ with $w=1/3$.}
\label{fig:vel_collapse_rad_z}
\end{figure}

In Figs.~\ref{fig:vel_collapse_rad_x} and \ref{fig:vel_collapse_rad_z},  
we plot 
the velocity $U^{x}$ and  $U^{z}$ on the $x$-$z$ plane,  
respectively, 
where
we observe a highly non-spherical distribution.
From the velocity plots of $U^{x}$ and  $U^{z}$,
we observe that the collapse initially progresses slightly faster along the $z$-axis compared to the $x$-axis, as indicated by the higher collapse velocity in $U^{z}$ (see the panels of $t/t_H\approx0.04$). 
However, presumably because of the larger pressure gradient along the $z$-axis, the 
contraction along the $x$-axis overtakes the contraction along the $z$-axis (see the panels of $t/t_H\approx40.04$ and compare the values of $U^x$ and $U^z$). 
Subsequently, the initial shape transitions from 
horizontally long to vertically long 
while preserving the ellipsoidal 
shape
and reducing the size of the overdensity region (see the panels of $t/t_H=90.04$ and those after that in Fig.~\ref{fig:energy_density_ratio_collapse_rad}). 
At very late times, the shape becomes nearly spherical.
%
%
%
%
%
%

In the velocity panels, we observe that, as the system approaches the formation of the apparent horizon at sufficiently late times, the fluid splits into two regions: one moving inward
and the other moving outward, 
creating an under-dense region. 
This behaviour is typical in spherical relativistic simulations when the fluctuation amplitudes are near their critical threshold (see \cite{escriva_solo} for comparison).

\begin{figure}[!htbp]
\centering
\includegraphics[width=1.9 in]{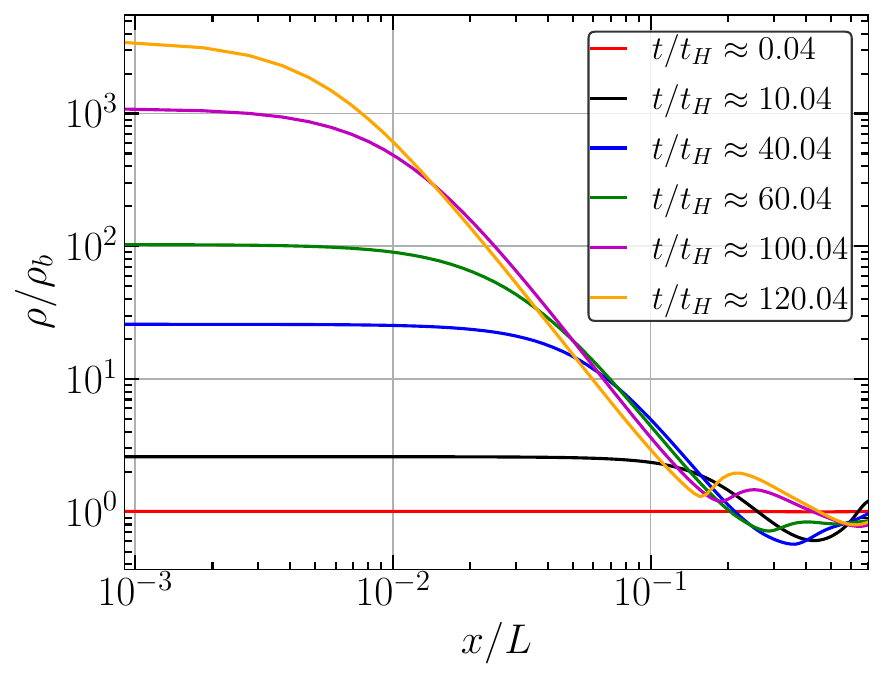}
\includegraphics[width=1.9 in]{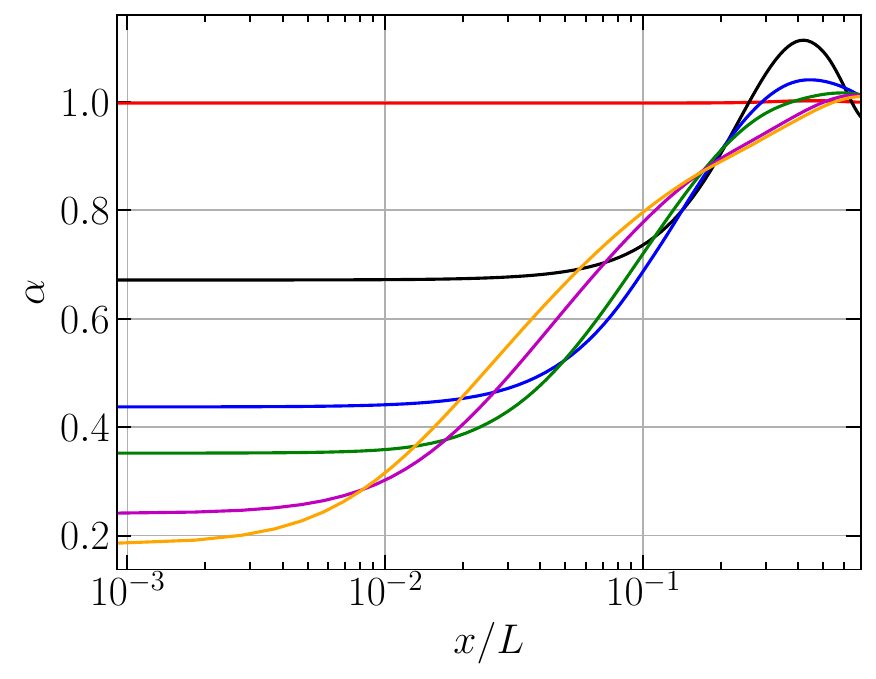}
\includegraphics[width=2.0 in]{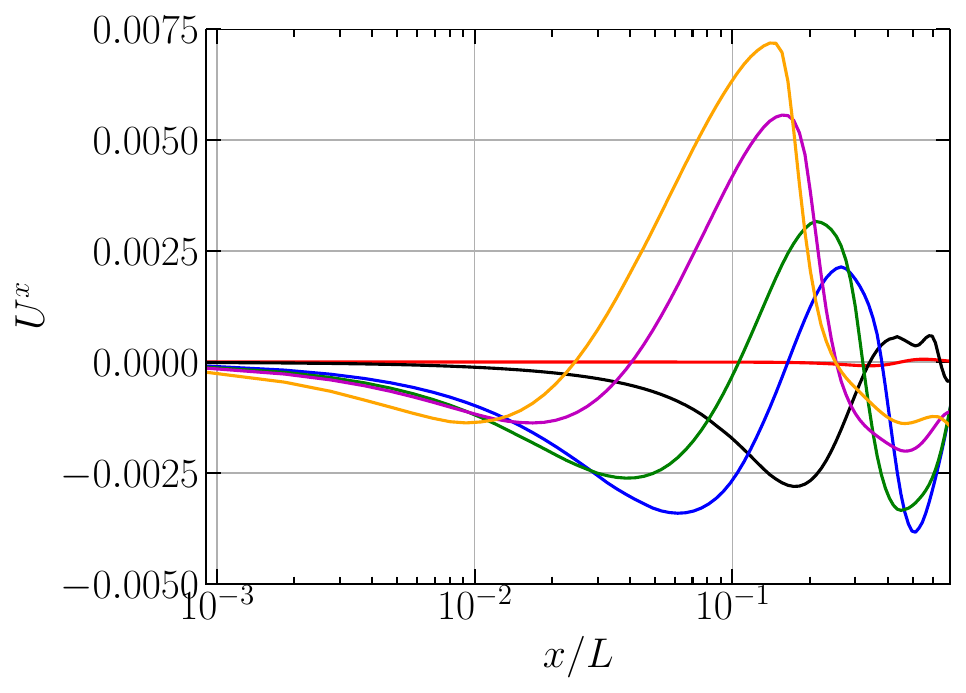}
\caption{Snapshots of the energy density $\rho/\rho_b$ (left-panel), lapse function $\alpha$ (middle-panel) and Eulerian velocity $U^{x}$ (right-panel)
on the $x$ axis ($y=z=0$) for 
$e=0.08$ and $p=0$. 
}
\label{fig:projection_variables_collapse_radiation}
\end{figure}

Figure \ref{fig:projection_variables_collapse_radiation} shows the 
the energy density, lapse function, and fluid velocity in the $x$ components 
on the $x$-axis. 
Similar behaviors are found for $y$ and $z$ axes, and we do not display them. 
We observe a continuous decrease in the lapse function at the centre, 
and we infer the formation of an apparent horizon. 





\begin{figure}[!htbp]
\centering
\includegraphics[width=1.5 in]{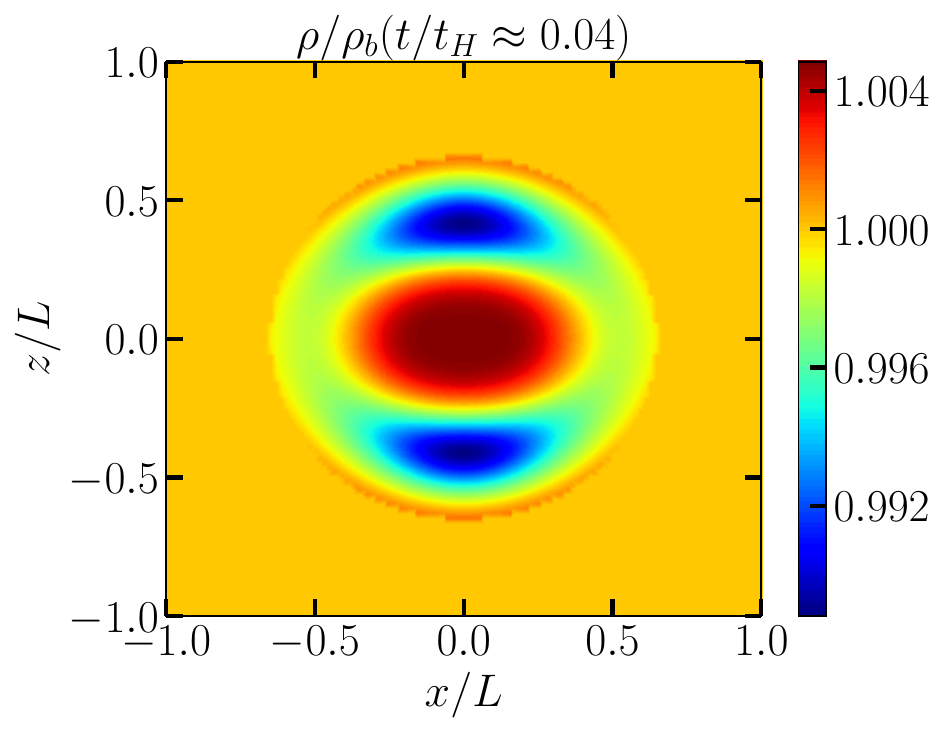}
\hspace*{-0.3cm}
\includegraphics[width=1.5 in]{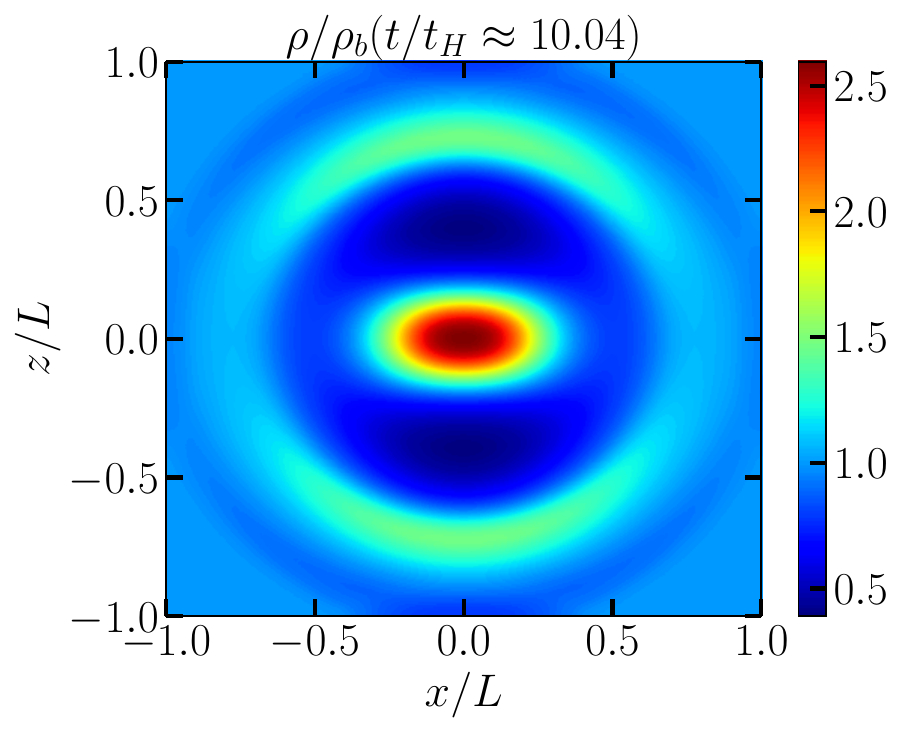}
\hspace*{-0.3cm}
\includegraphics[width=1.5 in]{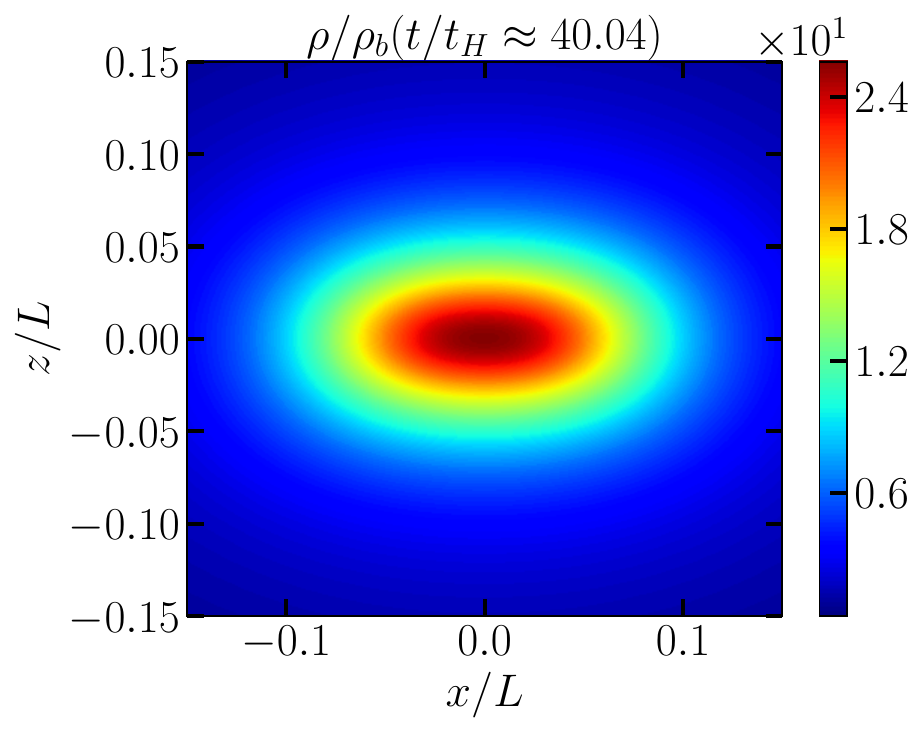}
\hspace*{-0.3cm}
\includegraphics[width=1.5 in]{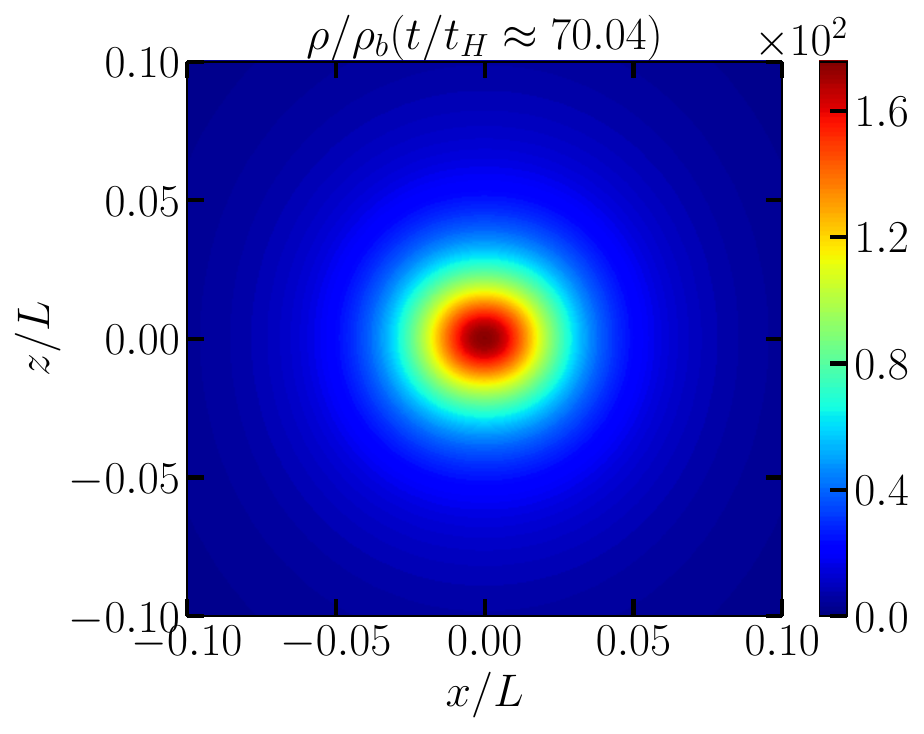}
\hspace*{-0.3cm}
\includegraphics[width=1.5 in]{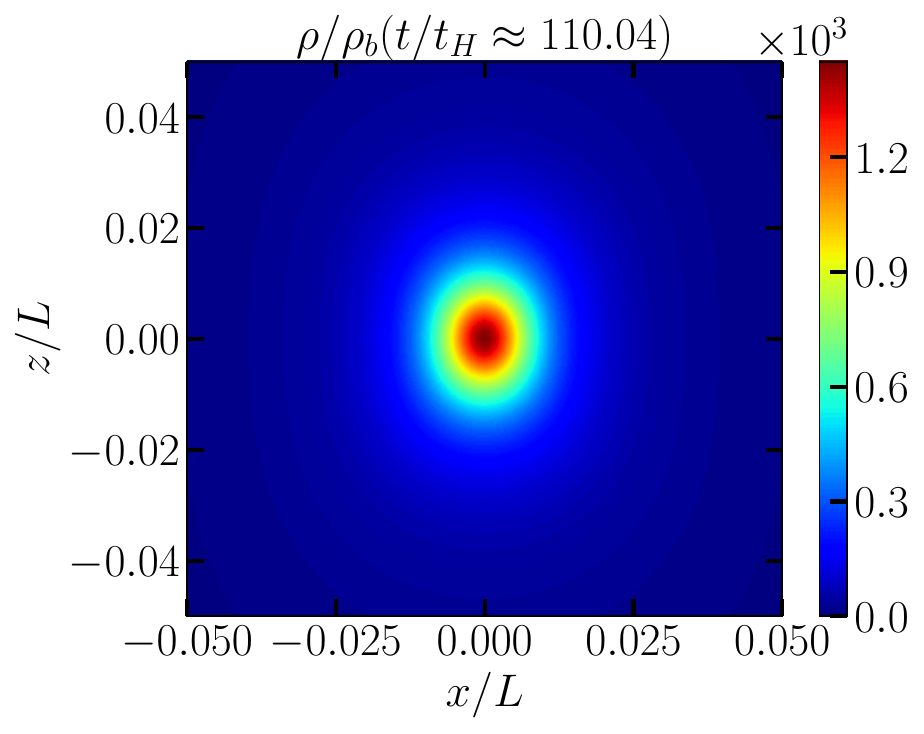}
\hspace*{-0.3cm}
\includegraphics[width=1.5 in]{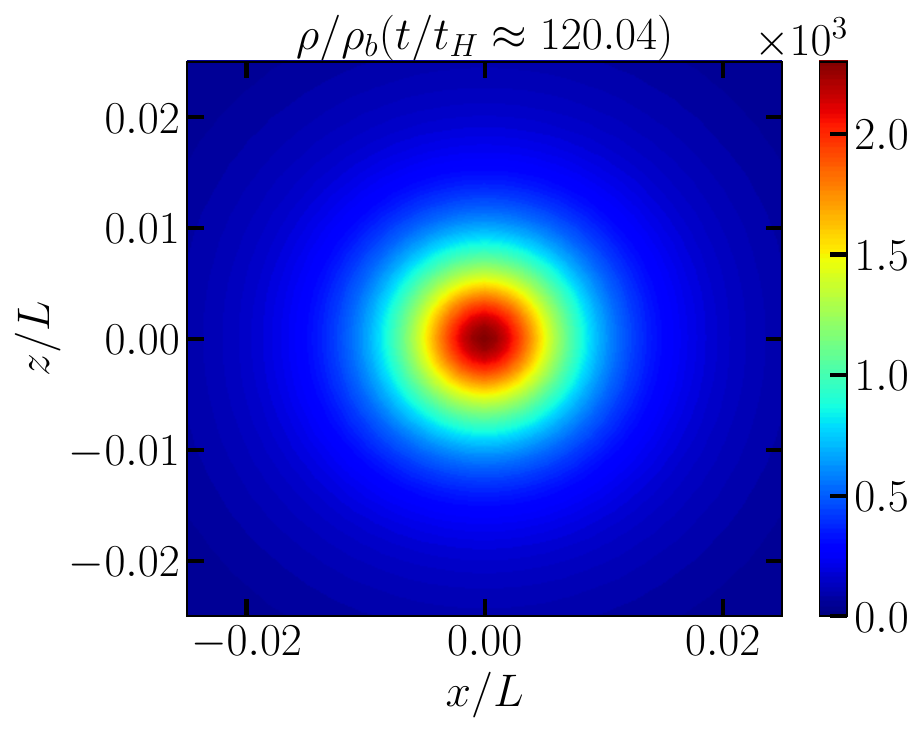}
\hspace*{-0.3cm}
\includegraphics[width=1.5 in]{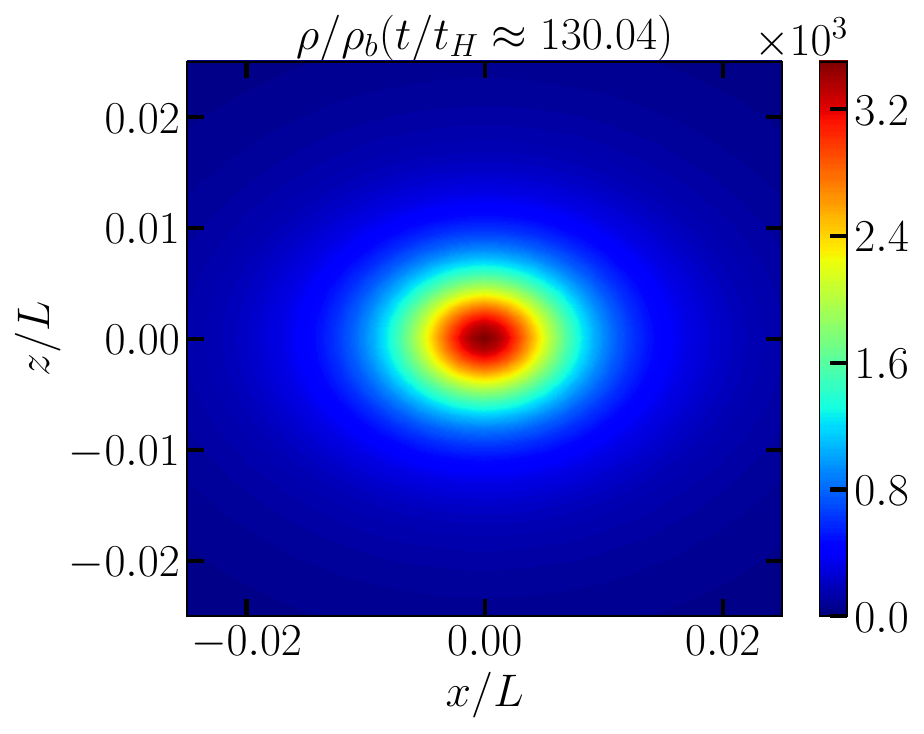}
\hspace*{-0.3cm}
\includegraphics[width=1.5 in]{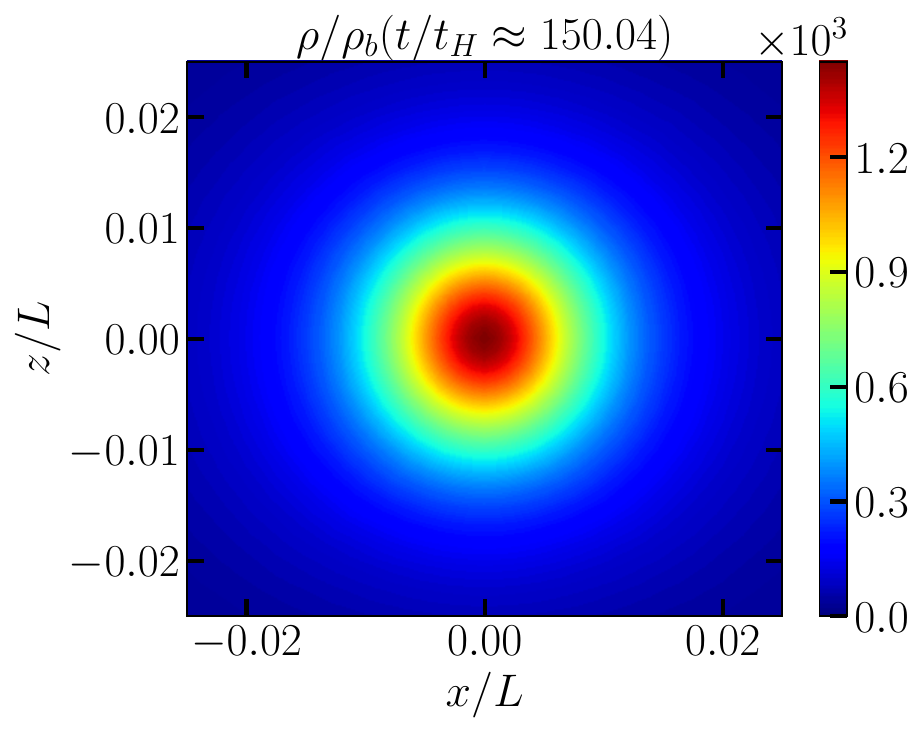}
\caption{Snapshots of the evolution of $\rho/\rho_b$ in the plane $y=0$ 
for
$e=0.06$ and $p=0.14$ with $w=1/3$.}
\label{fig:energy_density_ratio_dispersion_radiation}
\end{figure}

\begin{figure}[!htbp]
\centering
\includegraphics[width=1.5 in]{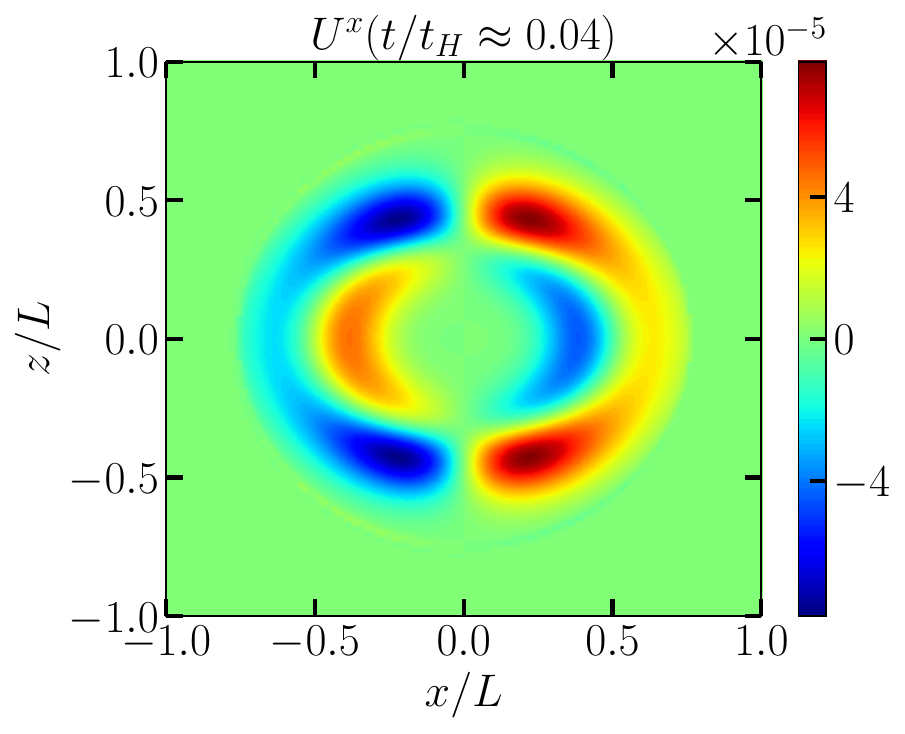}
\hspace*{-0.3cm}
\includegraphics[width=1.5 in]{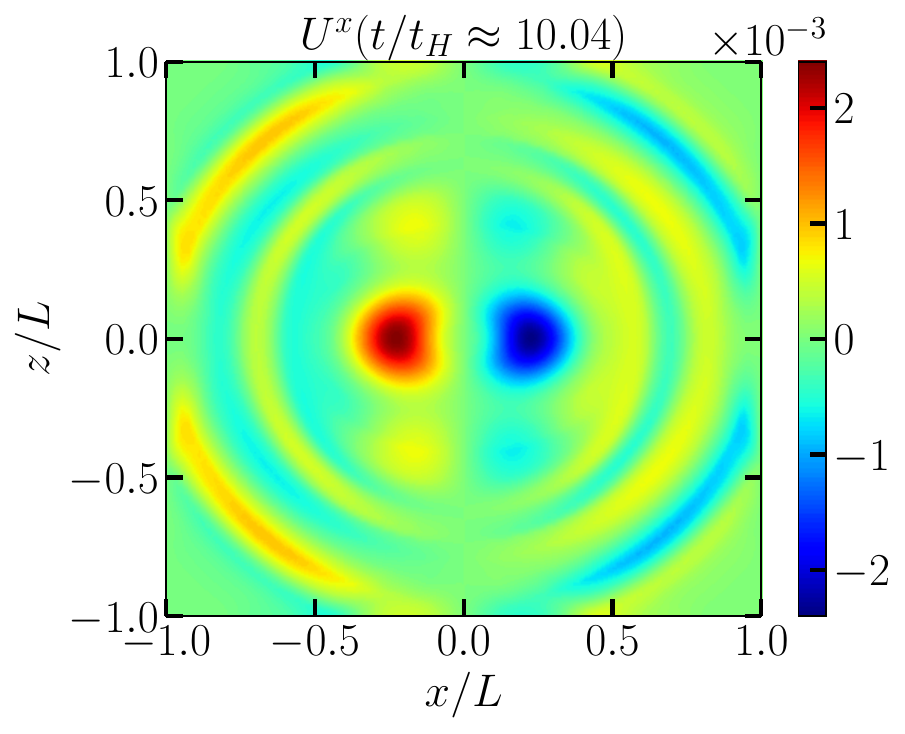}
\hspace*{-0.3cm}
\includegraphics[width=1.5 in]{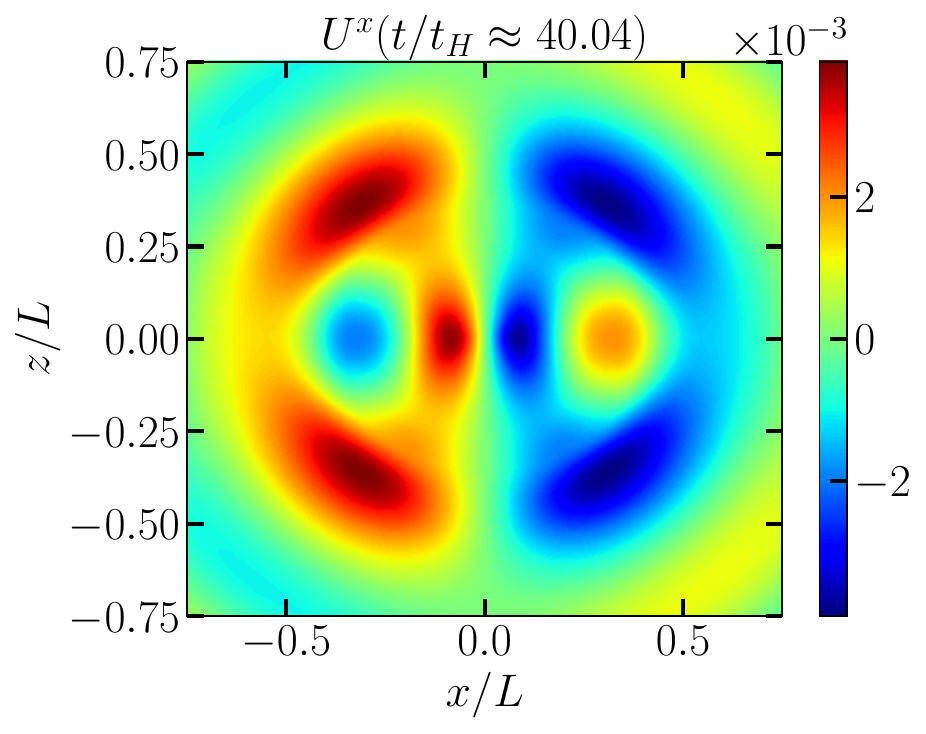}
\hspace*{-0.3cm}
\includegraphics[width=1.5 in]{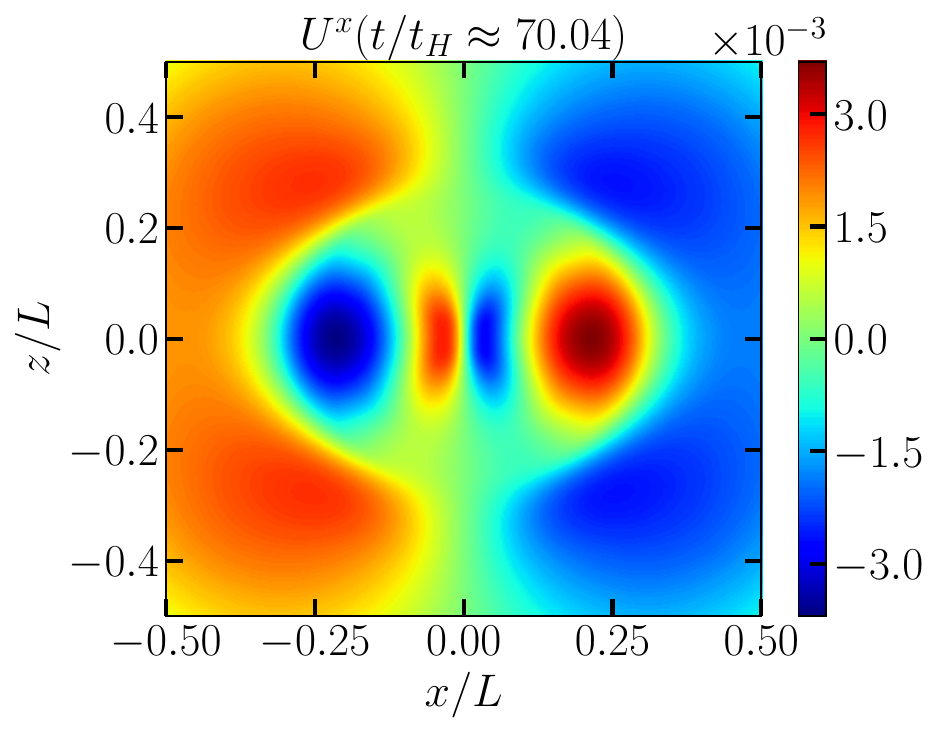}
\hspace*{-0.3cm}
\includegraphics[width=1.5 in]{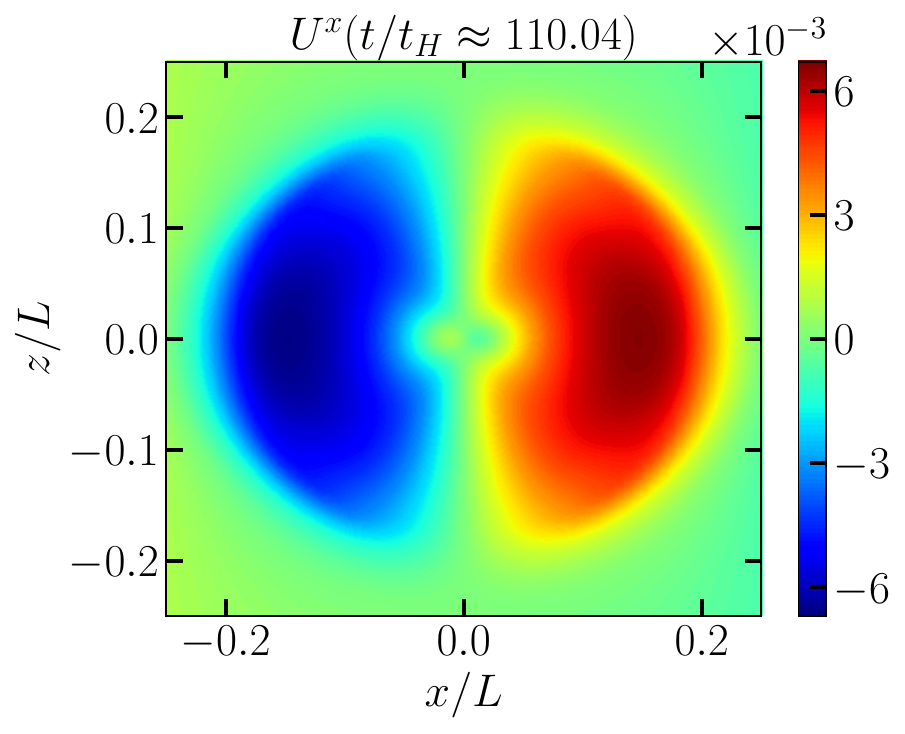}
\hspace*{-0.3cm}
\includegraphics[width=1.5 in]{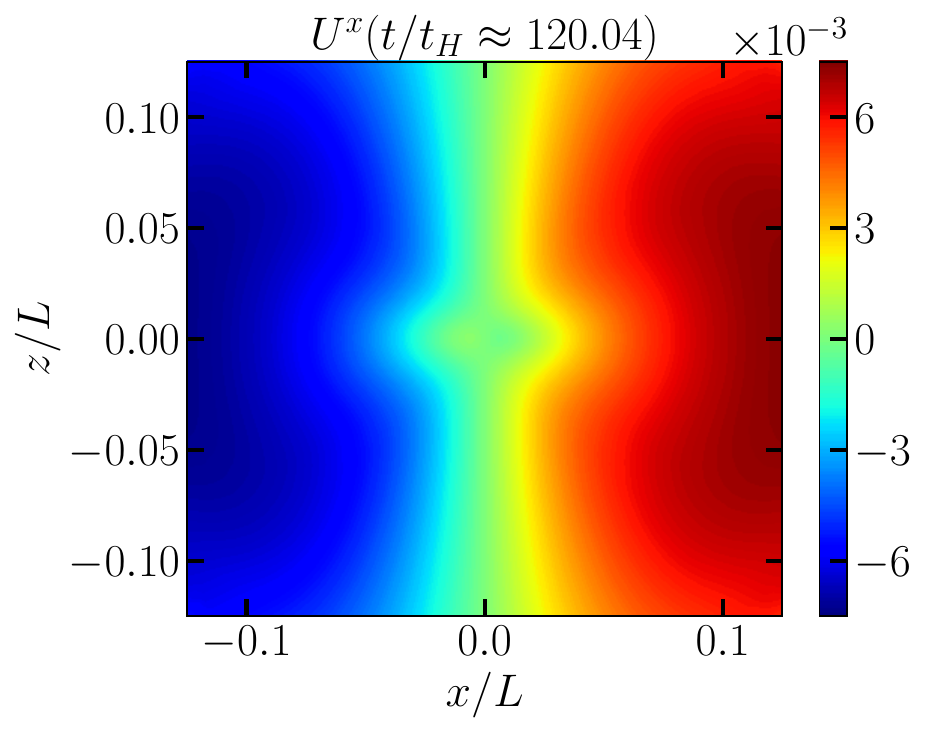}
\hspace*{-0.3cm}
\includegraphics[width=1.5 in]{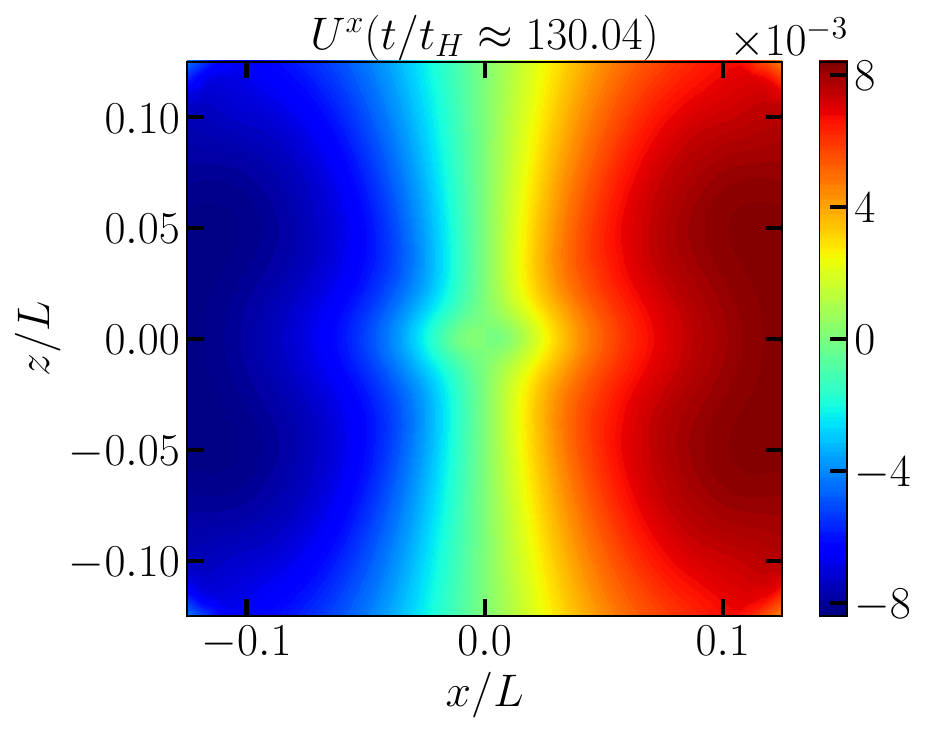}
\hspace*{-0.3cm}
\includegraphics[width=1.5 in]{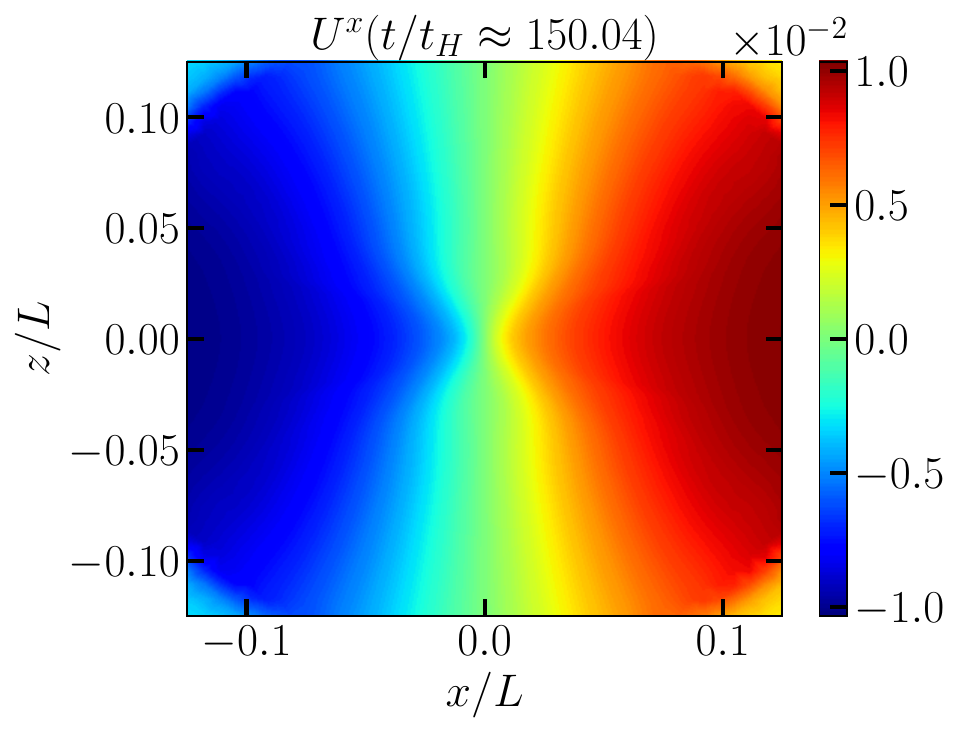}
\caption{Snapshots of the evolution of the fluid velocity $U^{x}$ in the plane $y=0$ 
for
$e=0.06$ and $p=0.14$ with $w=1/3$.}
\label{fig:vel_dispersion_rad_x}
\end{figure}

\begin{figure}[!htbp]
\centering
\includegraphics[width=1.5 in]{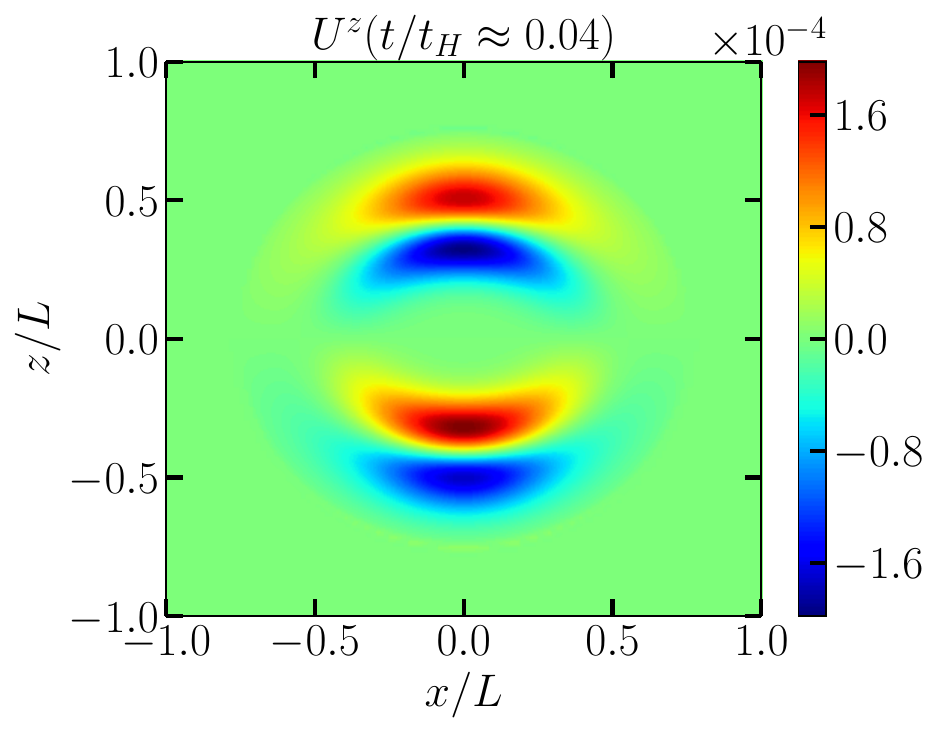}
\hspace*{-0.3cm}
\includegraphics[width=1.5 in]{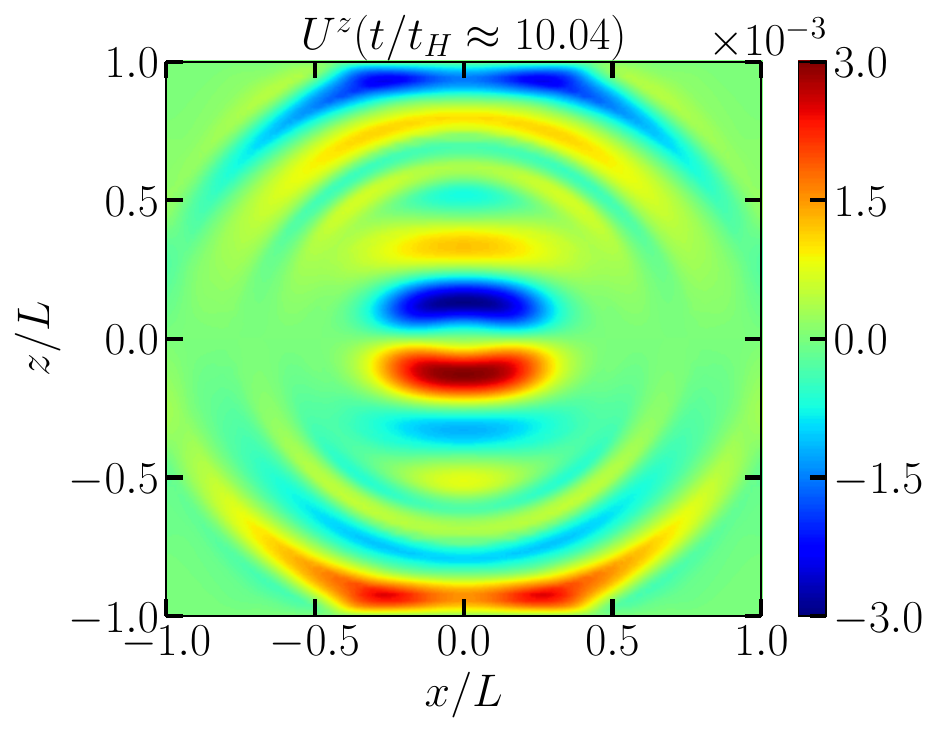}
\hspace*{-0.3cm}
\includegraphics[width=1.5 in]{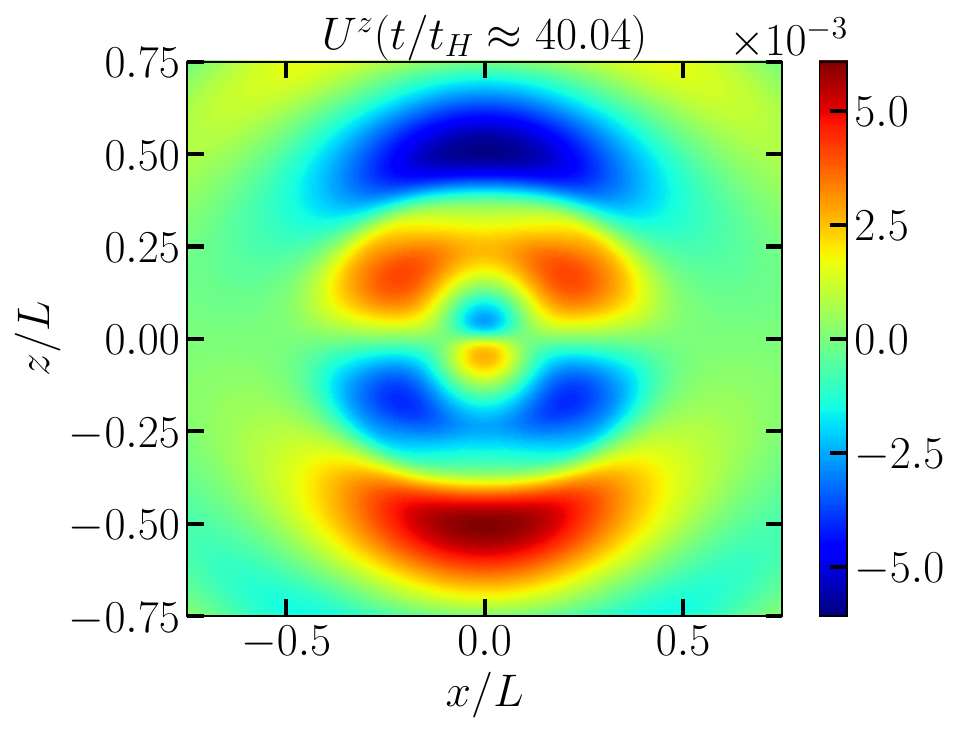}
\hspace*{-0.3cm}
\includegraphics[width=1.5 in]{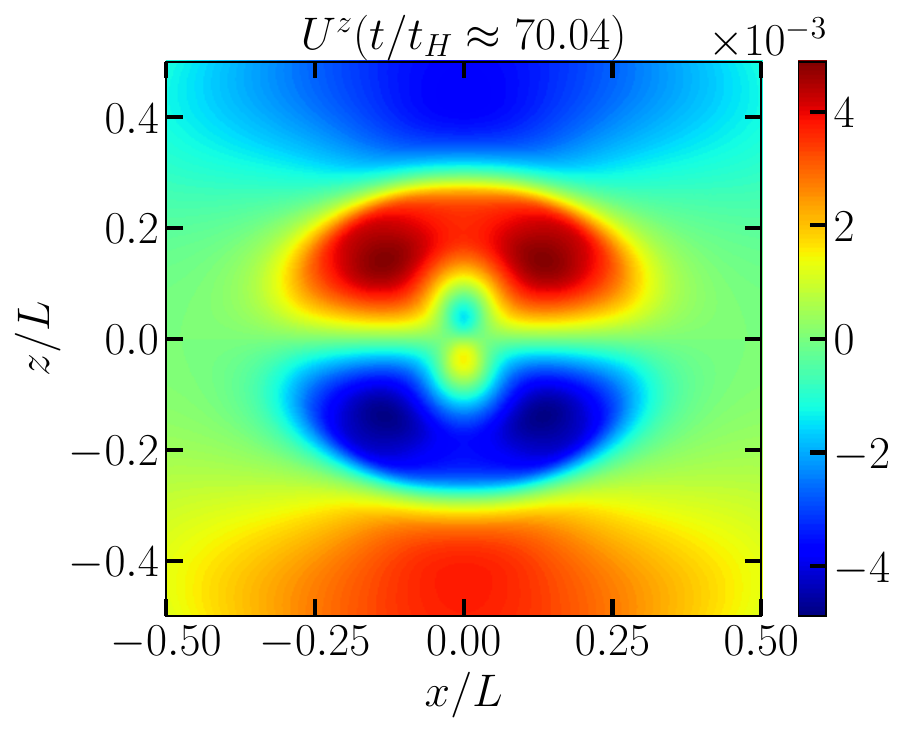}
\hspace*{-0.3cm}
\includegraphics[width=1.5 in]{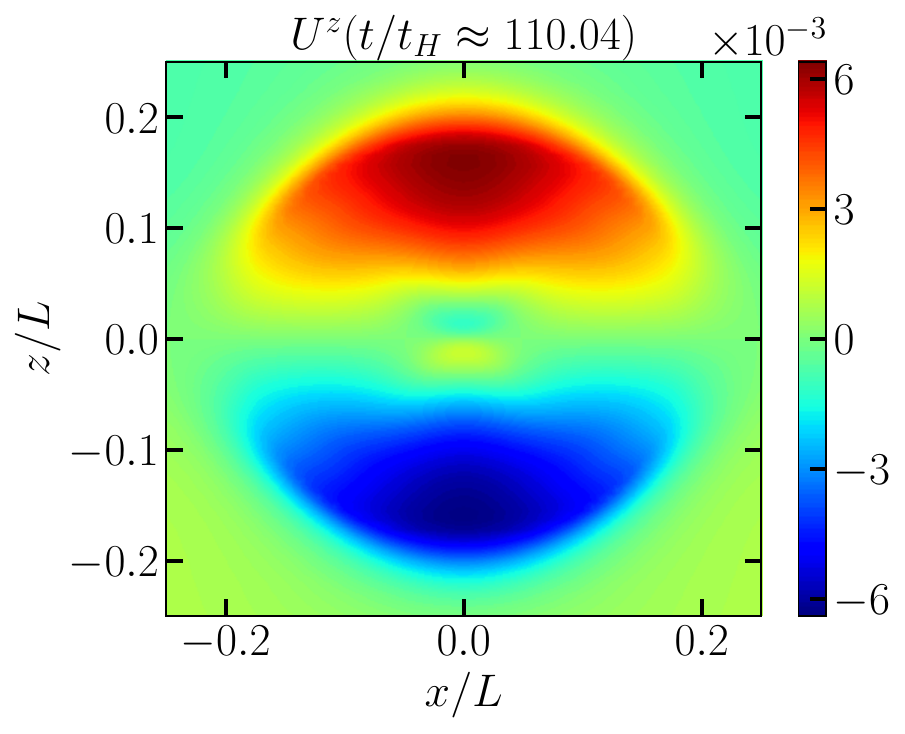}
\hspace*{-0.3cm}
\includegraphics[width=1.5 in]{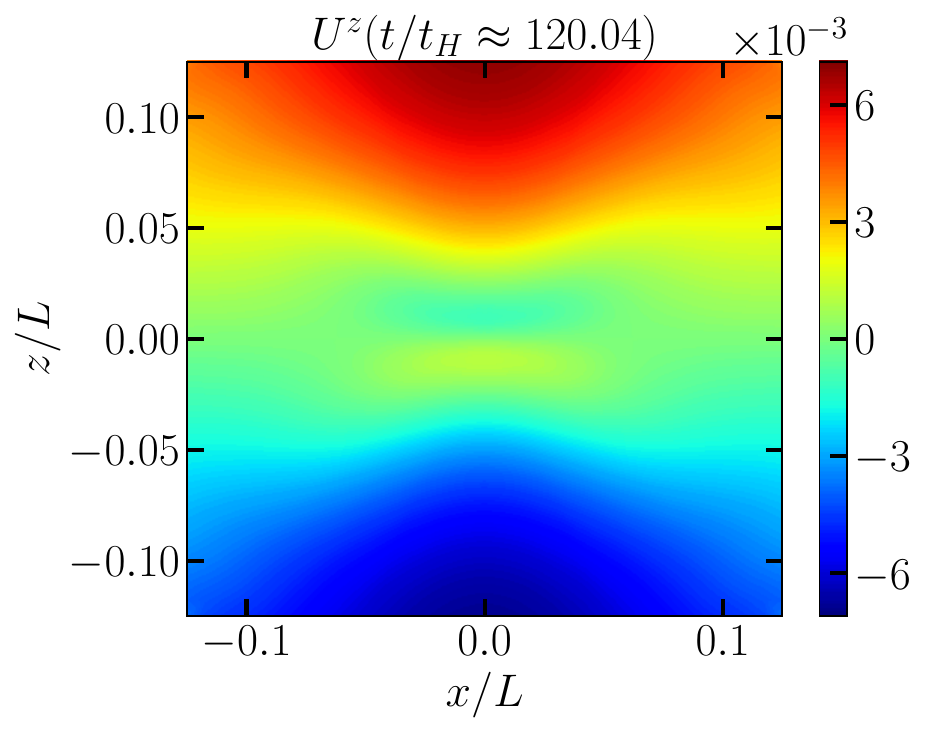}
\hspace*{-0.3cm}
\includegraphics[width=1.5 in]{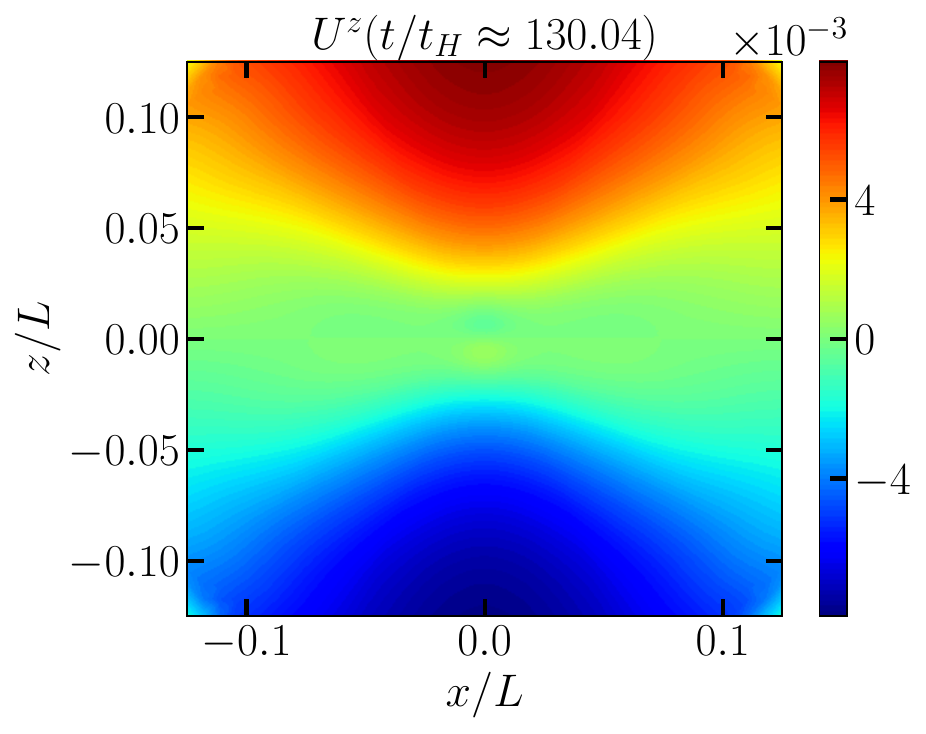}
\hspace*{-0.3cm}
\includegraphics[width=1.5 in]{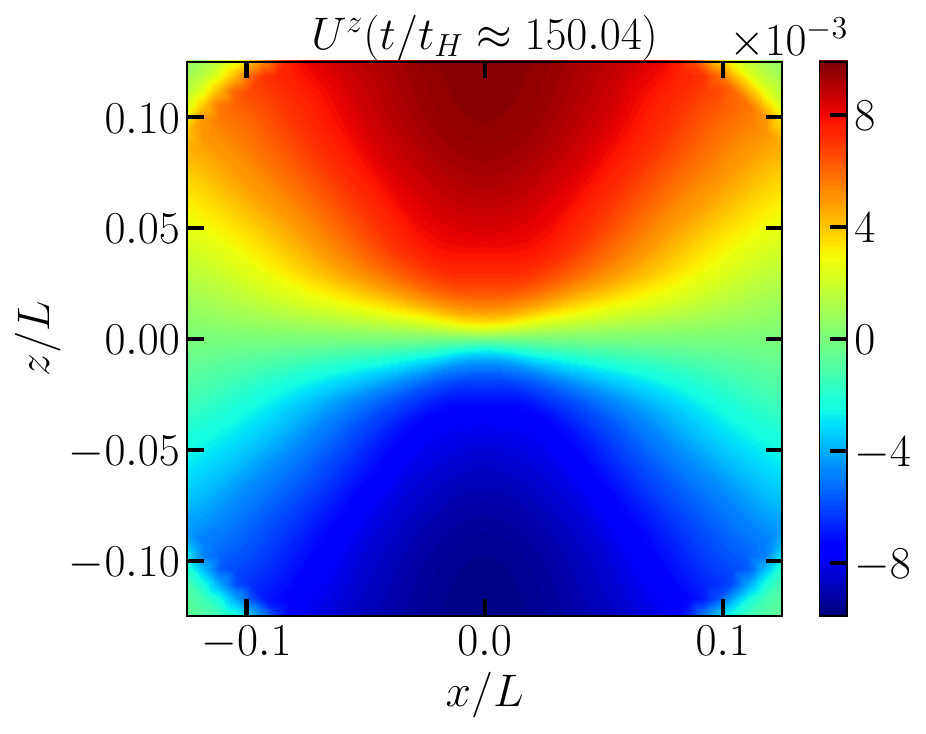}
\caption{Snapshots of the evolution of the fluid velocity $U^{z}$ in the plane $y=0$ 
for
$e=0.06$ and $p=0.14$ with $w=1/3$.}
\label{fig:vel_dipersion_rad_z}
\end{figure}

Let's now consider a case where 
the deviation from sphericity is sufficiently large, so that 
the fluctuation avoids black hole formation and disperses on the FLRW background.
Specifically, 
we choose $e=0.06$ and $p=0.14$, which correspond to 
the eigenvalues 
$\lambda_1/( \nu \sigma_2) = 1.32$ , $\lambda_2/( \nu \sigma_2)=0.72$ and $\lambda_3/( \nu \sigma_2) =0.96 $
following Eq.\eqref{eq:lambda_values_elipsoid}. 
In Figs.~\ref{fig:energy_density_ratio_dispersion_radiation}, \ref{fig:vel_dispersion_rad_x} and \ref{fig:vel_dipersion_rad_z}, 
we show the evolution of the energy density ratio $\rho/\rho_{b}$ and the velocities $U^{x}$ and $U^{z}$ for different times.

\begin{figure}[!htbp]
\centering
\includegraphics[width=1.9 in]{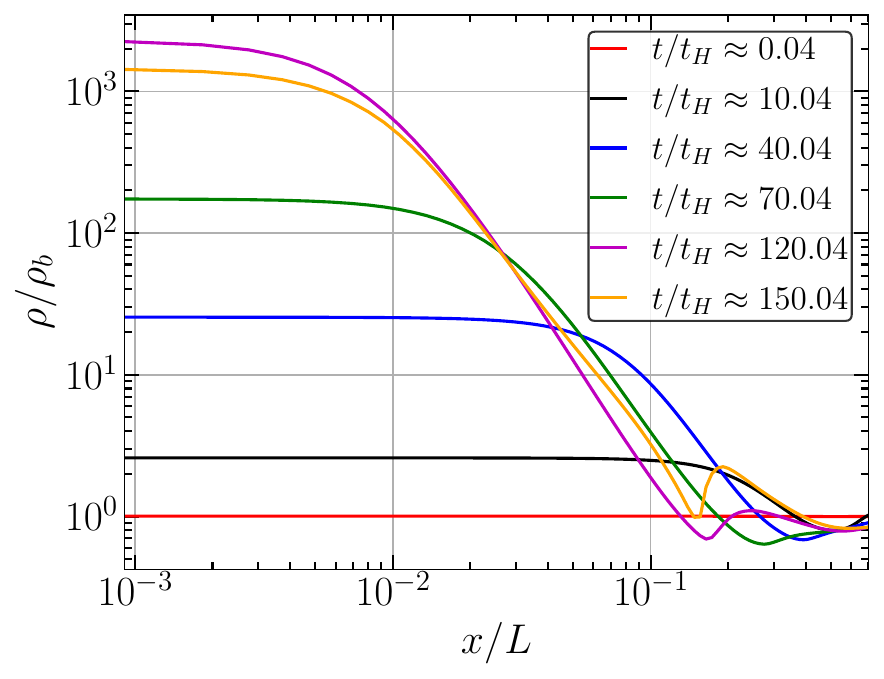}
\includegraphics[width=1.9 in]{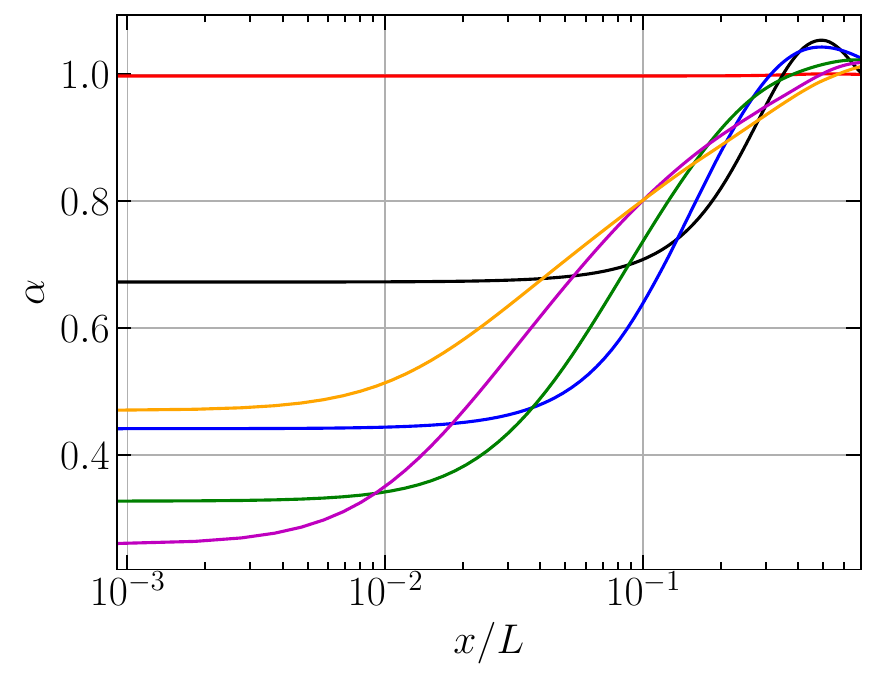}
\includegraphics[width=2.0 in]{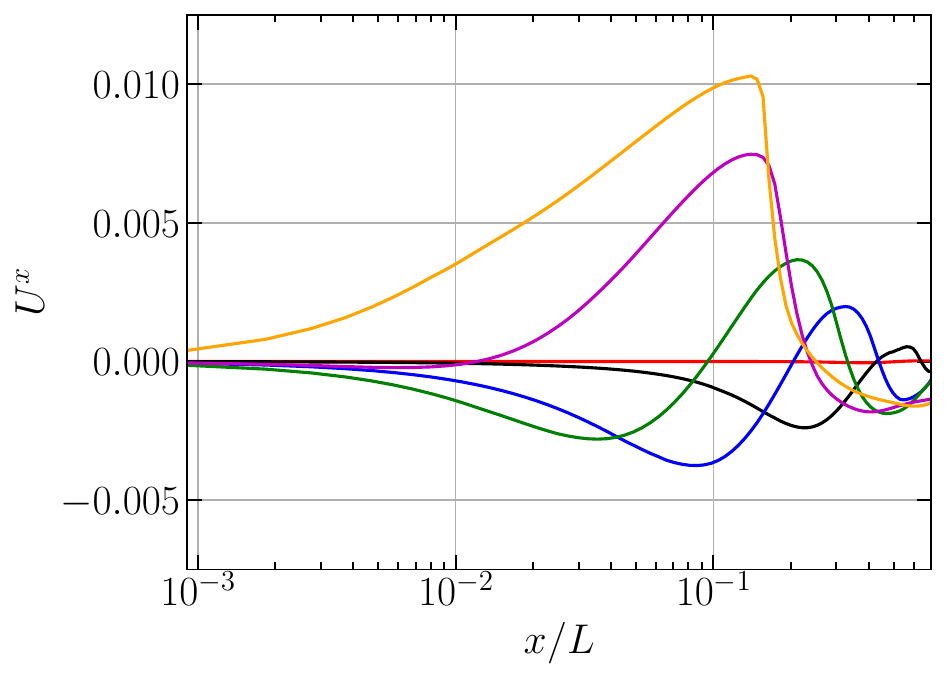}
\caption{Snapshots of the energy density $\rho/\rho_b$ (left-panel), lapse function $\alpha$ (middle-panel) and Eulerian velocity $U^{x}$ (right-panel)
on the $x$ axis ($y=z=0$) for
$e=0.06$ and $p=0.14$. 
}
\label{fig:projection_variables_dispersion_radiation}
\end{figure}

The initial 
shape
is similar to the previous case, with 
horizontally long
shape in the $z$-$x$ plane due to the same ordering of eigenvalues 
$\lambda_1>\lambda_2>\lambda_3$. 
We observe a similar behaviour regarding 
oscillatory behavior of the ellipsoidal shape
with a tendency to remain spherical at very late times. 
Compared to the previous case, the peak value of the energy density 
reaches
a maximum before subsequently decreasing. 
The velocities, $U^{x}$ and $U^{z}$, indicate an early-time contraction of the fluid's 
overdense region. 
However, at later times, no rarefaction waves are observed; instead, the fluid is simply dispersed within the region surrounding the fluctuation. 
In Fig.~\ref{fig:projection_variables_dispersion_radiation}, we plot 
the variables 
on the $x$-axis
as in the previous case. 
Notably, we observe that the lapse function at the centre experiences a bouncing behaviour as expected, indicating that the fluctuation will not form an apparent horizon.  


\begin{figure}[!htbp]
\centering
\includegraphics[width=3.0 in]{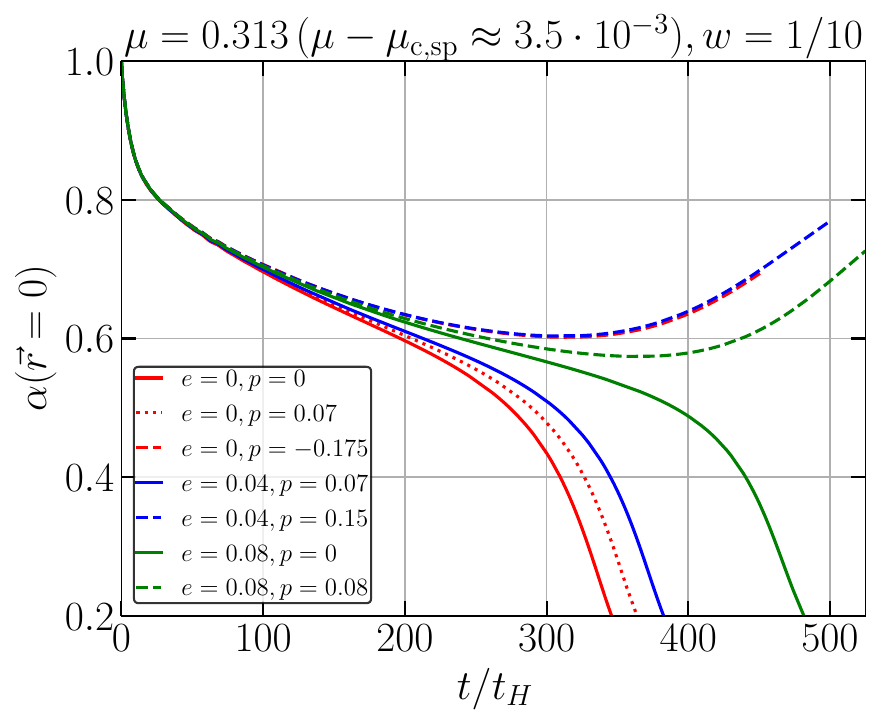}
\includegraphics[width=3.0 in]{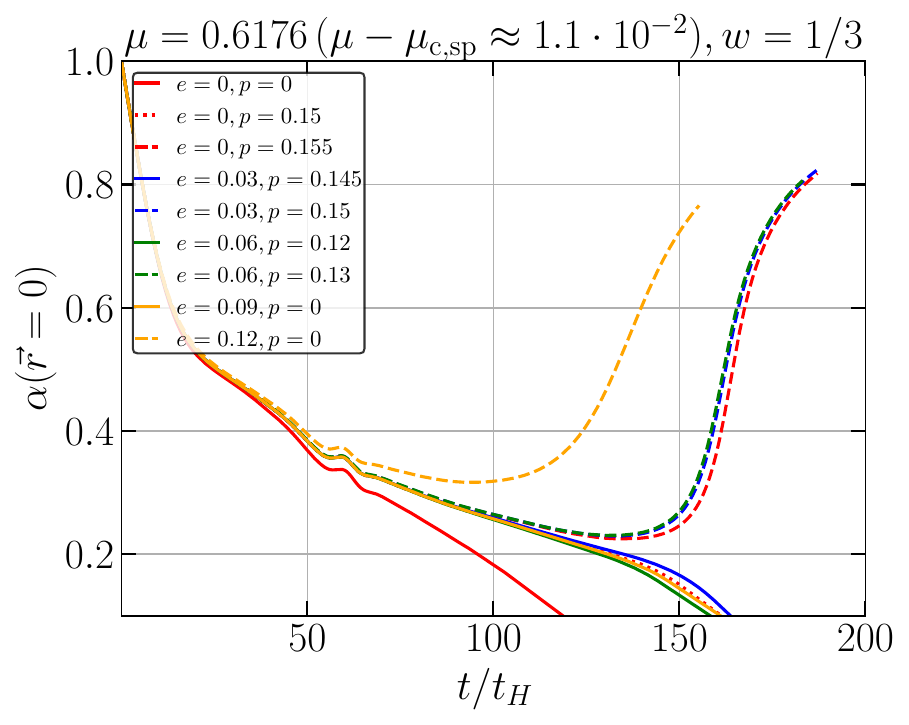}
\includegraphics[width=3.0 in]{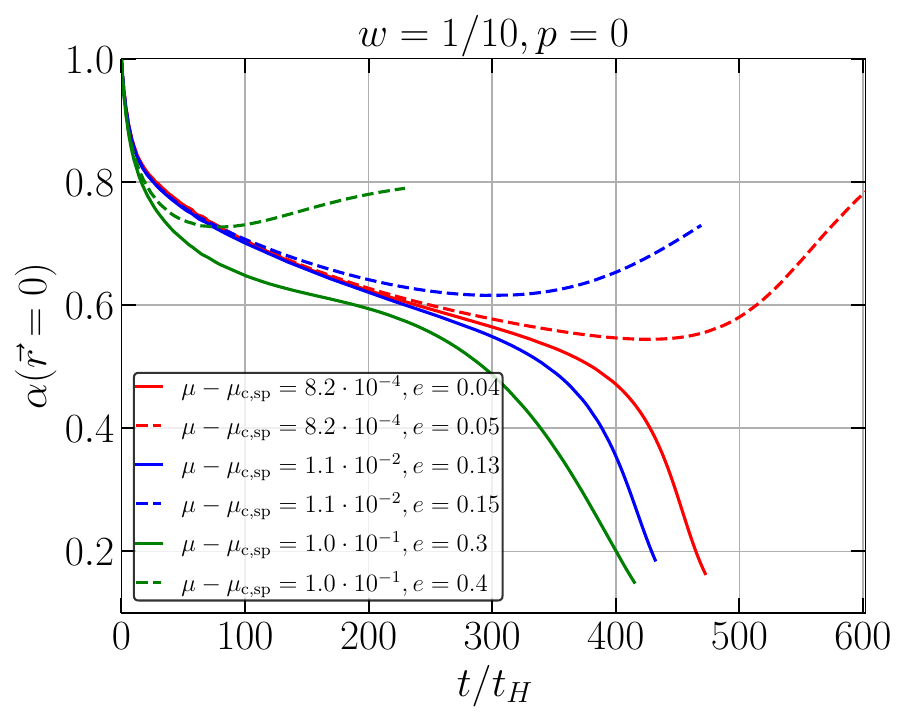}
\includegraphics[width=3.0 in]{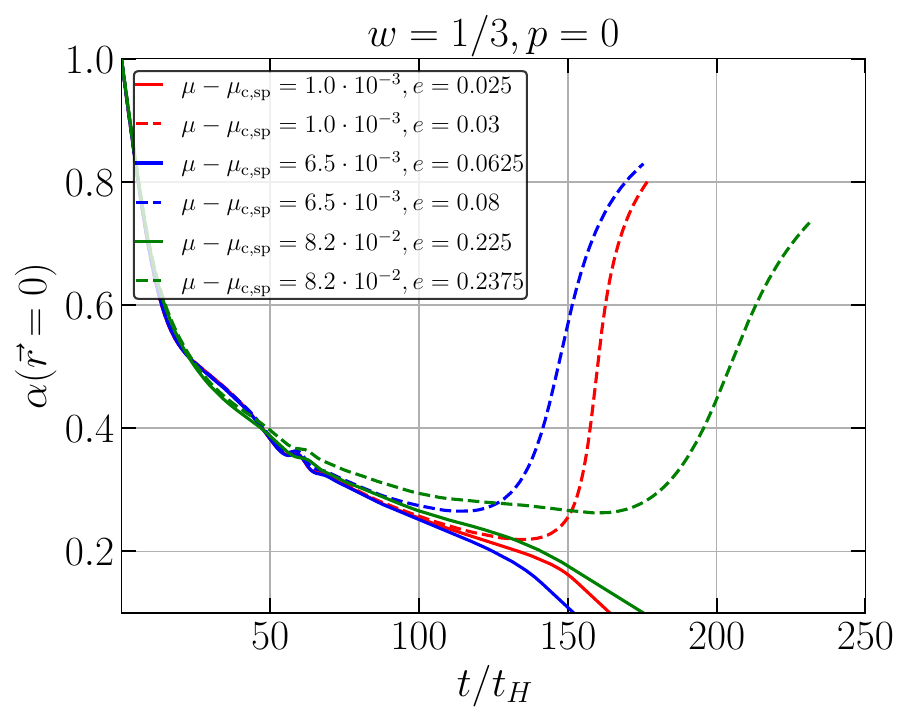}
\caption{Top-panels: Evolution of the lapse function at the origin for different configurations characterized by different values of $e$ and $p$ for $w=1/10$ with $\mu= 0.313$ (left-panel) and $w=1/3$ with $\mu=0.6176$ (right-panel).  
Bottom-panels: The same as the top-panel, but 
with $p=0$ for different values of $e$ and the amplitudes $\mu-\mu_{\rm c,sp}$. 
}
\label{fig:lapse_evolution_tipical_amplitude}
\end{figure}

Finally, Fig.~\ref{fig:lapse_evolution_tipical_amplitude} shows the time evolution of the lapse function at the centre $\alpha(\vec{r}=0)$ for different configurations. 
The figure highlights the detailed bouncing behaviour of the lapse function, which indicates 
apparent horizon formation when $\alpha(\vec{r}=0)$ continuously decreases and 
approaches to zero. 
We can observe that the time scale of the gravitational collapse is longer for a softer equation of state (smaller $w$). 
This can be understood as follows. In a simple model~\cite{Harada:2013epa}, the threshold of PBH formation is estimated by considering the critical situation in which the sound wave propagation time scale and the free-fall time scale are equal to each other. Since the sound wave speed is given by $\sqrt{w}$, the sound wave crossing time scale can be roughly estimated by $1/(\sqrt{w}H_{\rm ent})=3(1+w)/(2\sqrt{w})t_{\rm H}$. In the critical situation, this time scale coincides with the collapsing time scale.  
Therefore the gravitational collapse takes a much longer time with $w=1/10$ near the threshold \footnote{It should be noted that the values of the coordinate time shown in the figures are much larger than the estimated value because, with the gauge adopted in the numerical simulation, the time lapse is delayed in the central region.}. Another effect to consider is pressure as a gravitational source: while they initially oppose the collapse, once the collapse is triggered, they primarily favor it by reducing the collapse time of the fluctuation, as they represent a form of gravitational energy. See, for instance, Fig. 1 of \cite{Escriva:2020tak} for the case of spherical symmetry.

The top panels display cases with the fixed amplitude $\mu_{\rm t}$. 
Generally, we observe that non-spherical effects ($e, p \neq 0$) tend to slow down the collapse and can significantly increase the collapse time. 
The bottom panels show different configurations with varying amplitudes $\mu$ while fixing $p=0$. 
Here, we find that as $\mu$ increases beyond the critical value for spherical collapse, 
the deviation from sphericity required to prevent collapse also becomes larger, showing that non-spherical effects tend to prevent black hole formation. 
By running multiple simulations and analyzing the bouncing behaviour of the lapse function, we identified the threshold values, which will be discussed in detail in the next section.

One remarkable behavior
observed in this section from the non-spherical configuration 
is the damping oscillatory behaviour of the ellipticity. 
This suggests that the non-sphericity in our models 
decays over time, and remains small.  
This observation is
consistent with findings from non-spherical simulations (small deviations from sphericity) in asymptotically flat spacetimes \cite{Baumgarte:2015aza, Celestino:2018ptx}. 
More complex dynamics could arise with larger deviations from sphericity (see, for instance, \cite{Marouda:2024epb}) or with a misaligned deformation tensor, as seen in \cite{Yoo:2024lhp}. 


\subsection{Non-sphericity dependence of the thresholds with the typical amplitude $\mu_{\rm t}$. }

We now investigate the threshold for black hole formation in terms of the non-spherical parameters $e$ and $p$. 
We begin by fixing the amplitude $\mu$, to the typical value $\mu_{\rm t}$ 
used in the previous section for both cases of $w=1/10$ and $w=1/3$. 
With $\mu$ fixed, we perform a series of simulations to determine the critical values of $e$ and $p$ at which a fluctuation will either collapse to form a black hole or disperse. 
From these simulations, we identify the critical configurations, denoted as 
$(\mu_{\rm t}, e, p=\tilde{p}_{\rm c}(e))$, 
where we describe $\tilde p_c$ as a function of $e$ because 
the critical configurations draw a line in the space of $e$ and $p$. 
The results are presented in Fig.~\ref{fig:panel_thresholds_w_diagram}, where green and red dots indicate cases of black hole formation and dispersion, respectively. 
The threshold is estimated as the midpoint between these cases. 
The magenta and orange lines smoothly connect these points 
delineating the 
critical line 
in the $(e, p)$ parameter space 
for the cases with $w = 1/10$ and $w = 1/3$, respectively. 
Configurations 
inside these dashed lines 
will lead to black hole formation, as the deviations from sphericity are not large enough to prevent collapse. 
In contrast, configurations outside this region have sufficiently large deviations from sphericity, preventing the fluctuations from collapsing. 

We find that, 
from our numerical results, 
the critical line described by $p=\tilde{p}_{\rm c}(e)$
closely follow a \textit{superellipse}
expressed as 
%
%
%
%
%
%
%
%
\begin{equation}
 \left(\frac{\tilde{p}^{\pm}_{\rm c}(e)}{p^{\pm}_0}\right)^{n^{\pm}} + \left(\frac{e}{e_0}\right)^{n^{\pm}} =1 \Rightarrow   \tilde{p}^{\pm}_{\rm c}(e) =\pm p^{\pm}_0 \left[ 1- \left( \frac{e}{e_0}\right)^{n^{\pm}} \right]^{1/n^{\pm}}, 
\label{eq:superellipse}
\end{equation}
where $\tilde{p}_{\rm c}^\pm(e)$ is defined such that the critical line is given by $p=\tilde{p}_{\rm c}(e)=\tilde{p}_{\rm c}^\pm(e)$
for $p\gtrless0$, 
and $e_{0}$, $p^{\pm}_0$ and the exponent $n^{\pm}$ are 
the parameters characterising the superellipse. 
Making a non-linear fit of Eq.~\eqref{eq:superellipse} with our numerical results, we obtain $n^+ \approx 2.44 \pm 0.03$, $n^{-} \approx 1.70 \pm 0.02$ for $w=1/3$ and $n^+ \approx 2.53 \pm 0.10$, $n^{-} \approx 1.70 \pm  0.05$ for $w=1/10$. The behaviour of $\tilde{p}_{\rm c}(e)$ is not symmetric for 
the reflection along $e$ axis, that is, $\tilde{p}_{\rm c}^+(e)\neq-\tilde{p}_{\rm c}^-(e)$.  
This is expected since the set of eigenvalues $\lambda_l$ is not the same when considering 
$p\rightarrow -p$ with a fixed $e$. 
For the cases $e=0$, we found $\tilde{p}_{\rm c}^+(0)=p_0^+<|\tilde{p}_{\rm c}^-(0)|=|p_0^-|$ indicating that 
a slightly larger deviation from sphericity is required for the fluctuations to collapse
for prolate cases ($p<0$) compared with oblate cases ($p>0$). 
Note that 
the value of $\tilde{p}_{\rm c}(e)$ does not change much
in the region where the probability given by Eq.~\eqref{eq:P_ep} is non-zero. 
This supports the idea that 
we can ignore the $p$ dependence in the threshold estimation 
when we estimate the PBH mass function. 


Interestingly, we do not observe significant differences in the functional form of the critical line $p=\tilde{p}_{\rm c}(e)$
for the two values of $w$ considered. 
This suggests that the functional form of Eq.~\eqref{eq:superellipse} might have some ``universality" with similar values of $n^{\pm}$, indicating that the threshold of collapse is determined by the initial ellipticity irrespective of the parameter $w$ of the equation of state. 
However, simulations with other profiles and $\mu-\mu_{\rm c,sp}$ values would be needed to 
clarify the validity of
this hypothesis.

\subsection{Critical ellipticity as a function of the initial amplitude $\mu$}


\begin{figure}[!htb]
\centering
\includegraphics[width=4.2 in]{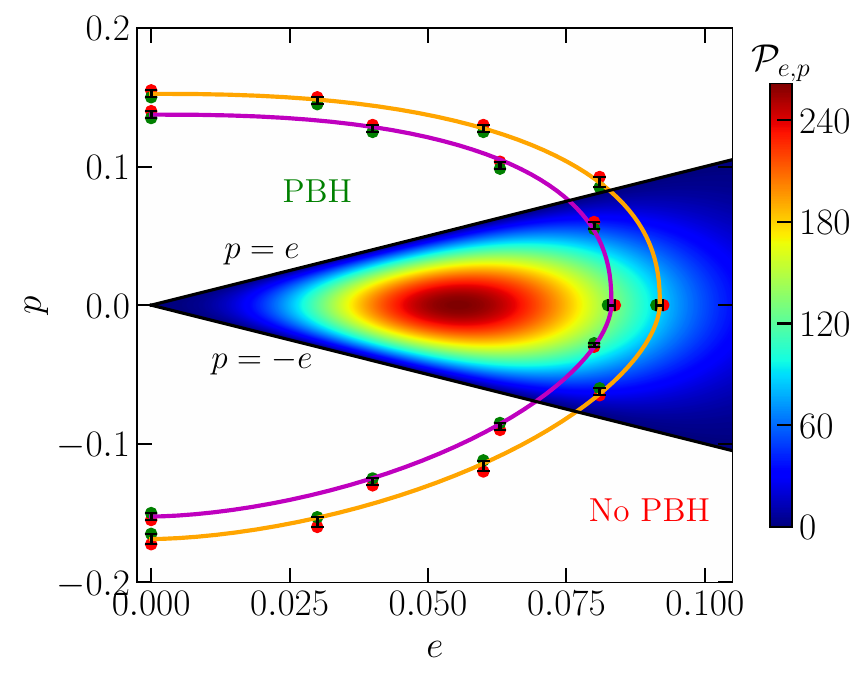}
\caption{
``Phase diagram" in the $(e,p)$ plane for the typical initial amplitude $\mu=\mu_t$. 
The red and green dots denote configurations of $e$ and $p$ that avoid and form black holes, respectively. 
The magenta and orange lines represent the numerical fitting of Eq.\eqref{eq:superellipse} to the numerical threshold values (infer as the middle point between the green and red points) for the case $w=1/10$ and $w=1/3$, respectively. The coloured region corresponds to the probability distribution $\mathcal{P}_{e,p}$ (Eq.\eqref{eq:P_ep}) fixing $\nu_c=8$. 
}
\label{fig:panel_thresholds_w_diagram}
\end{figure}



Here, let us focus on the cases with $p=0$ having the 
weak $p$-dependence of the threshold value in the parameter region relevant to the PBH mass function revealed in the previous section. 
Then, for a fixed value of $\mu>\mu_{\rm c,sp}$, 
we can find the critical value $e_{\rm c}=\tilde{e}_{\rm c}(\mu)$
as the threshold of $e$ for black hole formation. 
We 
describe this critical value as $\tilde{e}_{\rm c}(\mu)$ since it depends on the value of $\mu$. 
We conduct new simulations to find $\tilde{e}_{\rm c}(\mu)$ for $w=1/3$ and $w=1/10$.  
%
\begin{figure}[!htb]
\centering
\includegraphics[width=4.0 in]{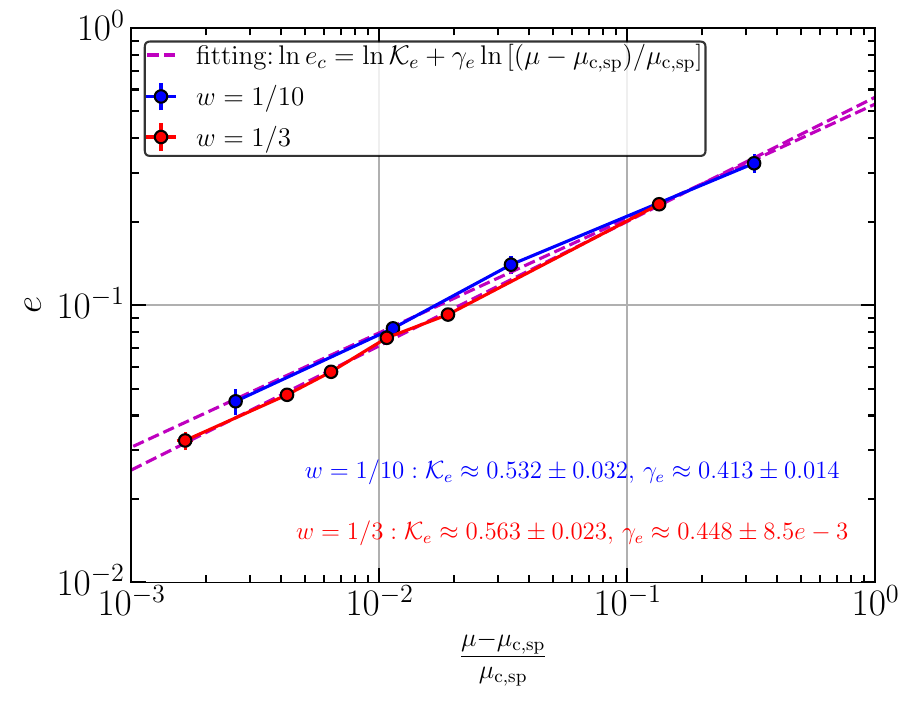}
\caption{Critical value $e_{\rm c}$ with $p=0$ is depicted as a function of $(\mu-\mu_{\rm c,sp})/\mu_{\rm c,sp}$ together with the error bars given by the intervals between the parameter values of $e$ for the corresponding two cases of the black hole formation and dissipation. 
The dashed magenta lines correspond to the fitting to the equation $\tilde{e}_c(\mu) = \mathcal{K}_e [(\mu-\mu_{\rm c,sp})/\mu_{\rm c,sp}]^{\gamma_e}$, whose values $\mathcal{K}_{e}$ and $\gamma_{e}$ are shown in the plot for the case $w=1/10$ (blue) and $w=1/3$ (red).}
\label{fig:ec_plot}
\end{figure}
In Fig.~\ref{fig:ec_plot}, we present the results of our simulations. 
Our numerical results closely follow a power-law behavior, described by,
\begin{equation}
\tilde{e}_{\rm c}(\mu) = \mathcal{K}_e \left(\frac{\mu-\mu_{\rm c,sp}}{\mu_{\rm c,sp}}\right)^{\gamma_{e}},
\label{eq:power_law_e}
\end{equation}
%
%
%
%
in the range 
$10^{-3} \lesssim (\mu-\mu_{\rm c, sp})/\mu_{\rm c, sp} \lesssim 10^{0}$. 
The values of the exponent $\gamma_e$ and the constant $\mathcal{K}_e$
are given by $\mathcal{K}_e\approx0.53$ and $\gamma_e\approx 0.41$ for $w=1/10$, and 
$\mathcal{K}_e\approx0.56$ and $\gamma_e\approx 0.45$ for $w=1/3$. 


Our numerical results 
have been done in
the range $10^{-3} \lesssim (\mu-\mu_{\rm c, sp})/\mu_{\rm c, sp} \lesssim 10^{0}$. 
Extending the analysis to smaller values would require significantly more computational time and higher resolution, which we leave for future research. 
However, in the small $\mu-\mu_{\rm c,sp}$ limit, the threshold configuration satisfying $e_{\rm c}=\tilde{e}_{\rm c}(\mu)$ approaches the spherically symmetric critical solution. Therefore,  if the power law behavior is totally characterized by the spherical critical solution and perturbation modes around that, the behavior is preserved for $(\mu-\mu_{\rm c,sp})/\mu_{\rm c,sp}\ll 10^{-3}$. 
This expectation, though, would need to be confirmed through detailed analyses.  

Taking into account Eqs.\eqref{eq:power_law_e} and \eqref{eq:superellipse}, 
we could incorporate the behaviour of $\tilde e_{\rm c}(\mu)$ into $\tilde p_{\rm c}(e)$. 
That is, for instance, assuming the exponent $n_\pm$ is unchanged, 
we may extend the parameters $p_0^\pm$ and $e_0$ 
as functions of $\mu$ written as $\tilde p_0^\pm(\mu)=\tilde p_0^\pm(\mu_t)\times \tilde e_c(\mu)/\tilde e_c(\mu_t)$ and $\tilde e_0(\mu)=\tilde e_c(\mu)$. 
Then we can draw contours in the $e$-$p$ plane. 
For a smaller value of $\mu$, the size of the contour shrinks toward the origin (spherical case) and vanishes in the limit $\mu\rightarrow\mu_{\rm c,sp}$. 
However, this proposal requires further validation through additional simulations, particularly those examining a wider range of profiles. 
This is an avenue left for future research.

Our numerical results in Fig.~\ref{fig:ec_plot} demonstrate that non-spherical effects make the gravitational collapse of fluctuations harder
compared to the spherical case. 
Moreover, due to the behaviour of Eq.\eqref{eq:power_law_e}, even a slight deviation from sphericity can prevent a fluctuation from collapsing into a black hole
if those fluctuations have an amplitude near 
the
threshold in the critical regime. 
Let us evaluate, due to the non-spherical effects, how large fraction is excluded 
from the total number of black holes 
which is supposed to be formed without taking into account the non-sphericity. 
The remaining fraction can be estimated by integrating the probability distribution 
of $e$ and $p$ in the region enclosed by $p=\pm e$ and $p=\tilde{p}_{\rm c}(e)$ with a fixed value of $\mu$. For simplicity, we approximate the line $p=\tilde{p}_{\rm c}(e)$ by
 $e=\tilde{e}_{\rm c}(\mu)$ of a vertical line in the $e$-$p$ plane and 
integrate the probability distribution 
$\mathcal{P}_{e,p}$ over the region $\mathcal{R}$ 
defined by
\begin{equation}
\label{eq:domain_R_integration}
\mathcal{R}(\mu) \in \{ 0 \leq e \leq \tilde{e}_{\rm c}(\mu) \, , \, -\tilde{e}_{\rm c}(\mu) \leq p \leq \tilde{e}_{\rm c}(\mu) \}
\end{equation}
in the parameter space $(e,p)$. 
%
%
%
Note that this approach provides an upper bound for the estimation. 
As shown in Fig.~\ref{fig:panel_thresholds_w_diagram}, 
although there are regions with $p>\tilde{p}^+_{\rm c}(e)$ or $p<\tilde{p}^-_{\rm c}(e)$ inside $\mathcal R$, which cause an overestimation of the number of black holes, 
the effects are negligible for a typical value of $\mu$. 

\begin{figure}[!htb]
\centering
\includegraphics[width=3.5 in]{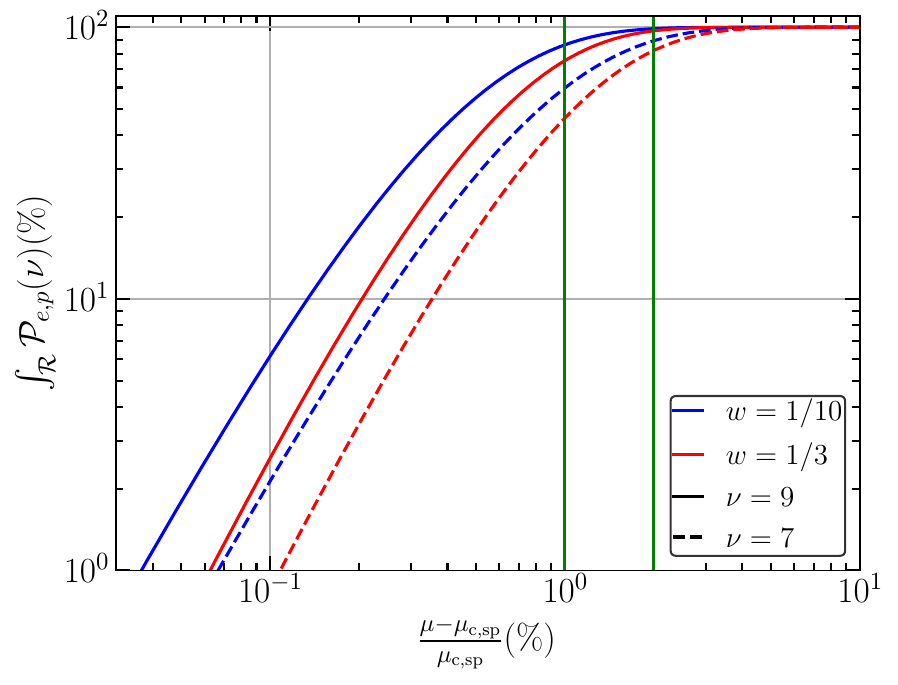}
\caption{Integration of the probability distribution of Eq.~\eqref{eq:P_ep} integrated over the domain $\mathcal{R}$ defined in Eq.\eqref{eq:domain_R_integration} in terms of $(\mu-\mu_{\rm c,sp})/\mu_{\rm c,sp}$ 
in percentatge value $\%$. Red and blue lines correspond to the cases with radiation and soft equations of state, respectively. 
The solid and dashed lines correspond to 
the cases for 
$\nu=9$ and $\nu=7$, respectively, 
which corresponds to putting the scale $k_p$ at different mass-scales; see 
Table\ref{table:parametters} and  Appendix \ref{sec:appendix_high_peaks} for details. The two green vertical lines indicate $1\%$ and $2\%$, respectively.}
\label{fig:P_ep_integrated}
\end{figure}

The integration of the probability distribution 
with the fitting function \eqref{eq:power_law_e}
can be found in Fig.~\ref{fig:P_ep_integrated} in percentatge ``$\%$". 
For a typical value of $\mu$, which gives $(\mu-\mu_{\rm c,sp})/\mu_{\rm c,sp}\sim1\%$, 
about $90 \%$ of the configurations will collapse depending on $w$ and $\nu$ chosen, overcoming the non-spherical effects. 
This supports the conclusion reached in \cite{Yoo:2020lmg} for $w=1/3$ with an exponential Gaussian profile considering $p = \pm e$, which found that non-spherical effects do not significantly impact the threshold by more than $\sim 1\%$\footnote{Notice that, in \cite{Yoo:2020lmg}, differently from this paper, another probability distribution, which is defined without considering the real space number density, is used for simplicity because this simplification does not cause any qualitative difference.}. 
Our findings suggest that the total number of black holes 
can be accurately estimated based on the threshold given by spherically symmetric simulations for typical settings. 

On the other hand, if we focus on the critical scaling regime, 
in which the amplitude is very close to $\mu_{\rm c,sp}$, 
the number of configurations collapsing into black holes can be significantly reduced. 
For instance, for $(\mu-\mu_{\rm c,sp})/\mu_{\rm c,sp}=10^{-3}$, most of the configurations are prevented from collapsing into black holes 
although it slightly depends on $w$ and $\nu$. 
Therefore the 
non-spherical effects can be highly significant in the critical regime, 
preventing a large fraction of configurations from collapsing into black holes. This behavior differs from the one when considering the Press-Schechter method \cite{1974ApJ...187..425P,Carr:1975qj} (see, for instance, \cite{Kuhnel:2016exn}), where all non-spherical configurations are considered equally probable, and therefore non-spherical effects are overestimated. Nevertheless, we report in the accompanying letter~\cite{companion} that, 
in terms of the PBH mass function, the effects of the ellipticity may not have a significant impact even on the power-law small mass tail 
originates from the critical behavior. 

We also conclude that non-spherical effects in a moderately soft equation of state ($w=1/10$) do not play a significant role. This finding contrasts with the expectations in the literature, where non-spherical effects were anticipated to have a more substantial impact in soft equations of state compared to radiation-dominated scenarios. One of the reasons for this result is the following. 
In general, the nearly spherical configuration is guaranteed by the validity of 
the high peak approximation $\nu\gg1$. 
For the soft equation of state, since the pressure gradient effects are 
weaker, typically expected non-sphericity
is larger than the case of the radiation fluid case for a fixed value of the bare amplitude $\mu$, and we naively expect the non-spherical effects to be more substantial for soft equations of state. 
However, because we fix the value of $\sigma_0$ to have $f_{\rm PBH}^{\rm tot}\simeq1$, the threshold value of $\nu=\mu/\sigma_0$, which is relevant for the high peak approximation, is not significantly different from the case with $w=1/3$ (see Table\ref{table:parametters}).  
Then the spherically symmetric assumption also works for the soft equation of state with $w=1/10$. 

The fact that we do not observe a significant impact of non-spherical effects on gravitational collapse for $w=1/10$ compared to $w=1/3$ (as shown in Fig.~\ref{fig:ec_plot}) could be attributed to the following reason. First, we should note that the collapsing dynamics of the ellipsoidal system is different from the dust case and the well-known instability shown by Lin-Mestel-Shu~\cite{1965ApJ...142.1431L} does not apply. Here let us try to understand our results following the Jeans criterion, which states that the system is unstable against gravitational collapse if the free-fall time scale is shorter than the sound wave crossing time scale. In our setting,  at least for a relatively small ellipticity, 
the free-fall time scale $\sim 1/\sqrt{\rho}$ would be still relevant, and it does not 
change much due to the small ellipticity $e$ because the product $\lambda_1 \lambda_2 \lambda_3$ is conserved at the linear order of $e$ with $p=0$. 
On the other hand, the sound wave crossing time scale is expected to be shorter 
because the sound wave along the short axis may propagate through the system in a shorter time. 
The earlier bounce due to the shorter time scale may cause a larger circumference in the last stage of the collapse and prevent the black hole formation in terms of the hoop conjecture~\cite{Thorne1972}. 
Therefore the gravitational collapse is expected to be harder. 
Although, with a given value of the initial amplitude $\mu$, 
this effect is expected to be more pronounced for larger $w$, 
the relative effect may be comparable since the threshold amplitude is smaller for a smaller value of $w$. 



Once non-spherical effects become dominant impeding factors against the gravitational collapse as $w\rightarrow 0$, 
the Jeans criterion does not apply, and 
we need to consider different criteria. 
Non-sphericities are expected to grow during collapse \cite{1970A&A.....5...84Z,1965ApJ...142.1431L}. 
This could lead to complex dynamics associated with the deformation~\cite{Harada:2016mhb} and rotation of over-densities~\cite{Harada:2017fjm}, and velocity dispersion potentially plays a significant role~\cite{Harada:2022xjp}.  
Further investigation is needed to clarify this aspect.



\section{Conclusions}

In this work, we employed $3+1$ relativistic numerical simulations to study the collapse of super-horizon curvature fluctuations with an ellipsoidal geometry, characterized by ellipticity ($e$) and prolateness ($p$), in line with peak theory \cite{peak_theory}
assuming a monochromatic power spectrum. 
When analyzing the dynamics
for the two cases of the equation of state with $w=1/10$ and  $1/3$, 
we observe a characteristic 
behaviour of the oscillating ellipsoidal shape 
between oblate and prolate configurations. 
This 
oscillation
persists until very late times, when the shape becomes nearly spherical, just before the formation of the apparent horizon. Although this implies that the assumption of an exactly spherical shape is not valid until the final stages of collapse, 
our results
indicate that, for the cases tested, the non-sphericity decays over time, in 
agreement 
with \cite{Baumgarte:2015aza,Celestino:2018ptx}, with relatively small non-sphericities in the initial data. On the other hand, non-spherical collapse will be accompanied by the emission of gravitational waves \cite{thorne}; however, a thorough investigation of this phenomenon is left for future research.


We have also examined how the threshold for black hole formation depends on the parameters $e$ and $p$. 
When we fixed the amplitude as a typical value $\mu_{\rm t}$, 
we found that the curve $p=p_{\rm c}(e)$, which describes the boundary of the region of black hole formation on the $e$-$p$ plane
fits well a superellipse curve (described by Eq.\eqref{eq:superellipse}) characterized by an exponent $n^{\pm}$, where the sign $\pm$ denotes the two branches 
for the region $p\gtrless0$. 
It would be interesting to investigate that functional form in terms of $\mu$, and its profile dependence 
for developing an analytical framework similar to spherically symmetric cases~\cite{universal1,Escriva:2020tak} or including non-Gaussianities~\cite{2020JCAP...05..022A}. 

In addition, we have shown that non-spherical effects can be highly significant for fluctuations with amplitudes very close to their threshold, $\mu_{\rm c}$, where even small deviations from sphericity can prevent black hole formation. This implies that, even for large peaks ($\nu \gg 1$), a substantial fraction of the configurations described by Eq.\eqref{eq:P_ep} can avoid black hole formation, as illustrated in Fig.\ref{fig:P_ep_integrated}. 
However, when considering the probability distribution of these peaks, we find that around $90\%$ of the non-spherical configurations 
cause
only a small shift in the threshold, $(\mu - \mu_{\rm c,sp})/\mu_{\rm c,sp}$, of less than $2\%$. This suggests that most configurations are unaffected by the small threshold shift. 
This conclusion aligns with the findings of \cite{Yoo:2020lmg} for a radiation-dominated Universe. Given that our study uses a different curvature profile from \cite{Yoo:2020lmg}, it may indicate that this conclusion could hold for other curvature profiles as well. However, a detailed study examining profile dependence is needed to clarify the significance of the non-spherical effects. 

Finally, when comparing our results for a radiation-dominated Universe to those for a softer equation of state, we do not observe significant differences. This suggests that non-spherical effects are not much more pronounced for a moderately soft equation of state compared to a radiation fluid, contrary to some expectations in the literature. 
It is likely that non-spherical effects only become dominant in the almost pressureless systems in which the non-spherical effects are dominant impeding factors against gravitational collapse. 
Therefore, we conclude that for a moderately soft equation of state (such as the cases 
$w\geq1/10$ we tested), 
non-spherical effects do not significantly alter the threshold for black hole formation compared to the radiation case, $w=1/3$. This implies that the results from spherical simulations should provide a 
sufficiently accurate 
threshold value for black hole formation for practical use. 
Our numerical results have been used in our companion \textit{letter} \cite{companion}, showing that, non-spherical configurations do not have a significant impact on the PBH mass function
although they become highly significant in the critical regime in terms of the probability distribution of the initial amplitude (see Fig.\ref{fig:P_ep_integrated}). 
\acknowledgments
A.E. acknowledges support from the JSPS Postdoctoral Fellowships for Research in Japan (Graduate School of Sciences, Nagoya University). C.Y. is supported in part by JSPS KAKENHI Grant Nos. 20H05850 and 20H05853.

\appendix
\section{Analytical formulas for the mean values of $e$ and $p$}
\label{sec:peak_theory_details_formulas}
The mean values of $e$ and $p$ for the probability distribution of Eq.\eqref{eq:P_ep} are given by
%
\begin{align}
\label{eq:mean_e_numerical}
\langle e \rangle (\xi) &=\frac{3 \, \exp{-5 \xi ^2/2}}{200 \sqrt{10 \pi} \, \xi \,f(\xi)} \Biggl[ \exp{\frac{15 \xi ^2}{8}} \left(\sqrt{30 \pi } \left(15 \xi ^2+4\right) \erf\left(\frac{1}{2} \sqrt{\frac{15}{2}} \xi \right)+45 \xi \left(5 \xi ^2+12\right)\right)\\ \nonumber
&+4 \exp{\frac{5 \xi ^2}{2}} \left(3 \sqrt{10 \pi } \left(\erf\left(\sqrt{\frac{5}{2}} \xi \right)-\erf\left(\frac{1}{2}
\sqrt{\frac{5}{2}} \xi \right)\right)+50 \xi ^3-180 \xi \right)+60 \xi \Biggr], 
\end{align}
%
%
%
%
\begin{align}
\label{eq:mean_p_numerical}
\langle p \rangle (\xi) &= \frac{9 \, \exp{-5 \xi ^2/2}}{200 \sqrt{10 \pi} \, \xi \, f(\xi)} \Biggl[4 \sqrt{10 \pi } \exp{\frac{5 \xi ^2}{2}} \left(\erf\left(\frac{1}{2} \sqrt{\frac{5}{2}} \xi \right)+\erf\left(\sqrt{\frac{5}{2}} \xi \right)\right)\\ \nonumber
&+\exp{\frac{15 \xi ^2}{8}} \left(\sqrt{30 \pi } \left(15 \xi ^2+4\right) \erf\left(\frac{1}{2} \sqrt{\frac{15}{2}} \xi \right)-15 \xi  \left(5 \xi ^2+12\right)\right)+60 \xi \Biggr]. 
\end{align}
%
%
%
%
In the large $\xi$ limit, the probability distribution Eq.\eqref{eq:P_ep} can be indeed approximated by a Gaussian and takes 
the
simpler form,
\begin{equation}
\label{eq:prob_e_p_large_nu}
    P_{\rm ep}(e,p) \approx P_{\rm e,p} \exp \left[ -\frac{(e-e_m)^2}{2 \sigma^2_e} - \frac{(p-p_m)^2}{2 \sigma^2_p}\right], 
\end{equation}
where the mean values $e_m$ and $p_m$ and dispersions $\sigma_e$ and $\sigma_p$  
are given by
\begin{align}
\label{eq:mean_values_dispersione}
    e_m = \frac{1}{\sqrt{5} \xi \sqrt{1+6/(5 \xi^2)}} , \, \,\, \, \, \, \sigma_e = \frac{e_m}{\sqrt{6}},  \\
    \label{eq:mean_values_dispersionp}
    p_m = \frac{1}{\sqrt{5} \xi^4 [1+6/(5 \xi^2)]^2} , \, \,\, \, \, \, \sigma_p = \frac{e_m}{\sqrt{3}}. 
\end{align}

\section{High peaks for PBH formation assuming spherical symmetry}
\label{sec:appendix_high_peaks}
In this Appendix, under the assumption of spherical symmetry, 
we quantify the height of the peaks $\nu_{\rm c}$ 
fixing 
the ratio of PBHs in dark matter $f_{\rm PBH}^{\rm tot}$.
We consider the case of a monochromatic power spectrum $\mathcal{P}_{\zeta} =\mathcal{A}_{\zeta} \delta(\ln(k/k_p))$ as in Sec.~\ref{sec:peak_theory_nonspherical}. To statistically compute the abundance of peaks leading to black hole formation, 
we follow the approach of \cite{Yoo:2018kvb,Yoo:2020dkz} based on the Gaussian statistics of $\zeta$ 
(see also \cite{Domenech:2024rks} where the approach was also used in the context of the PTA analysis with arbitrary $w$).

The typical profile for the monochromatic power spectrum (which is equivalent to the mean profile for that particular case) is given by
\begin{equation}
    \zeta_{\rm sp} = \mu \, \sinc(k_{p }  r). 
\end{equation}
Notice that for that case, the statistical procedure is simplified since only one relevant scale is involucrated, given by $k_p$. 
The normalized height of the peak is then given by $\nu= \mu /\sqrt{A_{\zeta}}$. 
The number of peaks in terms of $\nu$ is computed as,
\begin{equation}
    n_{\rm pk}(\nu) = \frac{1}{(2 \pi)^2} \frac{1}{3^{3/2}} k^{3}_p f\left(  \nu \right) \frac{1}{\sqrt{  \mathcal{A}_{\zeta}}} e^{-\frac{1}{2} \nu^2}, 
\end{equation}
where $f(\nu)$ is the function introduced in Eq.\eqref{eq:f}. 
For the monochromatic spectrum, the number density of PBHs $n_{\rm PBH}(M)$ for a fixed mass $M$ is simply given by \footnote{Notice that for the general case, the relation between $n_{\rm PBH}(M)$ will involve integration over the variable $\xi$ introduced in Sec.~\ref{sec:peak_theory_nonspherical}, but for the monochromatic power spectrum, this is simply given by $\xi = \nu$.}
\begin{equation}
    n_{\rm PBH}(M) d\ln M = n_{\rm pk}(\nu) \bigg| \frac{d \ln M}{d \mu}  \bigg|^{-1} d \ln M, 
\end{equation}
where the Jacobian reads
\begin{equation}
  \bigg| \frac{d \ln M}{d \mu} \bigg| = \bigg| \frac{3(1+w)}{1+3w} \frac{d \zeta_{\rm sp}(r_m)}{d \mu}+\frac{\gamma(w)}{\mu-\mu_{\rm c,sp}} \bigg|
  \label{eq:JacoblnMmu}
\end{equation}
with $r_m \approx 2.74/ k_p$. The PBH mass function, defined as the fraction of PBHs in the form of dark matter at the current time, is finally given by
\begin{equation}
    f_{\rm PBH}(M) = \frac{M \, n_{\rm PBH}(M)}{3 M^{2}_{\rm pl} H^2_0 \Omega_{\rm DM}}.  
\end{equation}
Integrating $f_{\rm PBH}(M)$, we can fix $\mathcal{A}_{\rm \zeta}$ for a desired value of $f_{\rm PBH,tot}$ and infer the value of $\nu_{\rm c}$ using the critical threshold value $\mu_{\rm c,sp}$. 
To obtain accurately the threshold values $\mu_{\rm c,sp}$, relativistic numerical simulations are necessary. We use the SPriBoSH code \cite{escriva_solo} to compute those values, which were already computed and are shown in Fig. 11 of \cite{Escriva:2022duf}.

\begin{figure}[!htbp]
\centering
\includegraphics[width=2.6 in]{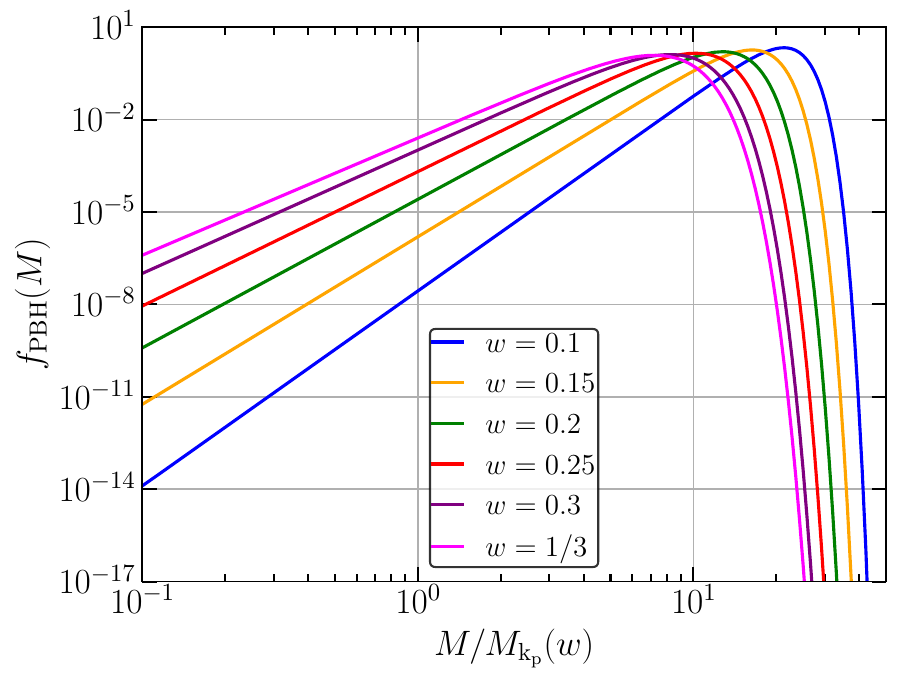}
\includegraphics[width=2.6 in]{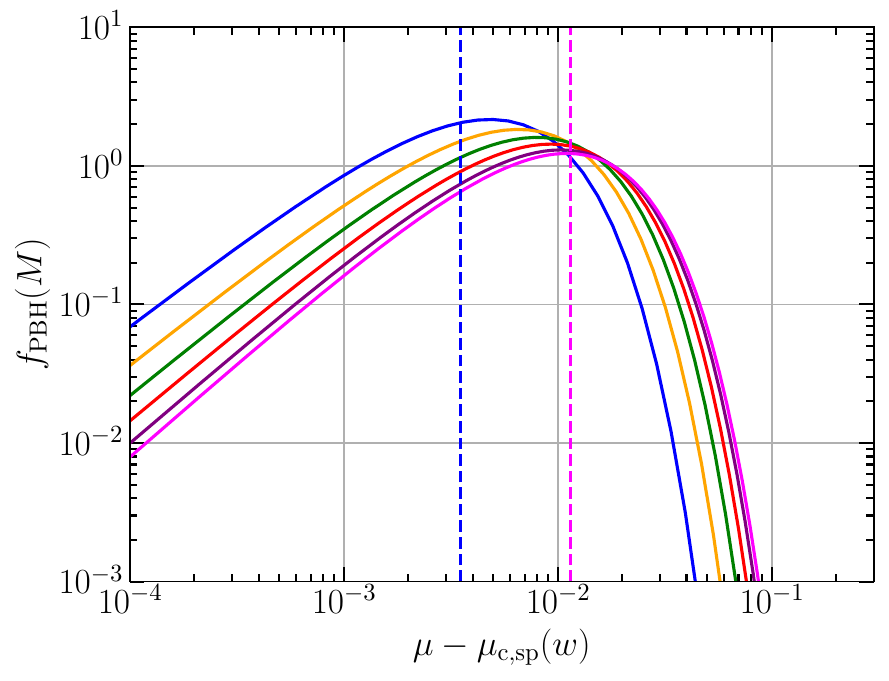}
\includegraphics[width=2.6 in]{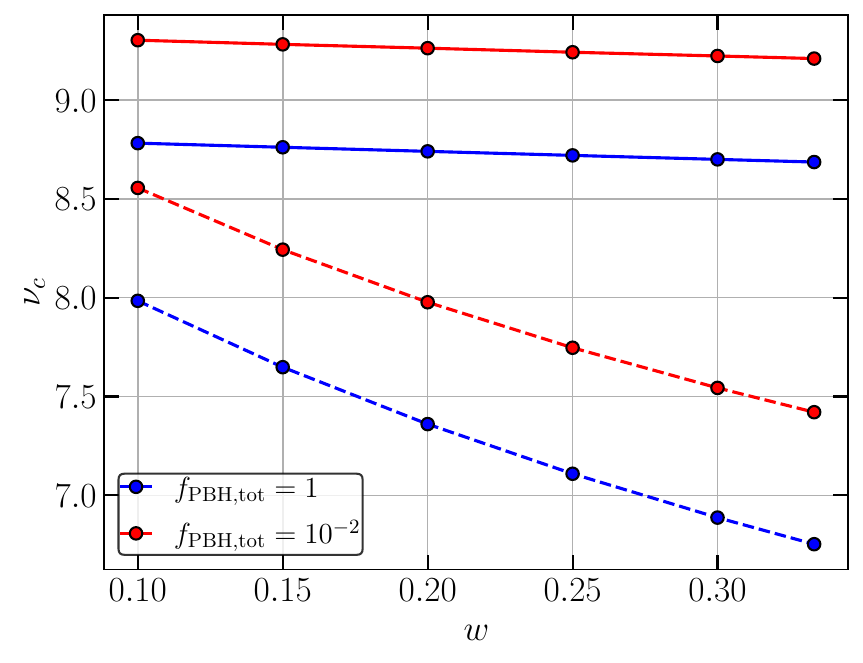}
\includegraphics[width=2.6 in]{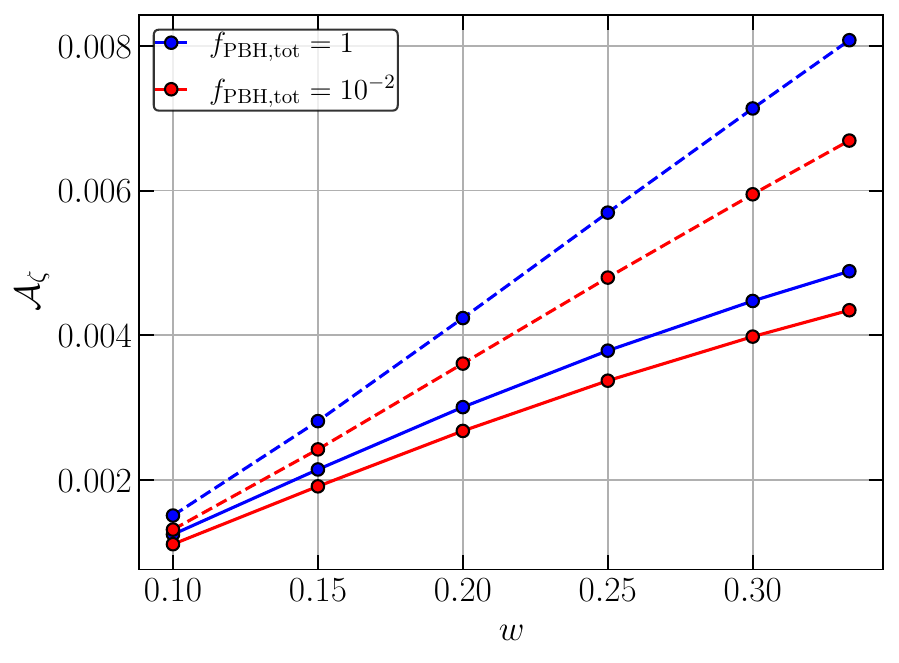}
\includegraphics[width=2.6 in]{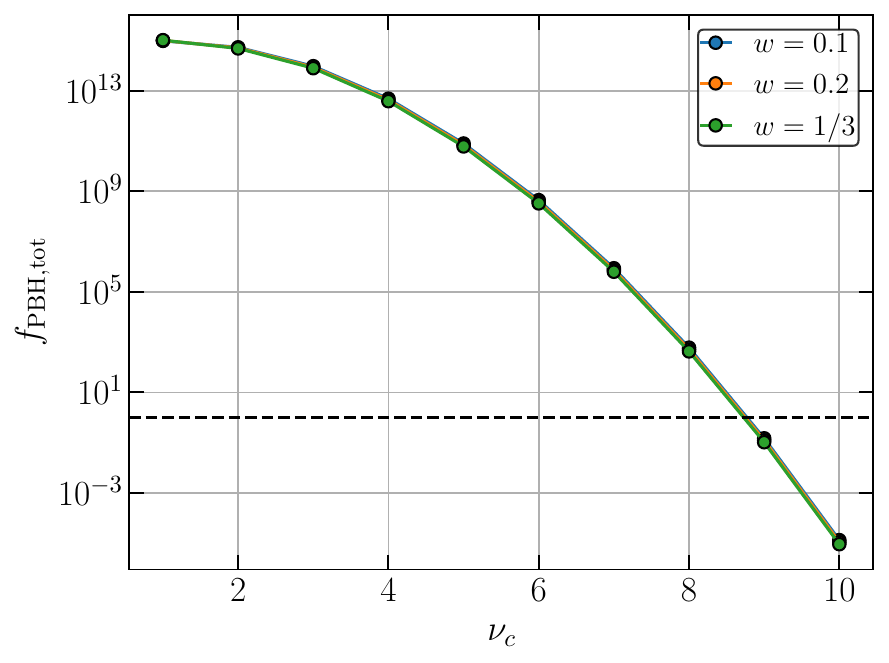}
\caption{Top panels: PBH mass functions for 
several values of $w$ with 
the monochromatic power spectrum 
as functions of 
$M/M_{k_p}$ (left) and $\mu-\mu_c$ (right) with $M_{k_p} = (g_{\star}/106.75)^{-1/6} \cdot 10^{20} \left(k_p/(1.56 \cdot 10^{13} \textrm{Mpc}^{-1})\right)^{-3(1+w)/(1+3w)} M_{\odot}$. 
We fix $k_p=10^{13.5}\textrm{Mpc}^{-1}$ (asteroid mass range) with $g_* = 106.75$ in both cases. 
The dashed vertical lines in the left panel denote the typical values $\mu_t$ for $w=1/10$ and $w=1/3$ used in the simulations. Middle panels: 
Critical values of $\nu_c$ (left) and the amplitude of the power spectrum $\mathcal{A}_{\zeta}$ (right) as functions of $w$. 
The solid and dashed lines correspond to $k_p=10^{13.5}$ (asteroid mass range) and $k_p=10^{7}$ (solar mass range), respectively, 
with $g_* = 10.75$. 
Bottom panel: Value of $f_{\rm PBH}^{\rm tot}$ 
as a function of $\mu_c$ for 
$k_p = 10^{13.5}\textrm{Mpc}^{-1}$.}
\label{fig:mass_functions_appendix}
\end{figure}


In Fig.\ref{fig:mass_functions_appendix} we show 
the results for several values of $w$. 
In the top-left panel, we show the mass functions $f_{\rm PBH}(M)$ 
with
$f_{\rm PBH}^{\rm tot}=1$. 
Notice that the slope of the mass function 
in
small $M$ 
originates from 
the critical exponent $\gamma$ of the critical collapse since the mass functions will grow as $f_{\rm PBH}(M) \sim M^{1+1/\gamma(w)}$ due to the Jacobian term (see Eq.~\eqref{eq:JacoblnMmu} with $M\sim (\mu-\mu_{\rm c,sp})^{\gamma(w)}$ from the critical collapse regime \cite{choptuik,Evans:1994pj,Koike:1995jm,Niemeyer:1999ak,Musco:2012au}), where the term $\gamma(w)/(\mu-\mu_{\rm c,sp})$ will become dominant when the fluctuations are in the critical regime $\mu \rightarrow \mu_{\rm c,sp}$. In the top-right panel, we show the same $f_{\rm PBH}(M)$ but 
as functions of
$\mu-\mu_{\rm c,sp}(w)$. 
Notice that the mass function peaks when $\mu-\mu_{\rm c,sp} \approx 10^{-2}$; we consider these as typical values of the amplitude of fluctuation, 
where the maximal number of PBHs is produced. 
In the middle left panel, we show the critical $\nu_{\rm c}$ (when $\mu \rightarrow \mu_{\rm c,sp}$), where it is explicitly shown that we need $\nu \gg 1$ (a large peak) to have $f_{\rm PBH}=1$. The fact that $\nu_{\rm c}$ is increased when the equation of state parameter becomes softer can be understood from the fact that in this scenario, the production of PBHs will be higher due to the threshold reduction, then for the same $f_{\rm PBH}$, the needed amplitude of the power spectrum will be reduced accordingly (see the middle right panel). Notice that we need $\nu_{\rm c} \gtrsim 7$ to have $f_{\rm PBH}^{\rm tot} \leq 1$ for $w \leq1/3$. 
Finally, in the bottom panel, we show an example of how PBHs are overproduced if we consider a lower value $\nu_{\rm c}$ (larger value of $\mathcal{A}_{\zeta}$)
for the monochromatic power spectrum. 


\section{Suplemental figures of the gravitational collapse for $w=1/10$}
\label{sec:appendix_suplemental_figures}

\begin{figure}[!htbp]
\centering
\includegraphics[width=1.5 in]{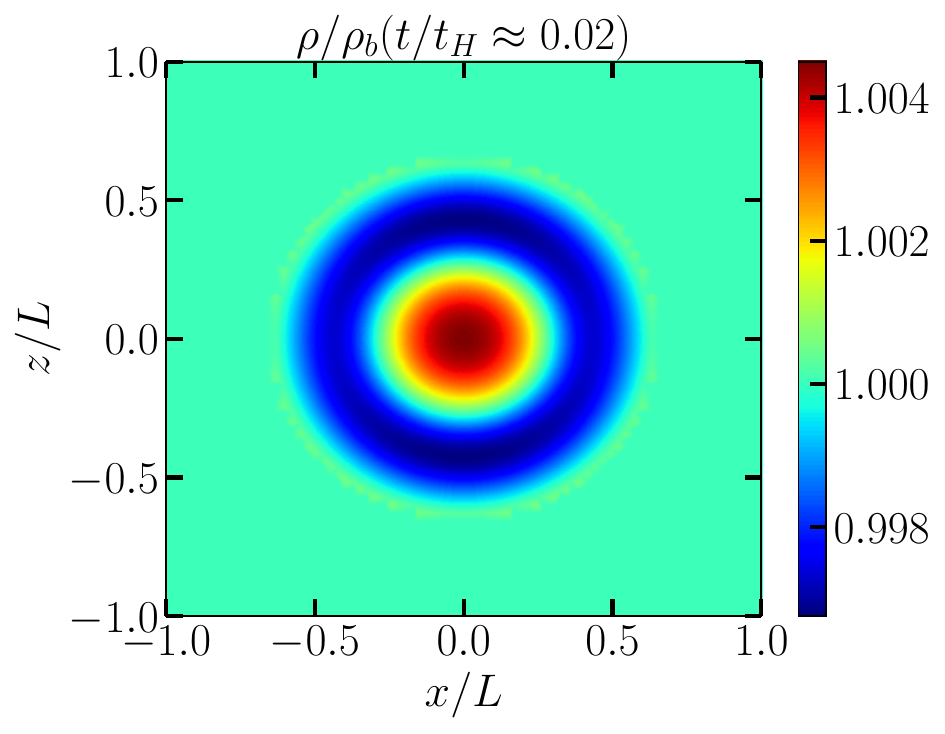}
\hspace*{-0.3cm}
\includegraphics[width=1.5 in]{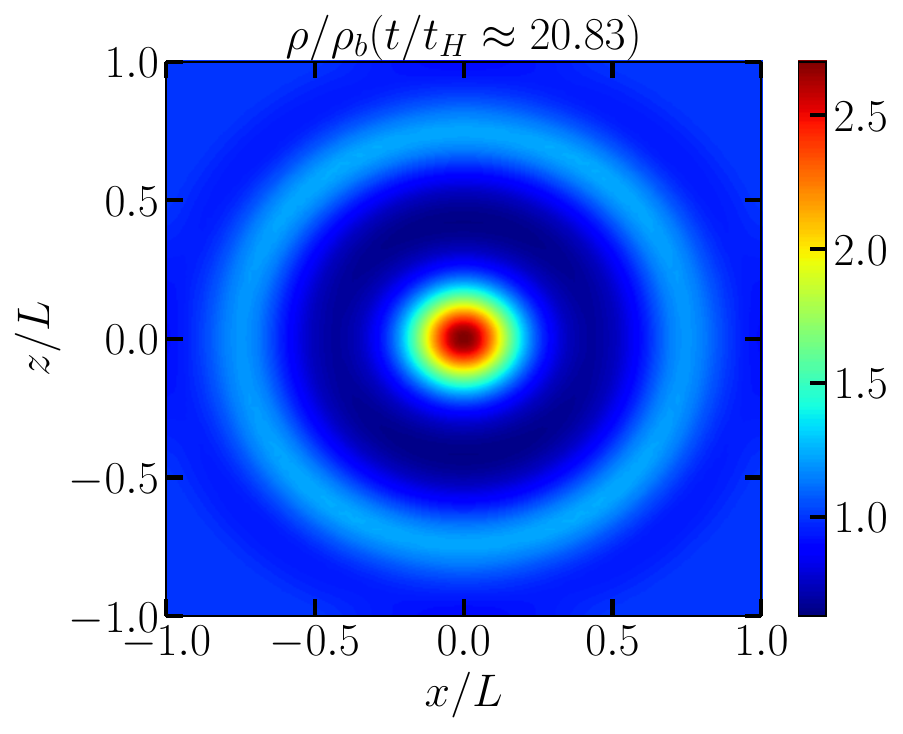}
\hspace*{-0.3cm}
\includegraphics[width=1.5 in]{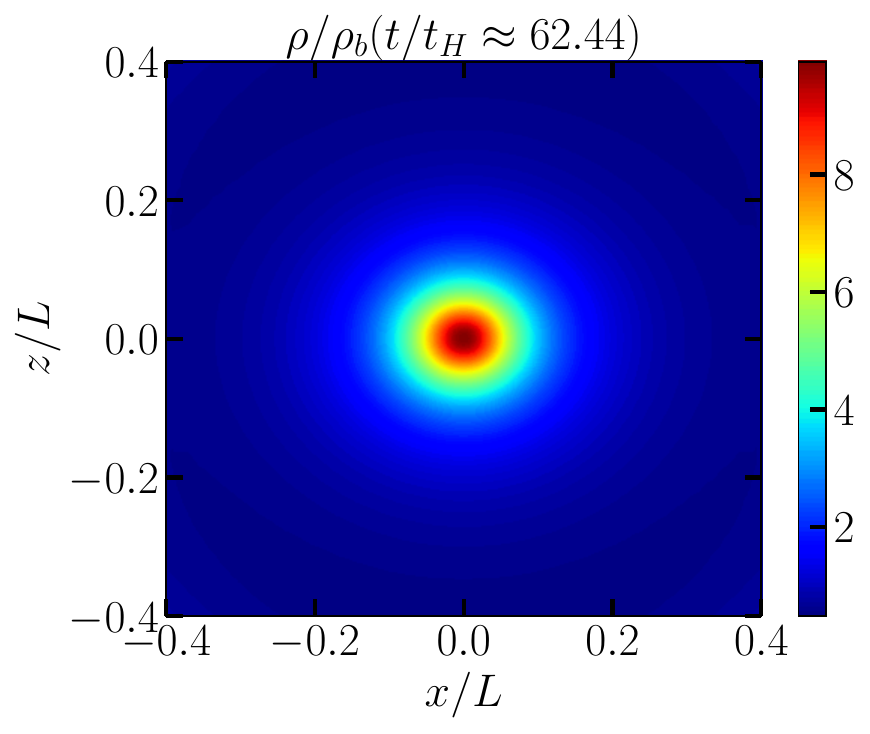}
\hspace*{-0.3cm}
\includegraphics[width=1.5 in]{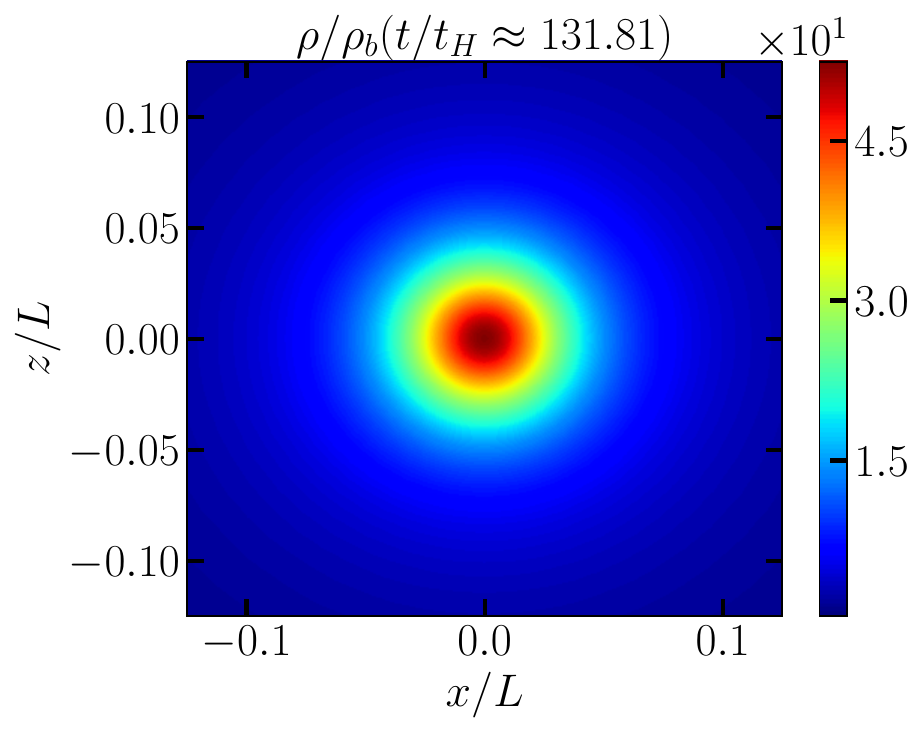}
\hspace*{-0.3cm}
\includegraphics[width=1.5 in]{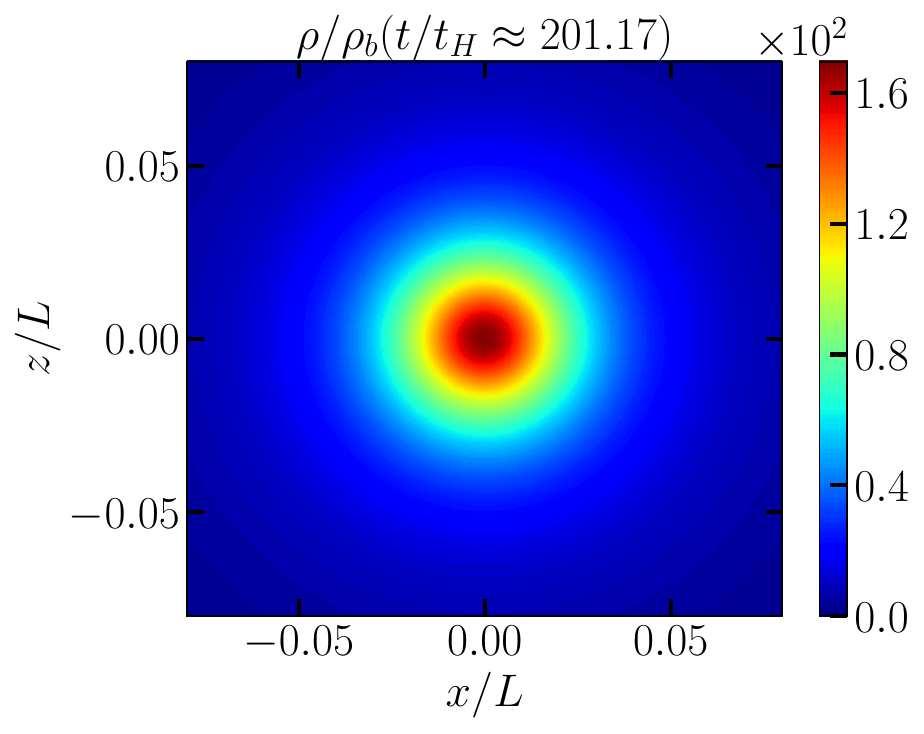}
\hspace*{-0.3cm}
\includegraphics[width=1.5 in]{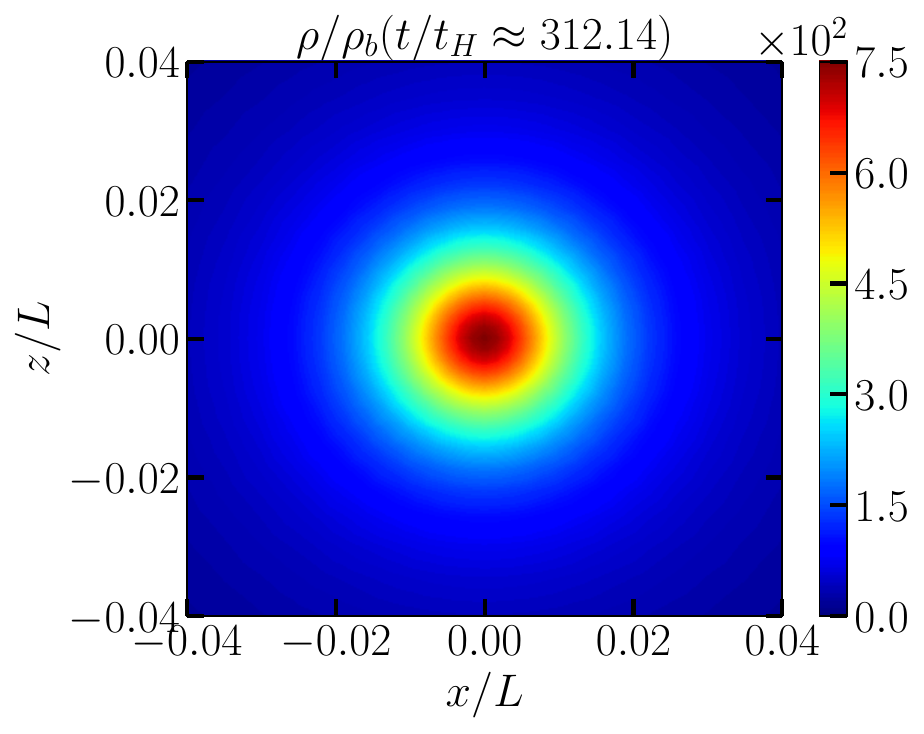}
\hspace*{-0.3cm}
\includegraphics[width=1.5 in]{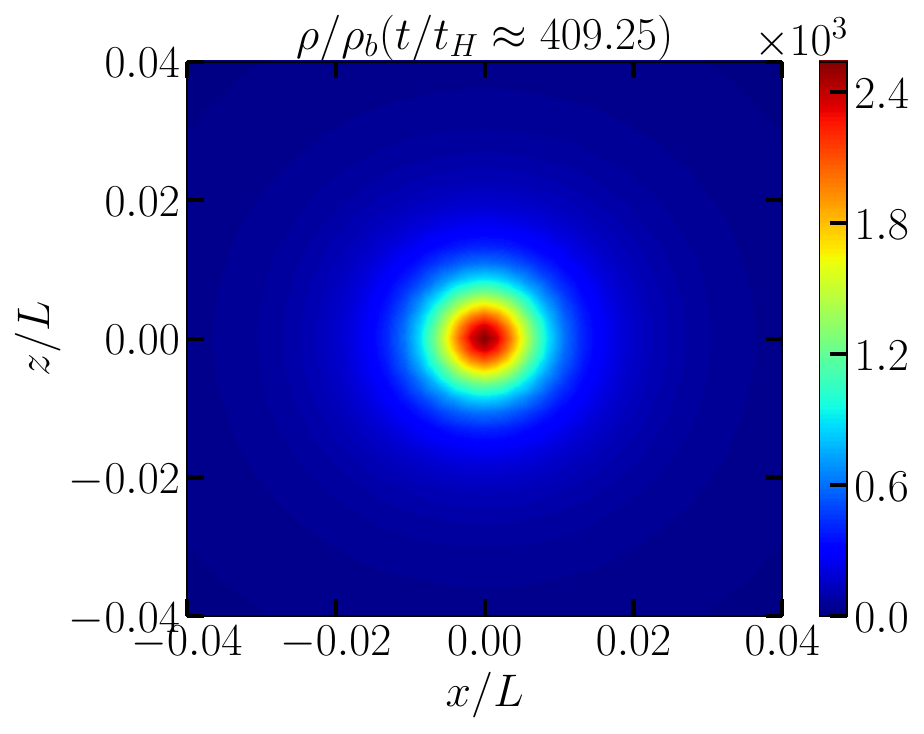}
\hspace*{-0.3cm}
\includegraphics[width=1.5 in]{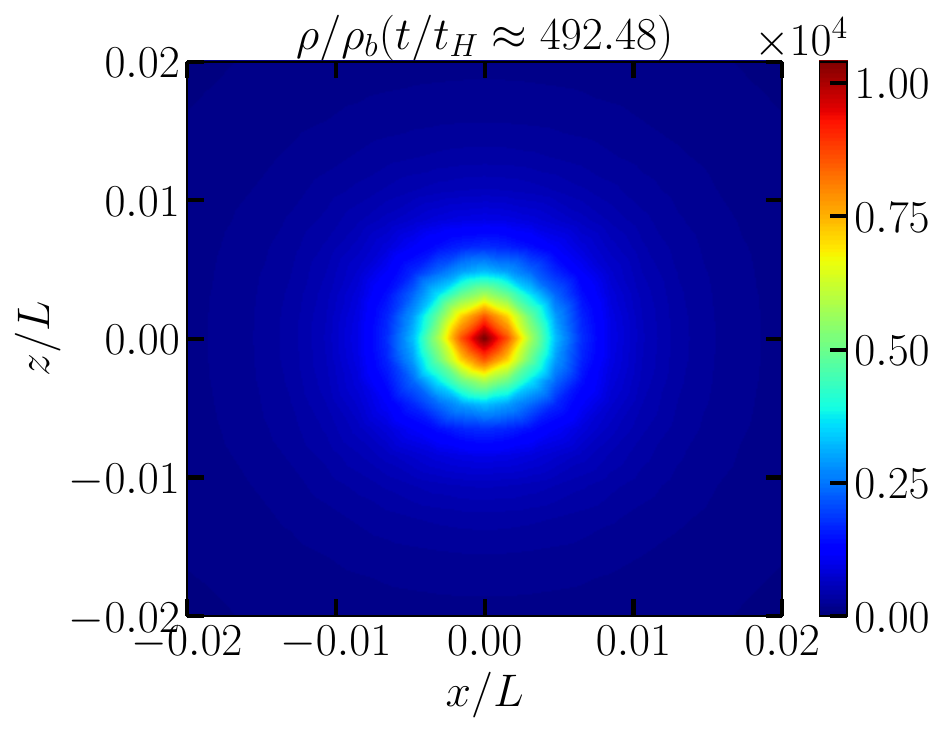}
\caption{Snapshots of the evolution of $\rho/\rho_b$ in the plane $y=0$ 
for
$e=0.075$ and  $p=-0.05$ with $w=1/10$.}
\label{fig:energy_density_ratio_soft_collapse}
\end{figure}

\begin{figure}[!htbp]
\centering
\includegraphics[width=1.5 in]{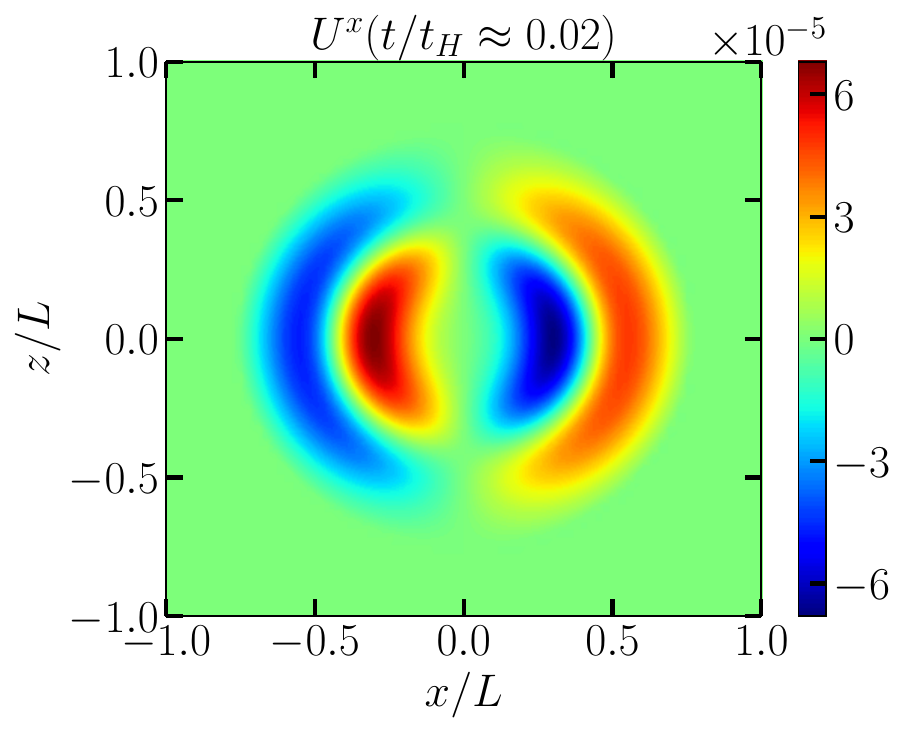}
\hspace*{-0.3cm}
\includegraphics[width=1.5 in]{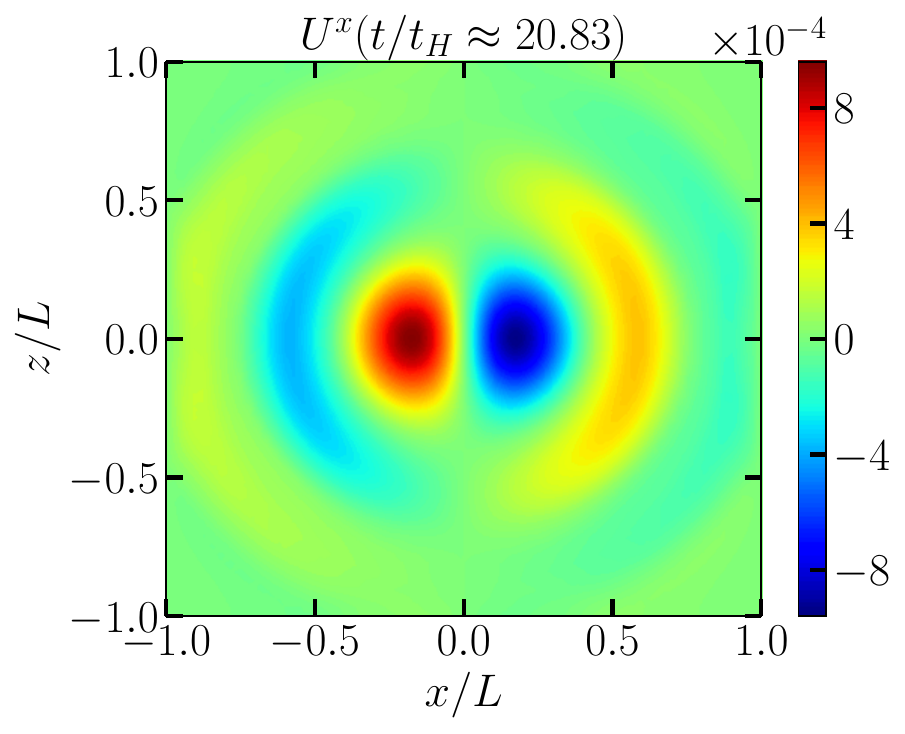}
\hspace*{-0.3cm}
\includegraphics[width=1.5 in]{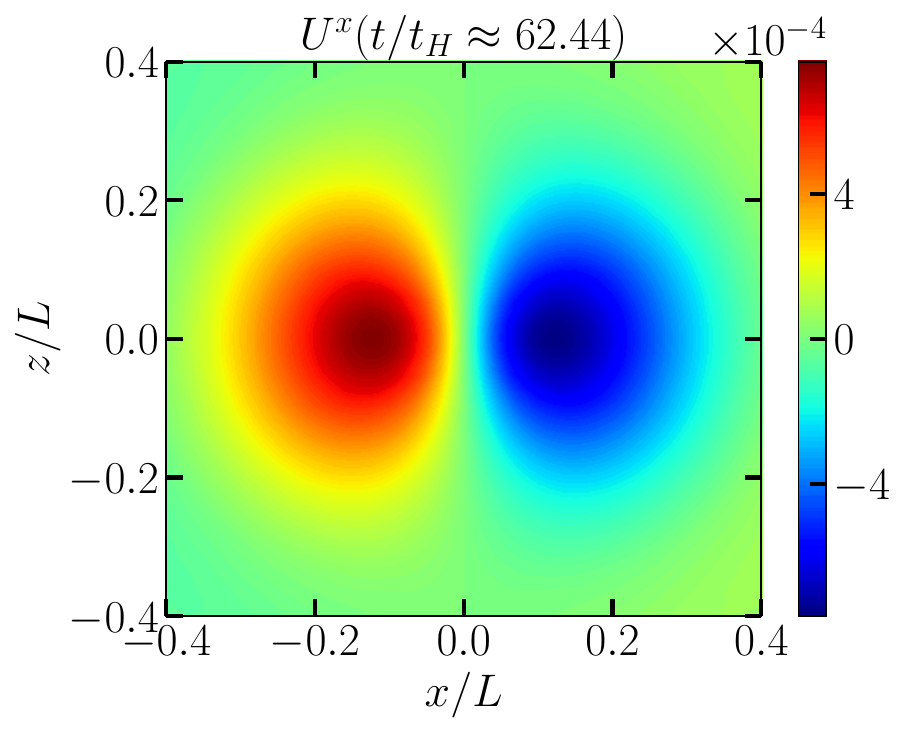}
\hspace*{-0.3cm}
\includegraphics[width=1.5 in]{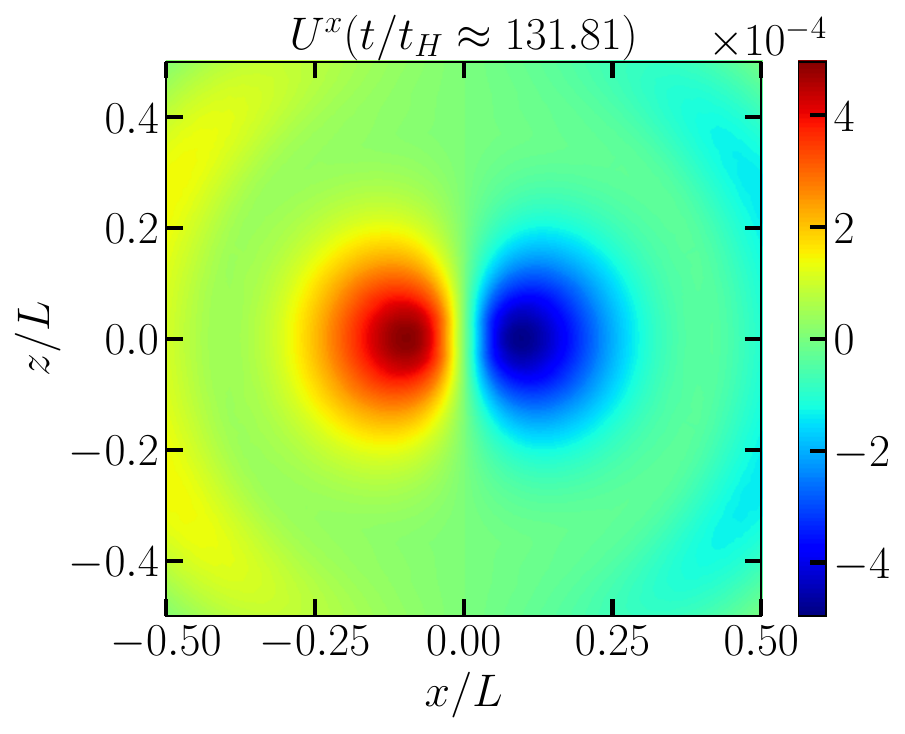}
\hspace*{-0.3cm}
\includegraphics[width=1.5 in]{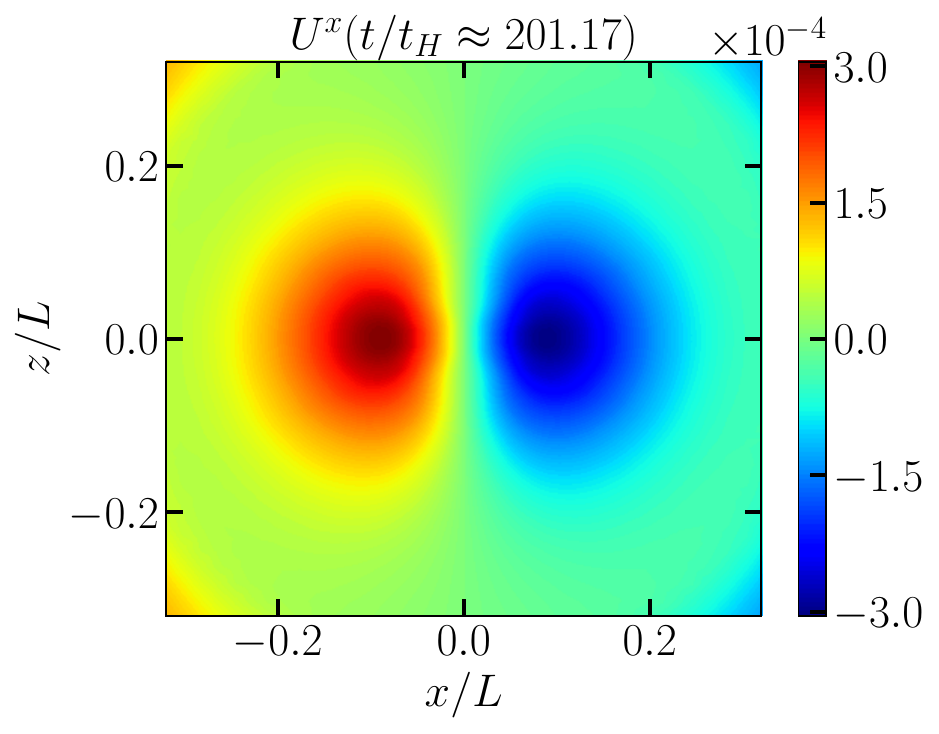}
\hspace*{-0.3cm}
\includegraphics[width=1.5 in]{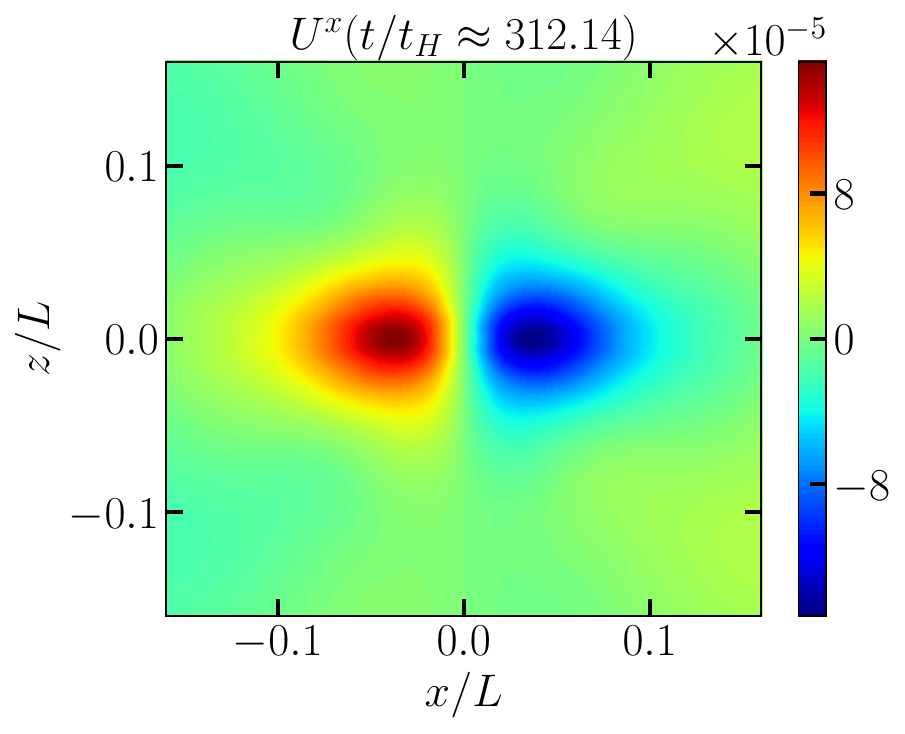}
\hspace*{-0.3cm}
\includegraphics[width=1.5 in]{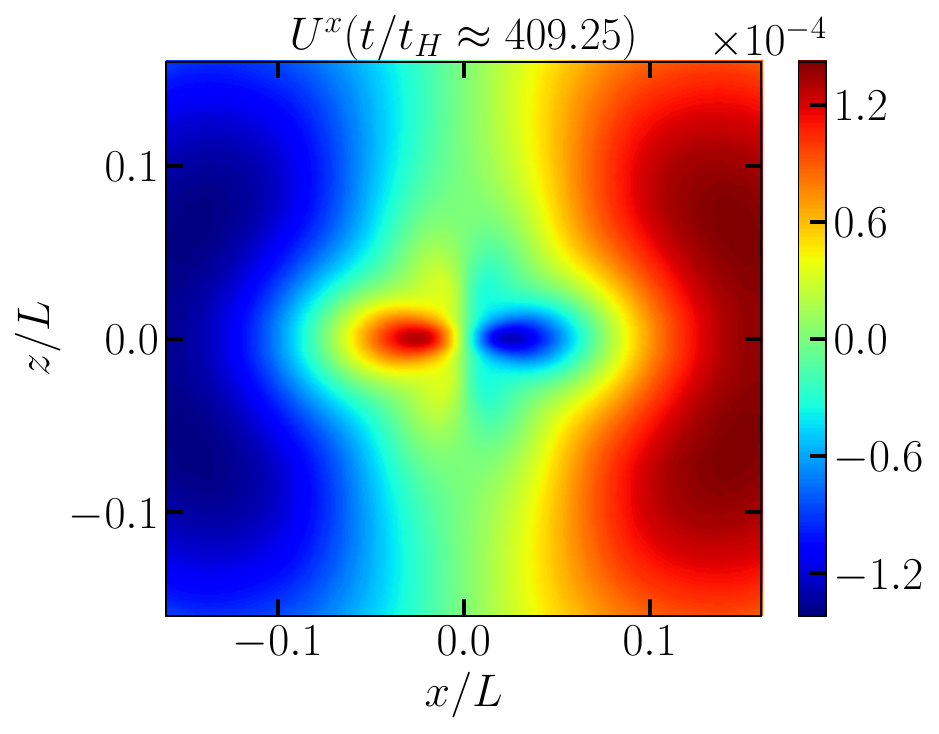}
\hspace*{-0.3cm}
\includegraphics[width=1.5 in]{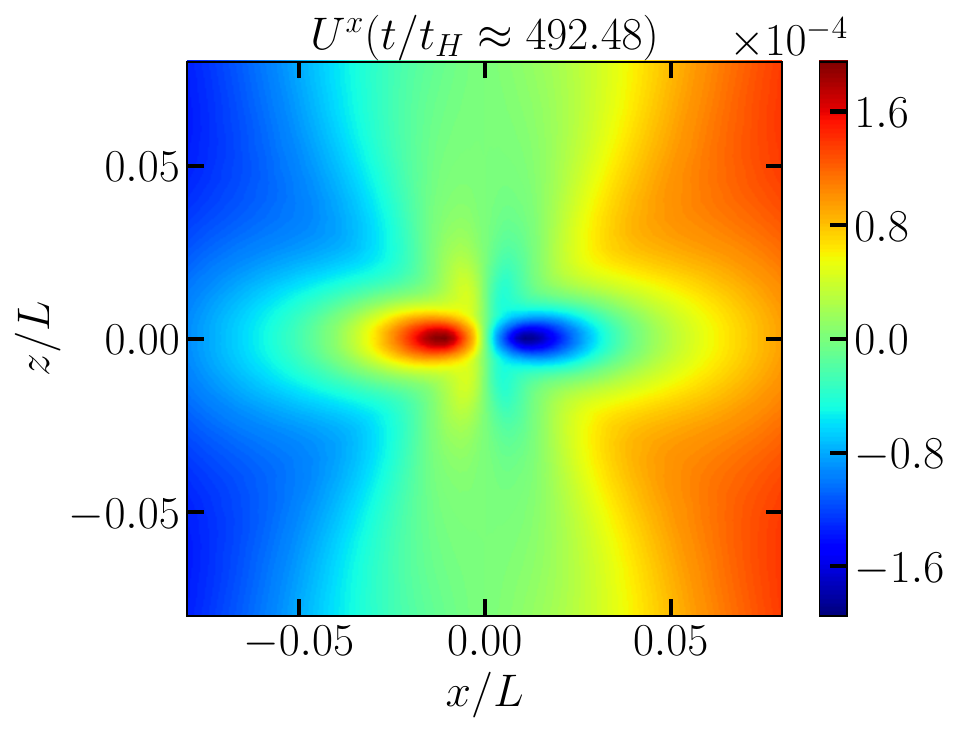}
\caption{Snapshots of the evolution of the Eulerian velocity $U^{x}$ in the plane $y=0$
for
$e=0.075$ and $p=-0.05$ with $w=1/10$.}
\label{fig:vel_soft_x_collapse}
\end{figure}

\begin{figure}[!htbp]
\centering
\includegraphics[width=1.5 in]{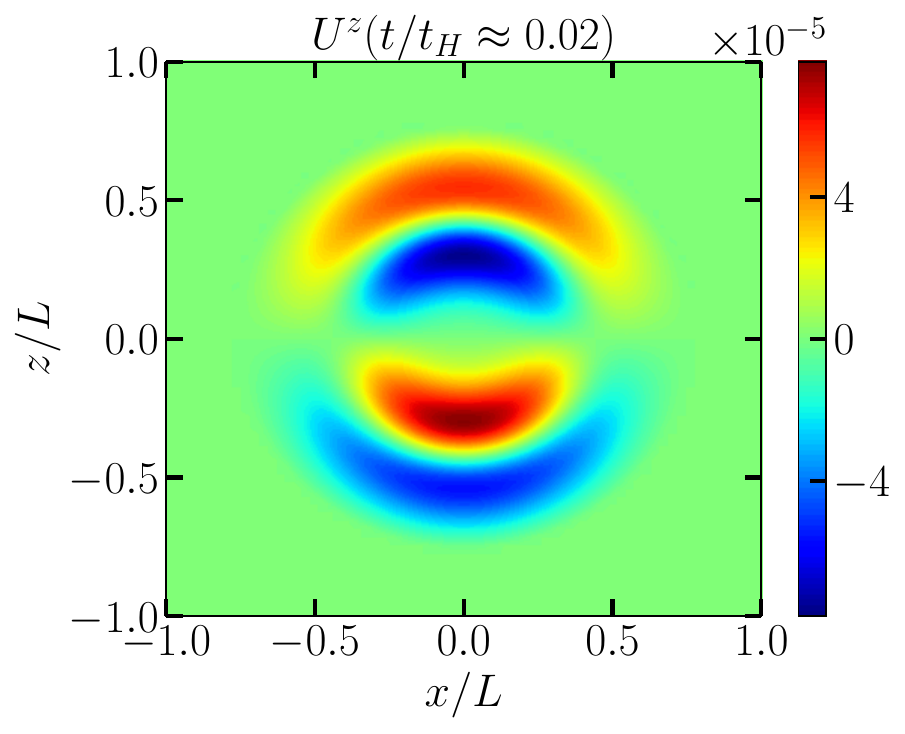}
\hspace*{-0.3cm}
\includegraphics[width=1.5 in]{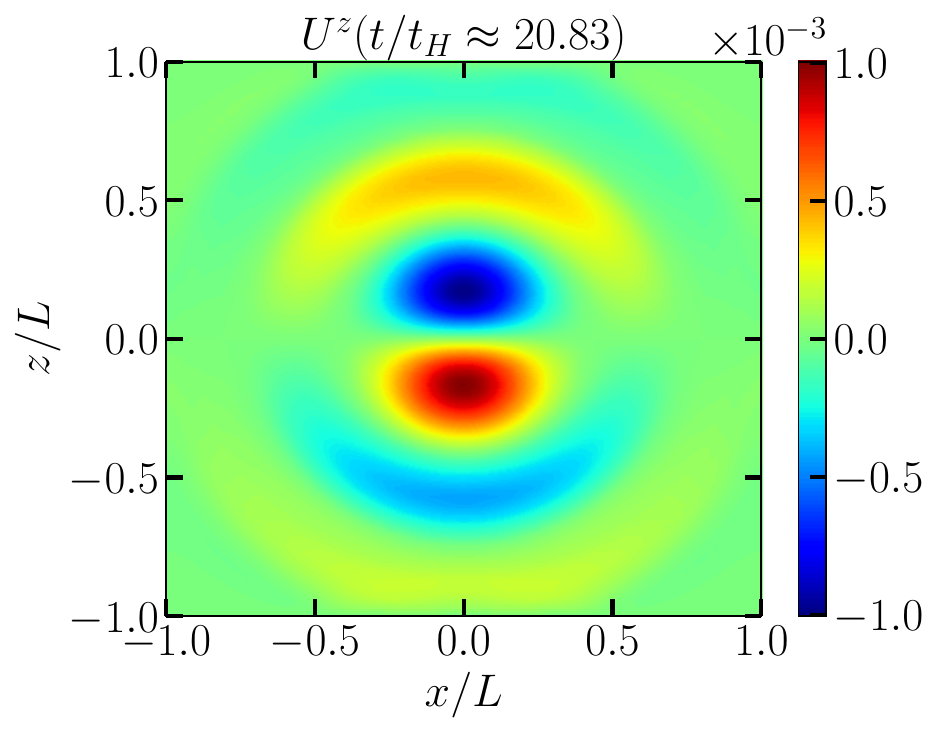}
\hspace*{-0.3cm}
\includegraphics[width=1.5 in]{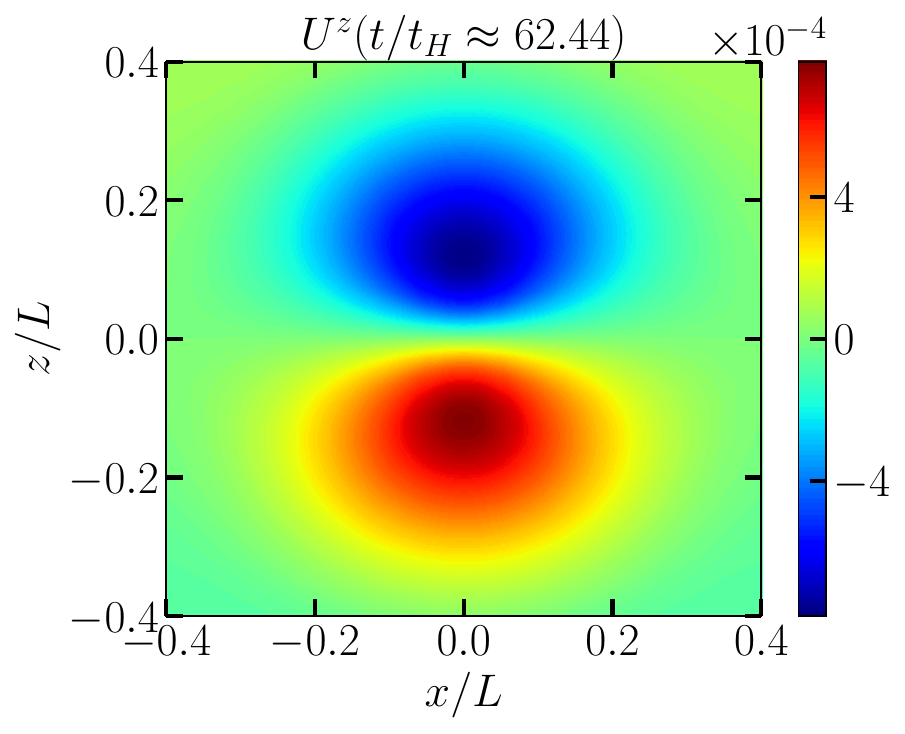}
\hspace*{-0.3cm}
\includegraphics[width=1.5 in]{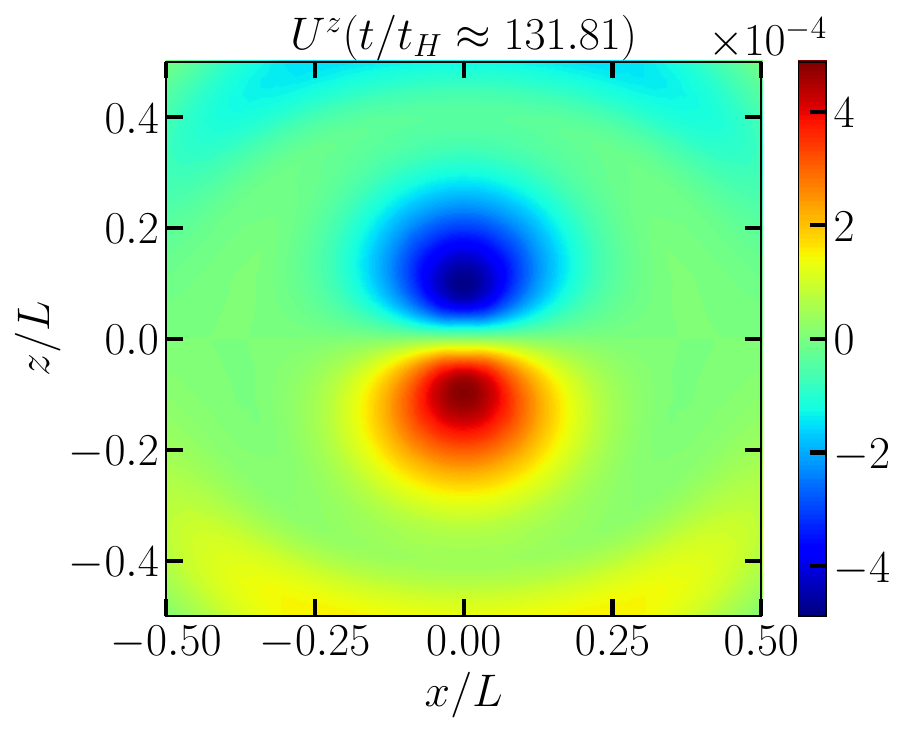}
\hspace*{-0.3cm}
\includegraphics[width=1.5 in]{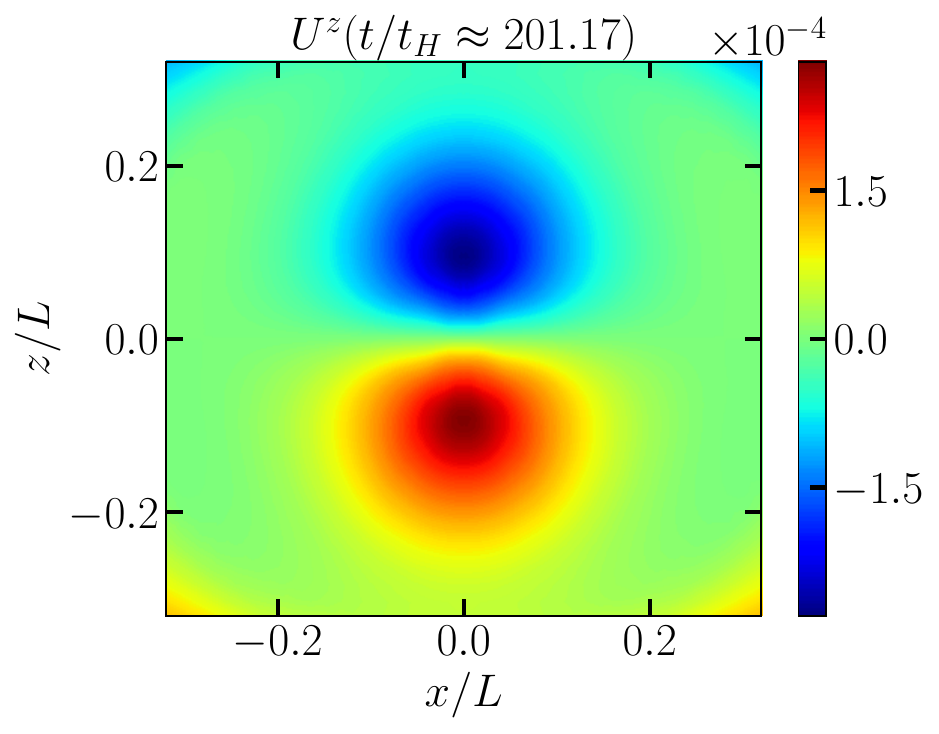}
\hspace*{-0.3cm}
\includegraphics[width=1.5 in]{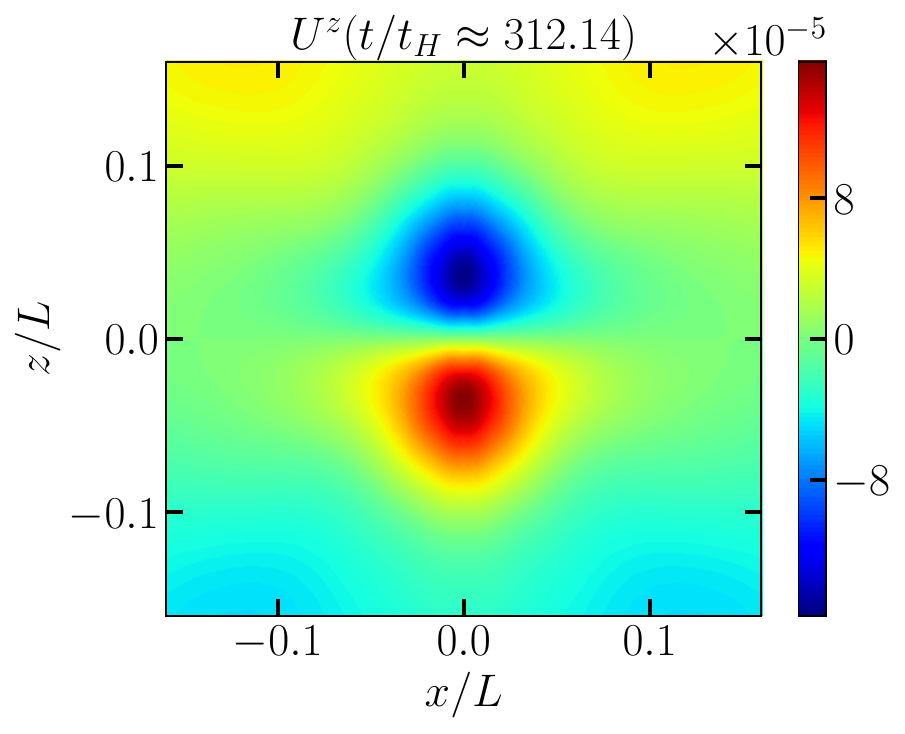}
\hspace*{-0.3cm}
\includegraphics[width=1.5 in]{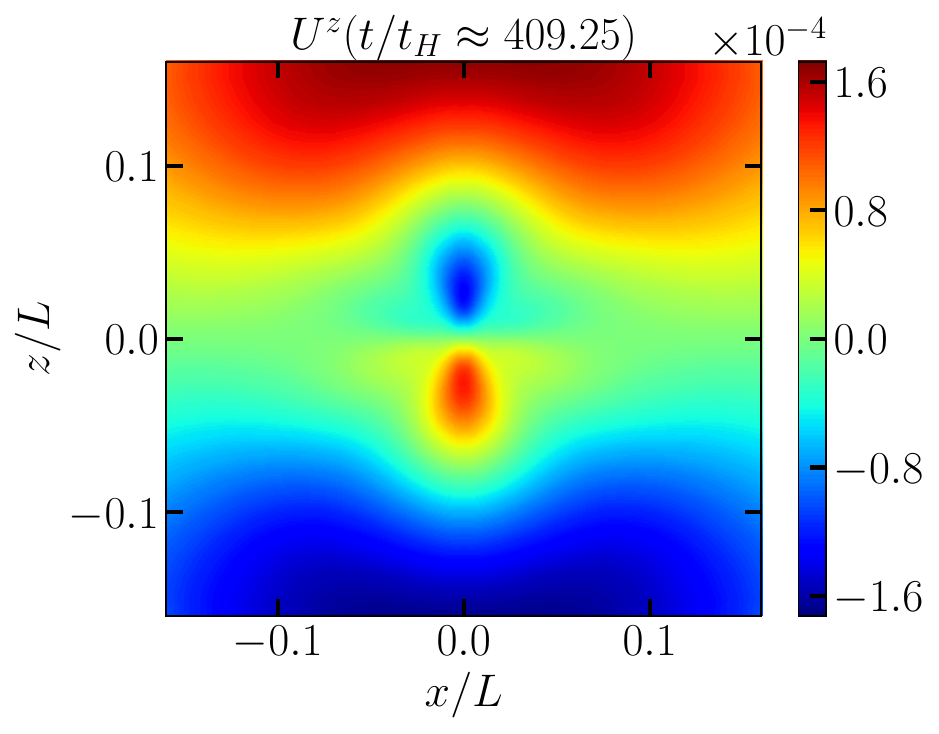}
\hspace*{-0.3cm}
\includegraphics[width=1.5 in]{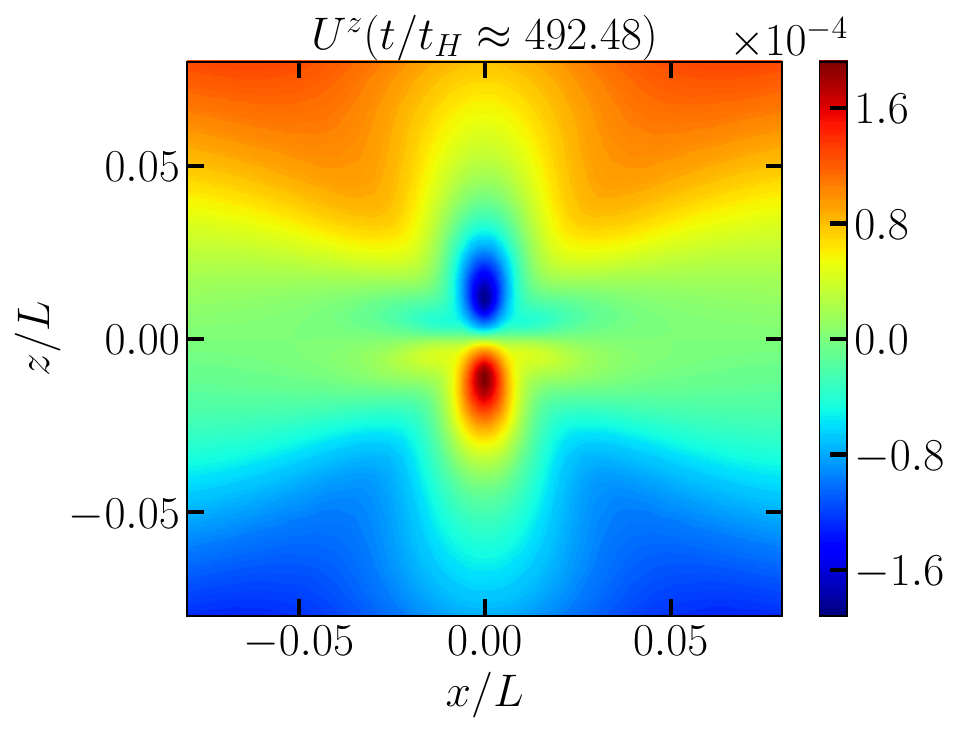}
\caption{Snapshots of the evolution of the Eulerian velocity $U^{z}$ in the plane $y=0$
for
$e=0.075$ and $p=-0.05$ with $w=1/10$.}
\label{fig:vel_soft_z_collapse}
\end{figure}

\begin{figure}[!htbp]
\centering
\includegraphics[width=1.9 in]{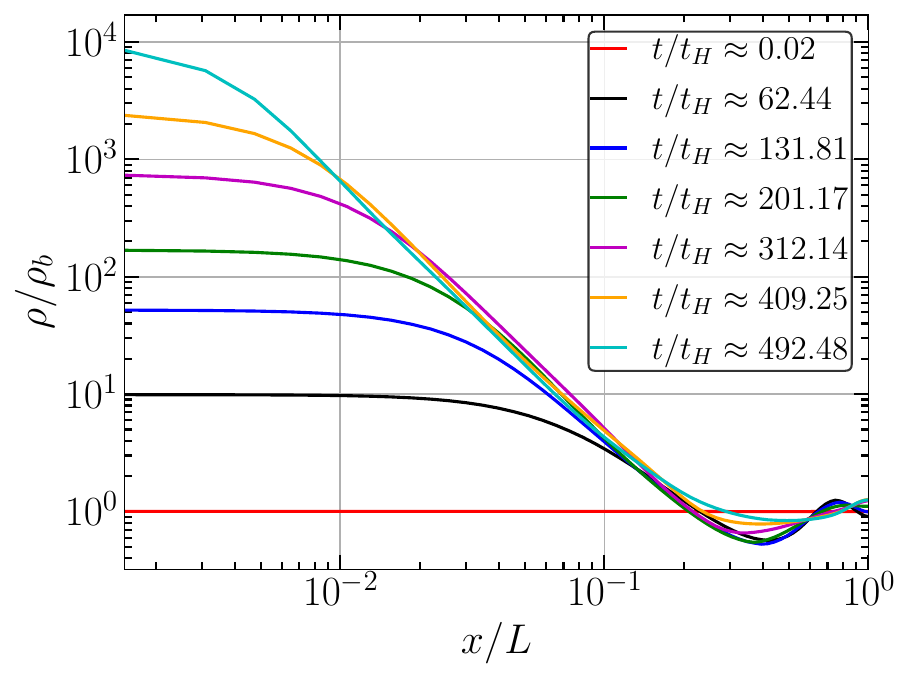}
\includegraphics[width=1.9 in]{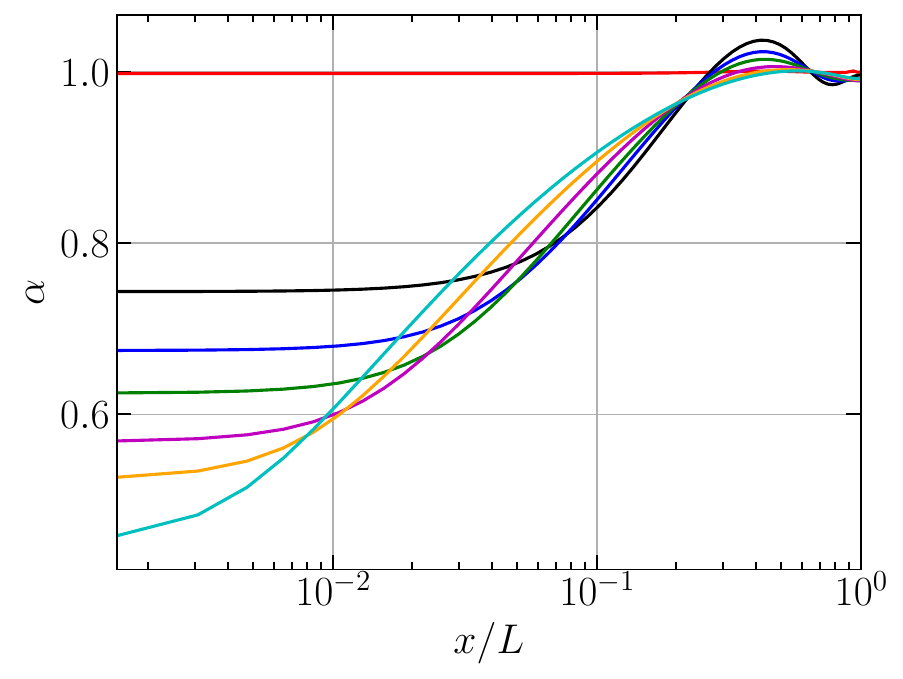}
\includegraphics[width=2.0 in]{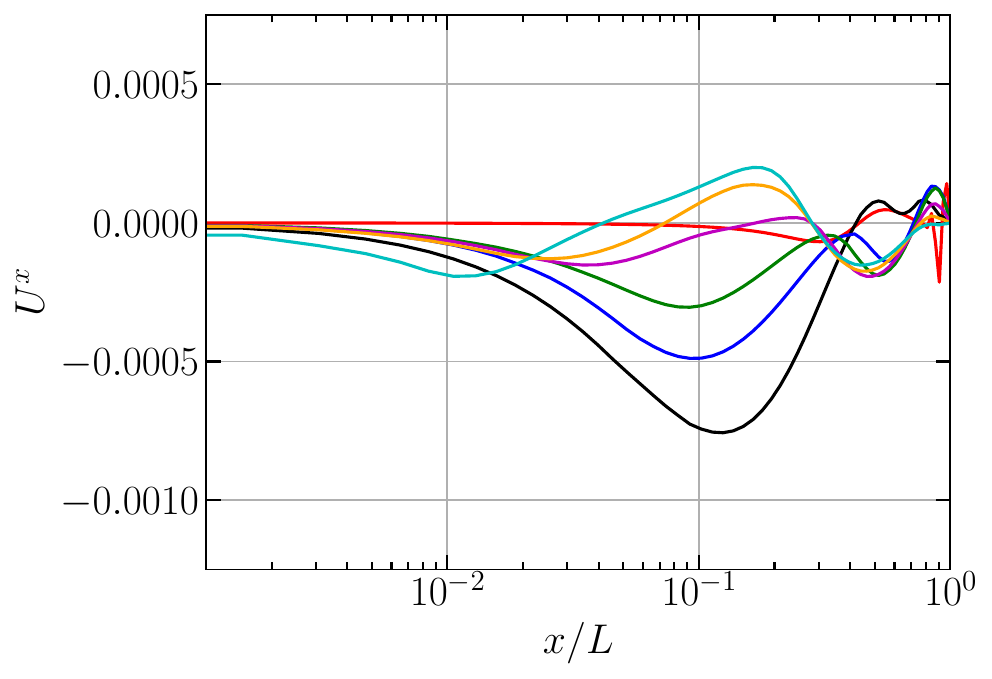}
\caption{
Snapshots of the energy density $\rho/\rho_b$ (left-panel), lapse function $\alpha$ (middle-panel) and Eulerian velocity $U^{x}$ (right-panel)
on the $x$ axis ($y=z=0$) for 
$e=0.075$ and $p=-0.05$ with $w=1/10$. 
}
\label{fig:projection_variables_collapse_soft}
\end{figure}

\begin{figure}[!htbp]
\centering
\includegraphics[width=1.5 in]{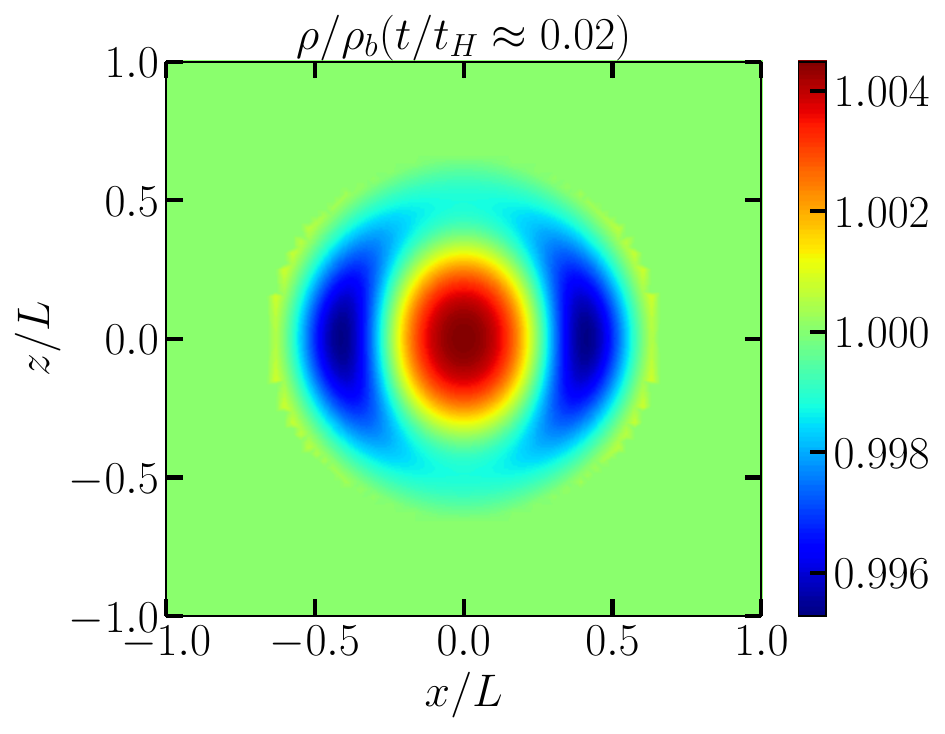}
\hspace*{-0.3cm}
\includegraphics[width=1.5 in]{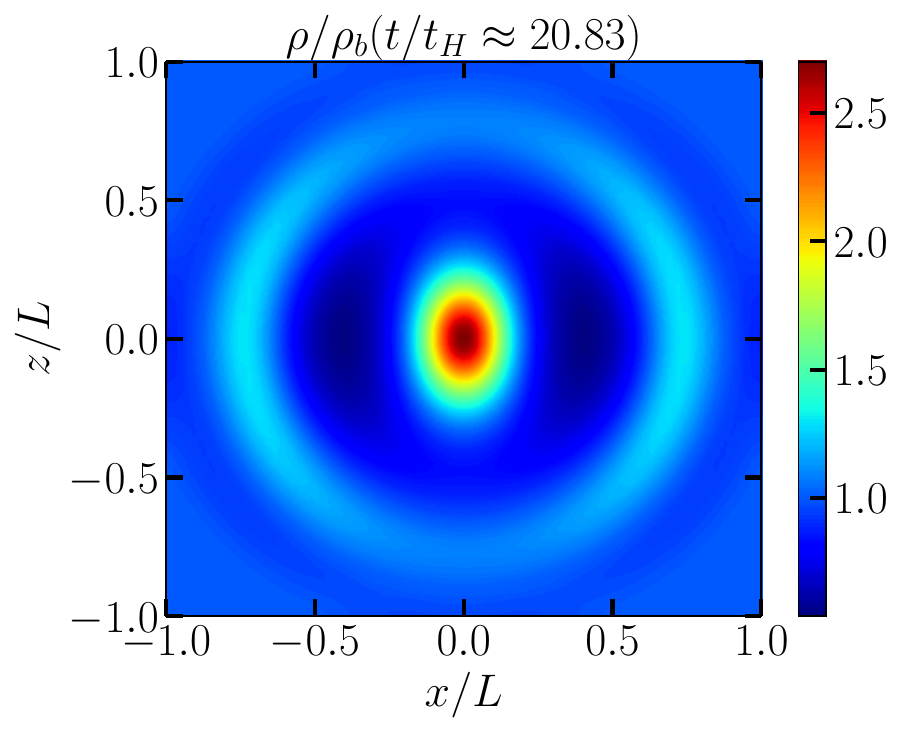}
\hspace*{-0.3cm}
\includegraphics[width=1.5 in]{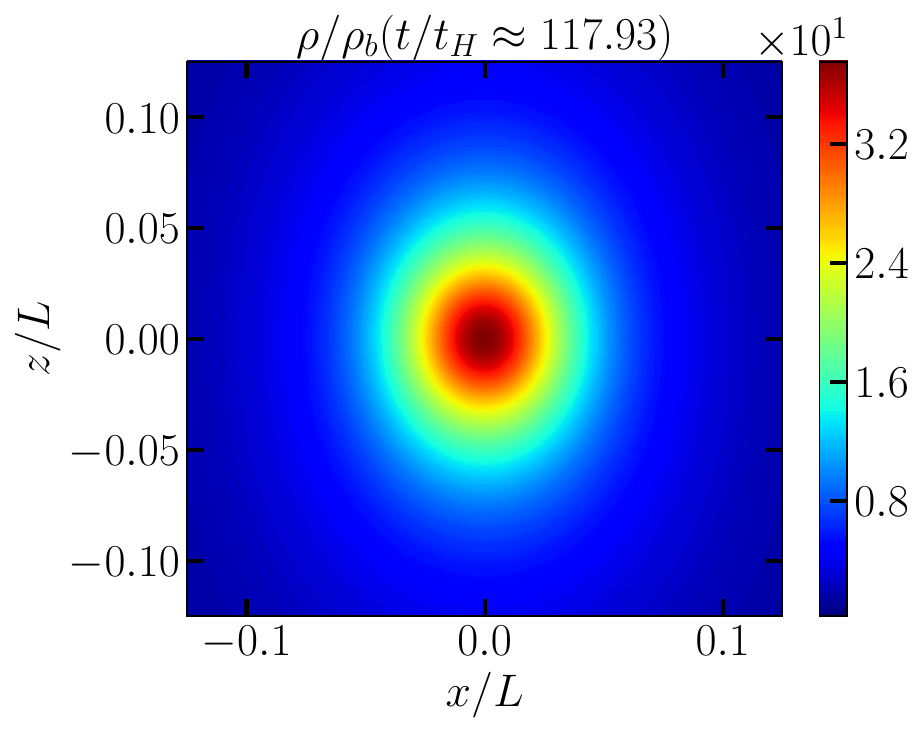}
\hspace*{-0.3cm}
\includegraphics[width=1.5 in]{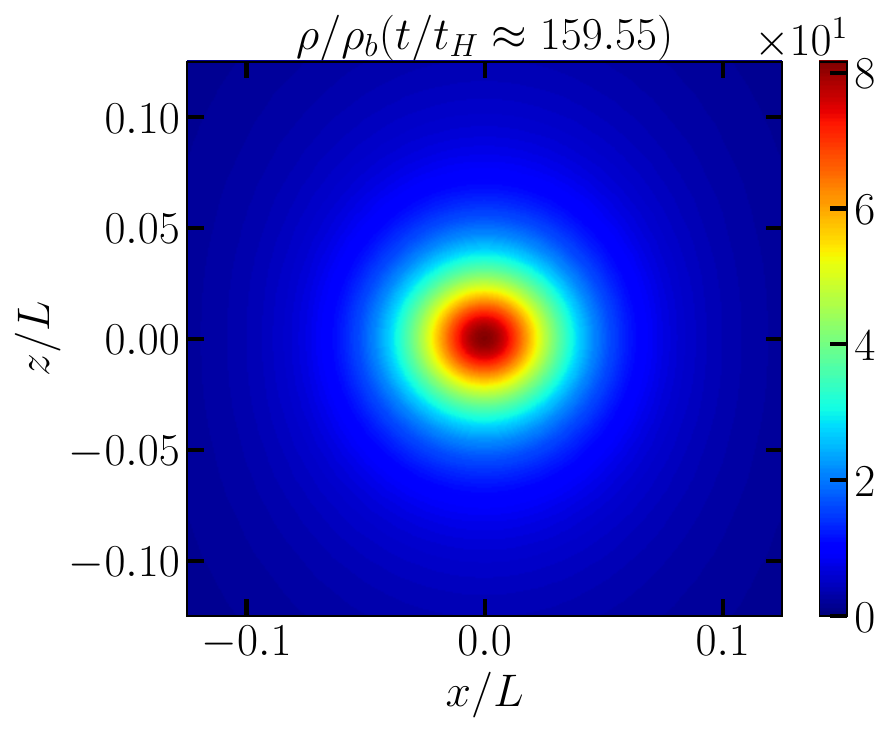}
\hspace*{-0.3cm}
\includegraphics[width=1.5 in]{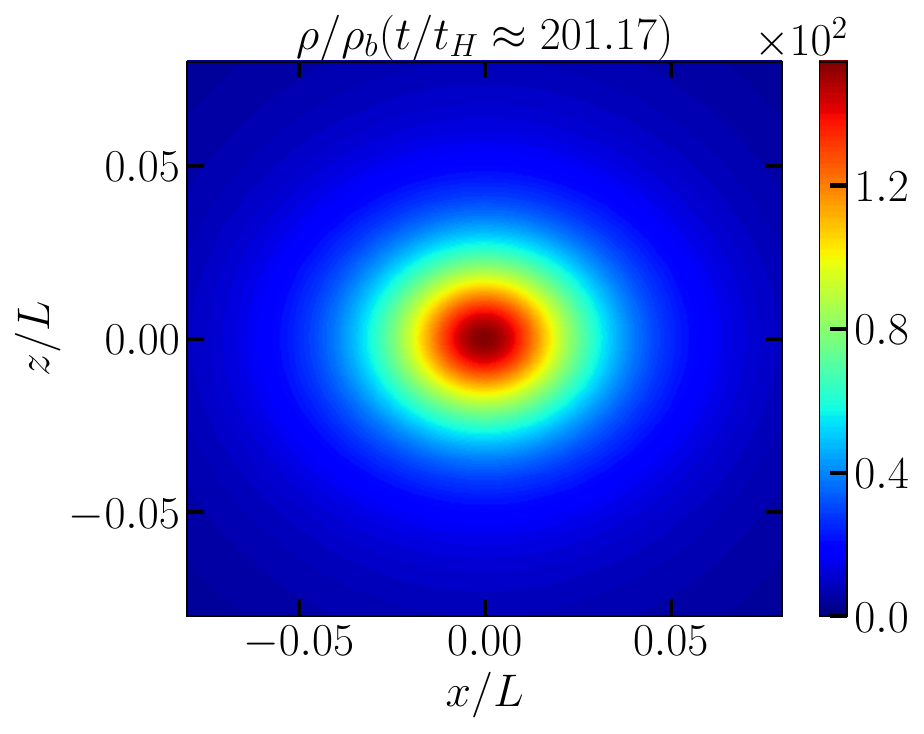}
\hspace*{-0.3cm}
\includegraphics[width=1.5 in]{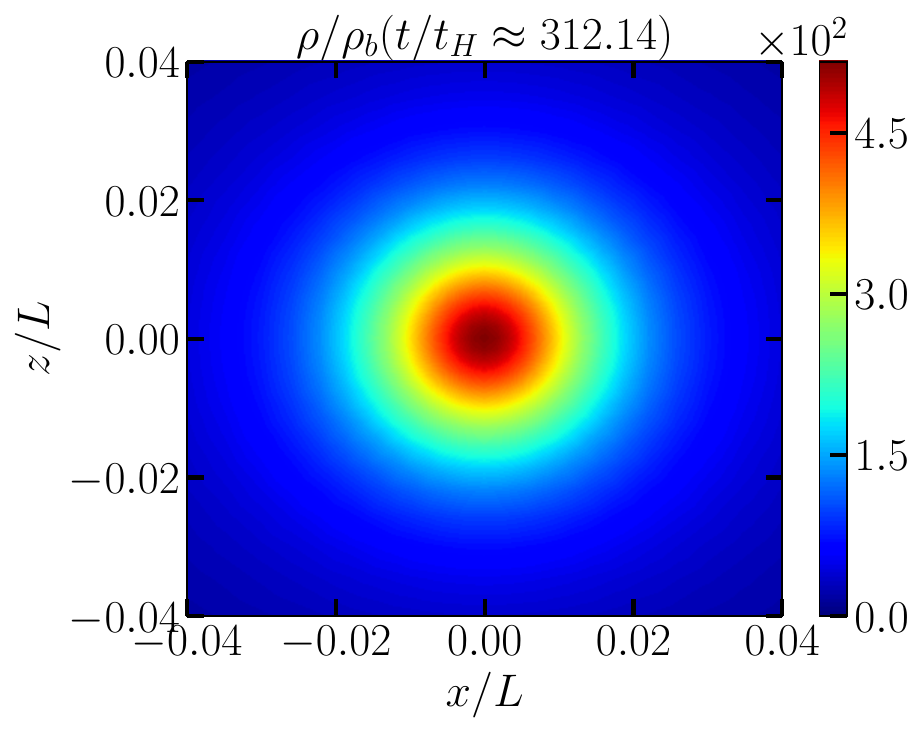}
\hspace*{-0.3cm}
\includegraphics[width=1.5 in]{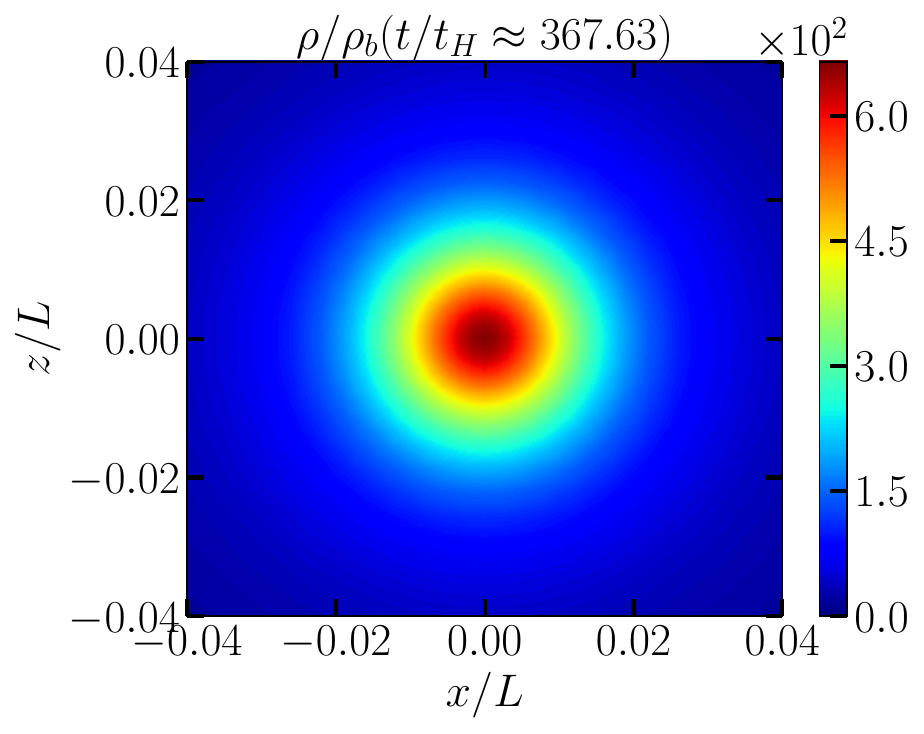}
\hspace*{-0.3cm}
\includegraphics[width=1.5 in]{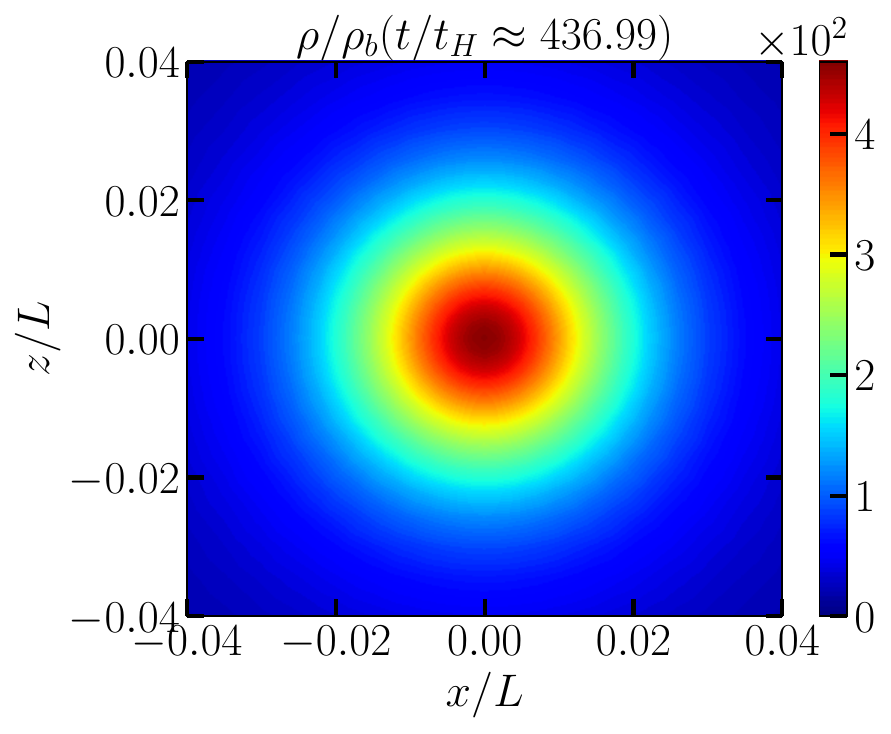}
\caption{Snapshots of the evolution of $\rho/\rho_b$ in the plane $y=0$ 
for
$e=0$ and $p=-0.175$ with $w=1/10$.}
\label{fig:energy_density_ratio_soft_dispersion}
\end{figure}

\begin{figure}[!htbp]
\centering
\includegraphics[width=1.5 in]{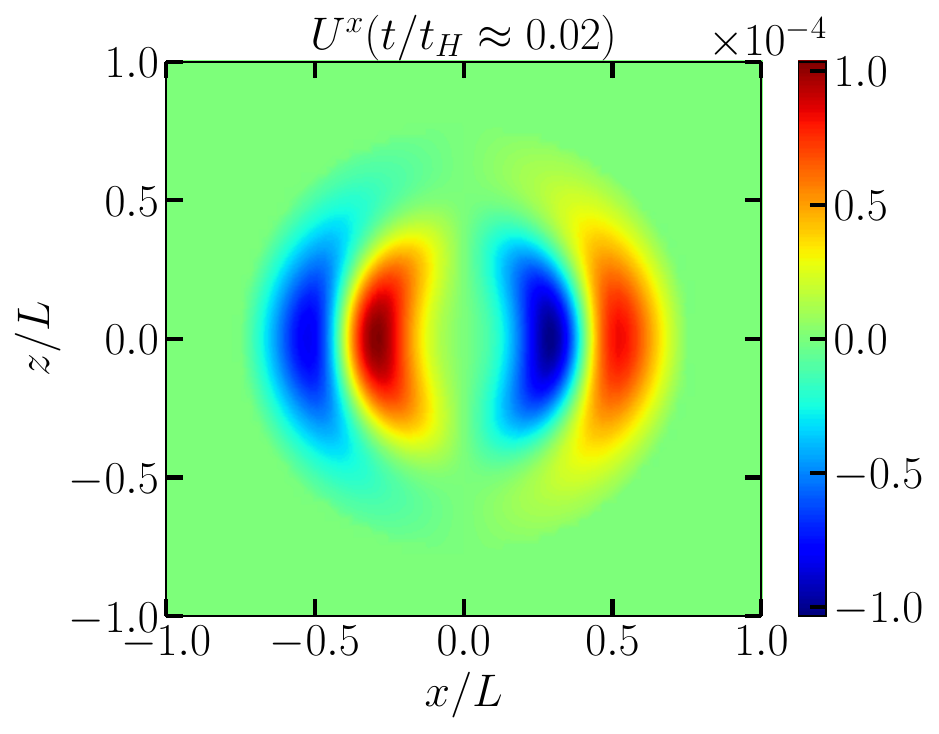}
\hspace*{-0.3cm}
\includegraphics[width=1.5 in]{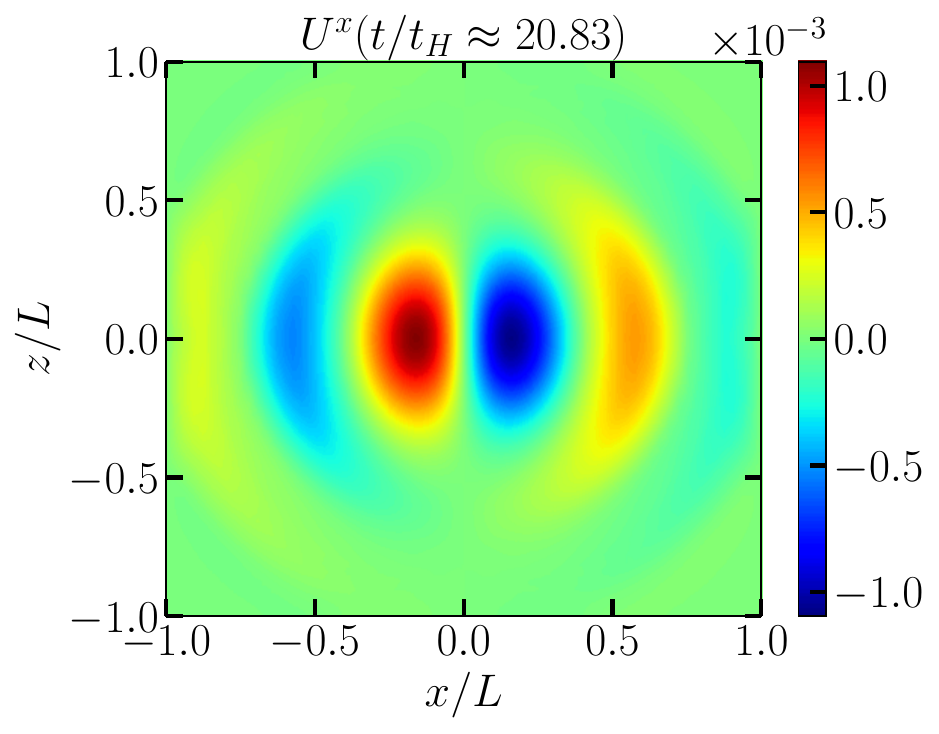}
\hspace*{-0.3cm}
\includegraphics[width=1.5 in]{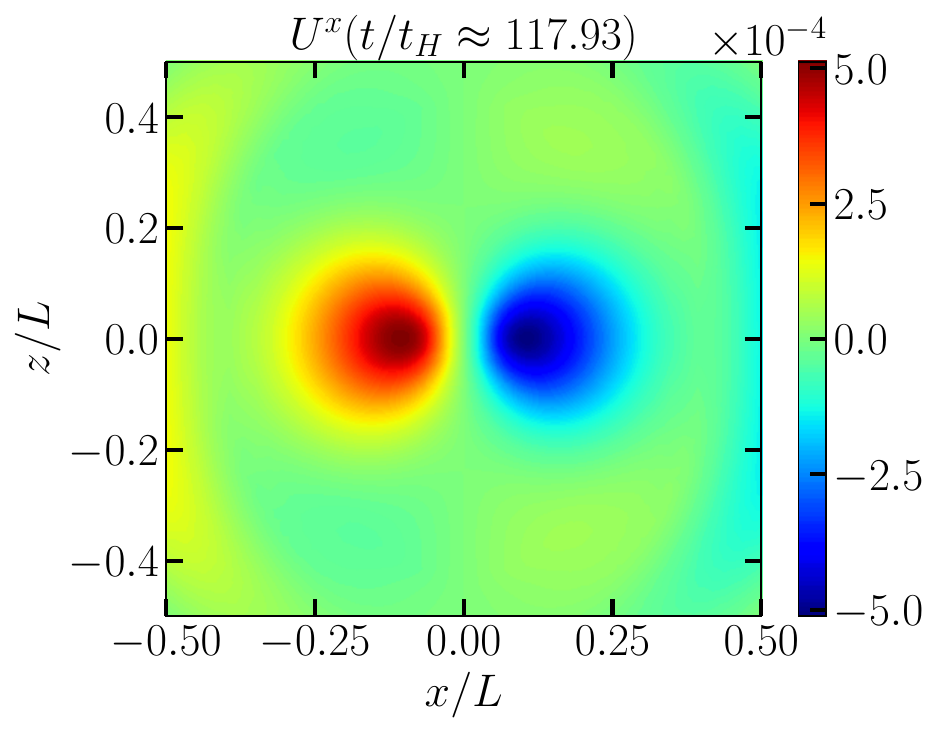}
\hspace*{-0.3cm}
\includegraphics[width=1.5 in]{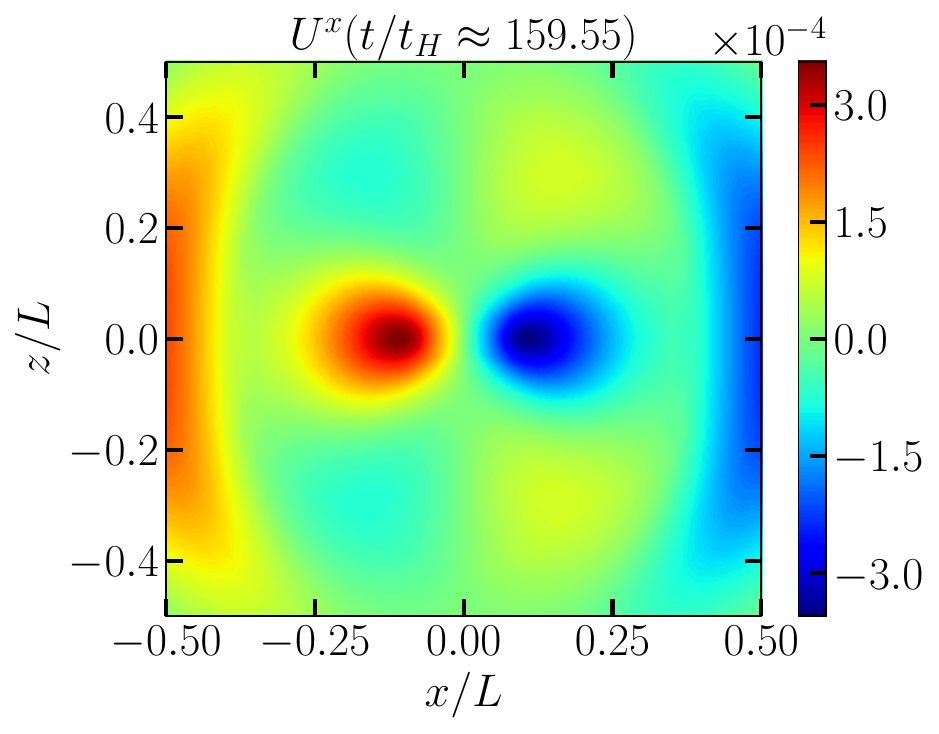}
\hspace*{-0.3cm}
\includegraphics[width=1.5 in]{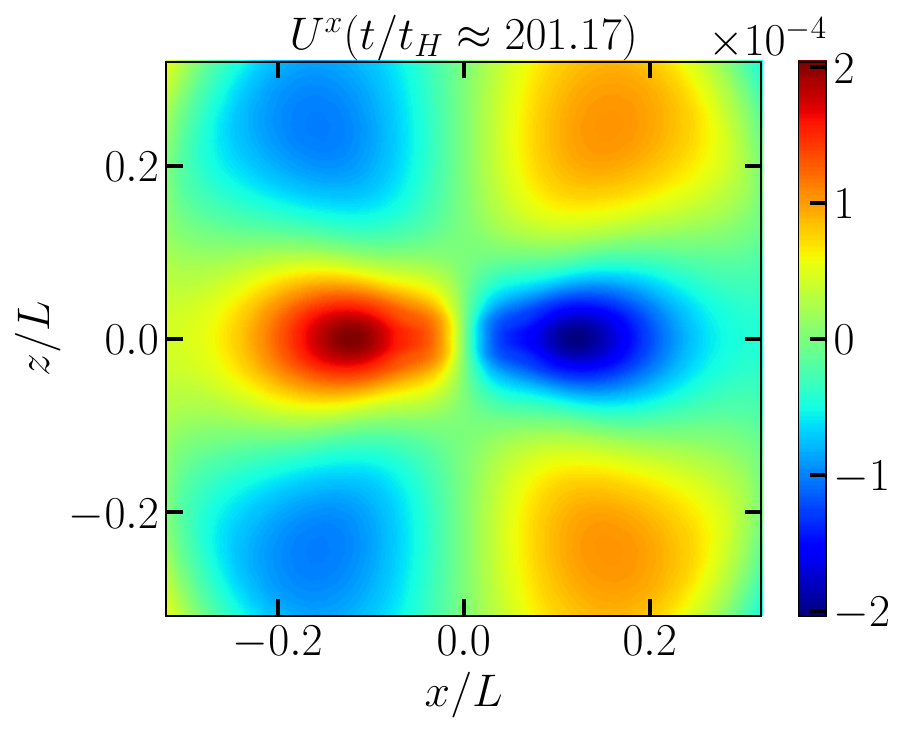}
\hspace*{-0.3cm}
\includegraphics[width=1.5 in]{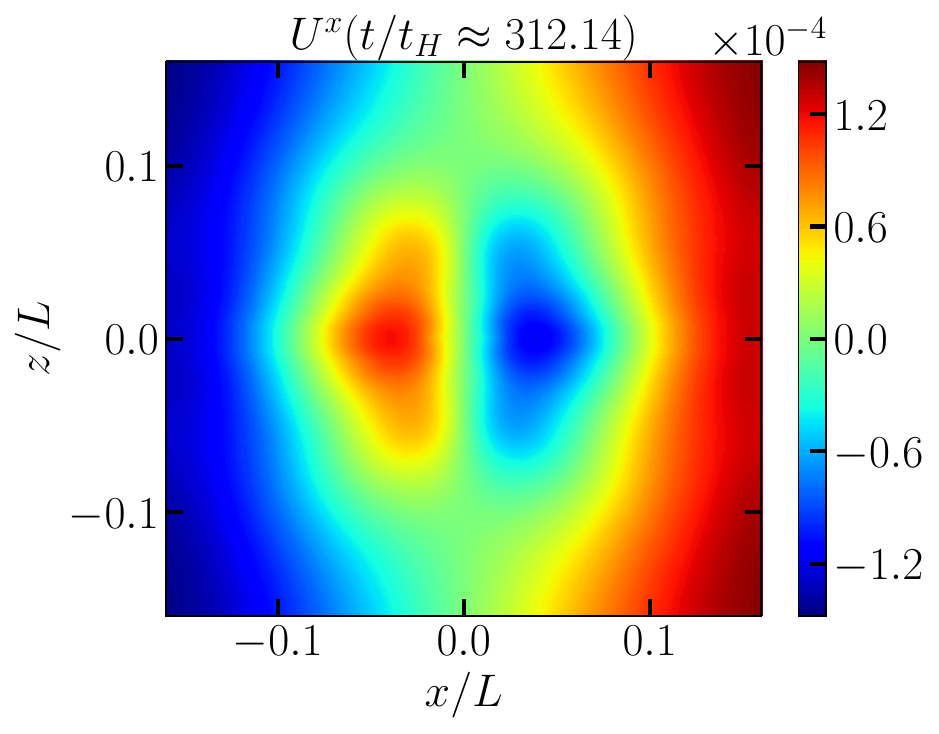}
\hspace*{-0.3cm}
\includegraphics[width=1.5 in]{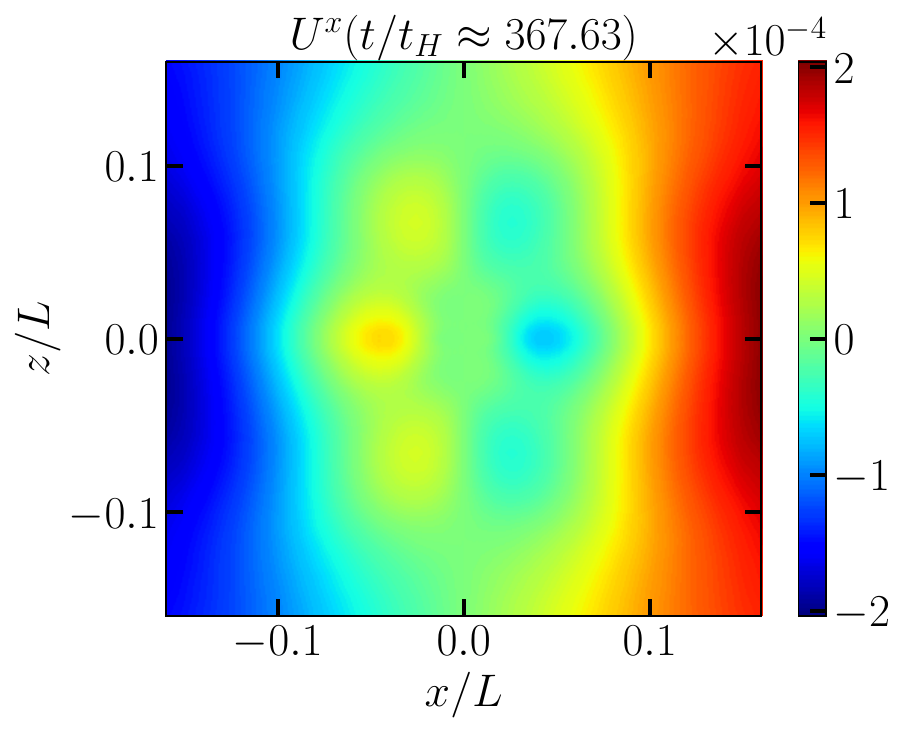}
\hspace*{-0.3cm}
\includegraphics[width=1.5 in]{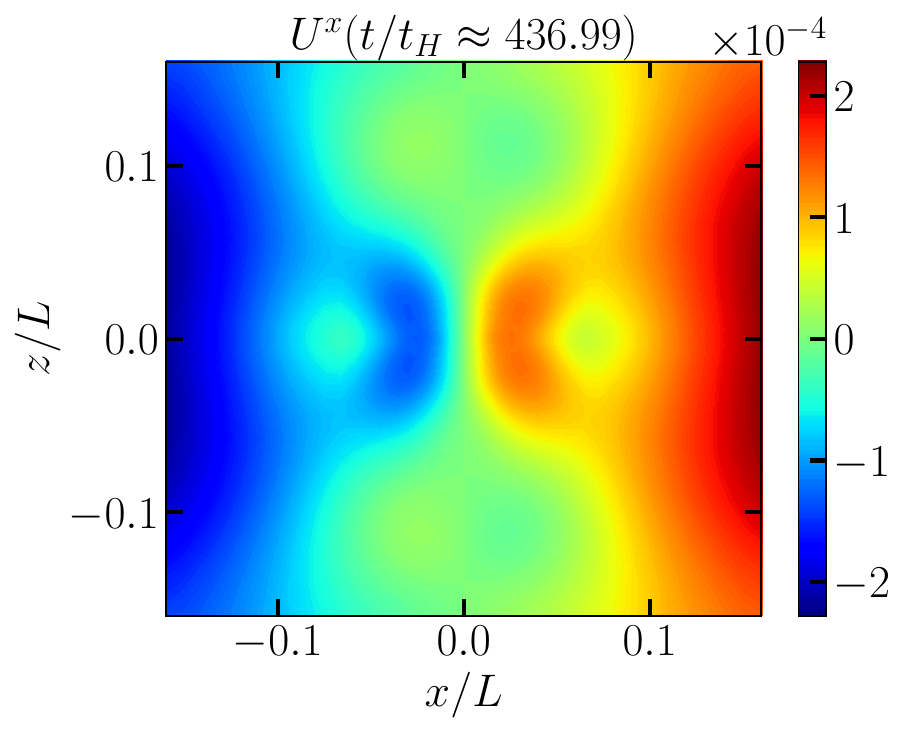}
\caption{Snapshots of the evolution of the Eulerian velocity $U^{x}$ in the plane $y=0$
for
$e=0$ and $p=-0.175$ with $w=1/10$.}
\label{fig:vel_dispersion_soft_x}
\end{figure}

\begin{figure}[!htbp]
\centering
\includegraphics[width=1.5 in]{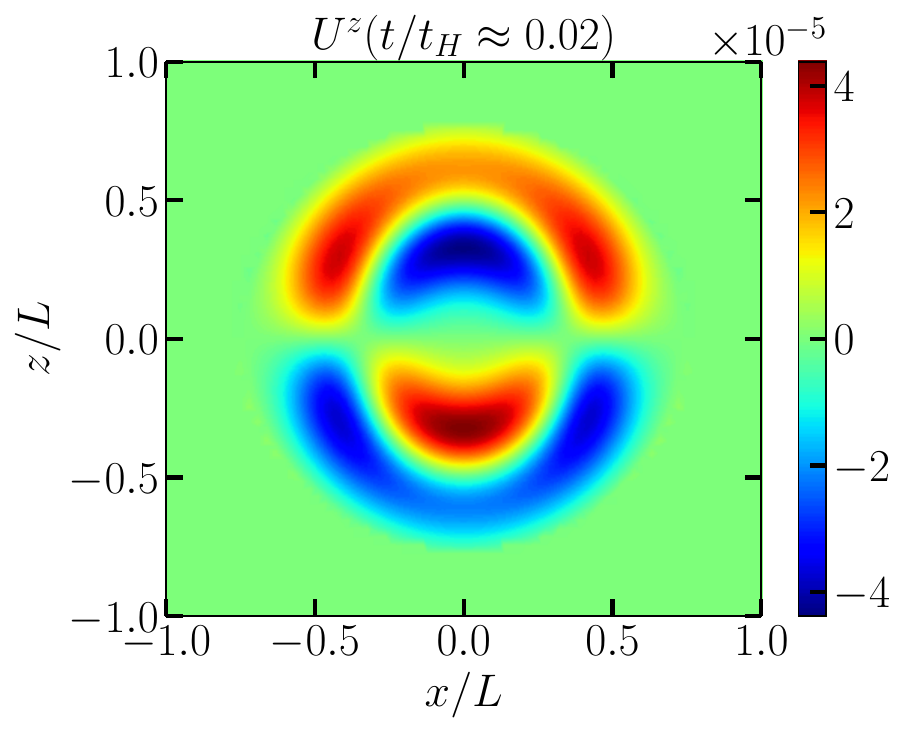}
\hspace*{-0.3cm}
\includegraphics[width=1.5 in]{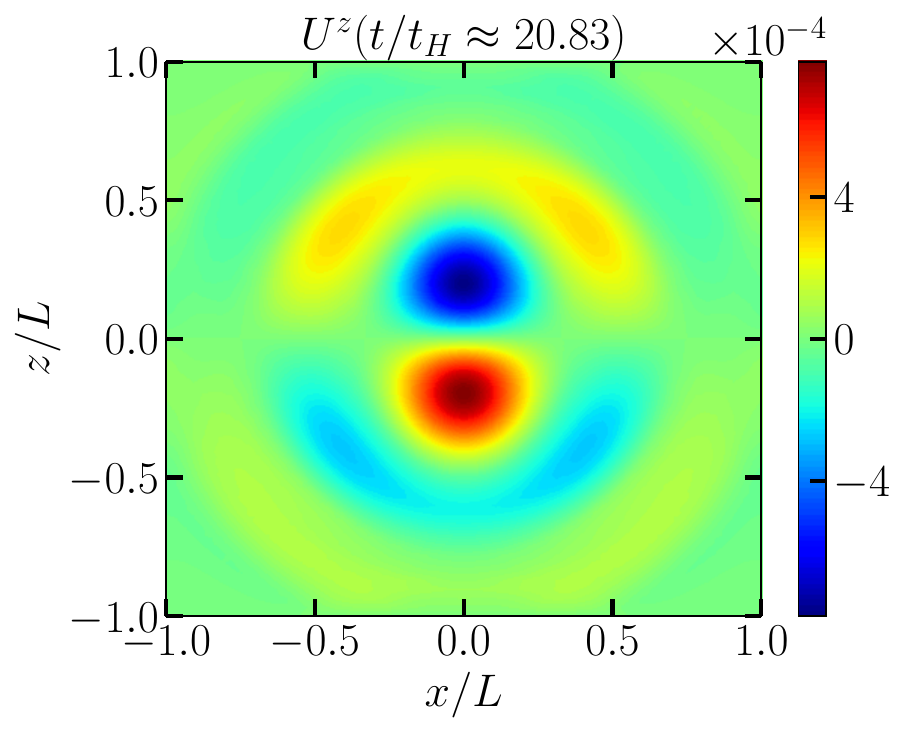}
\hspace*{-0.3cm}
\includegraphics[width=1.5 in]{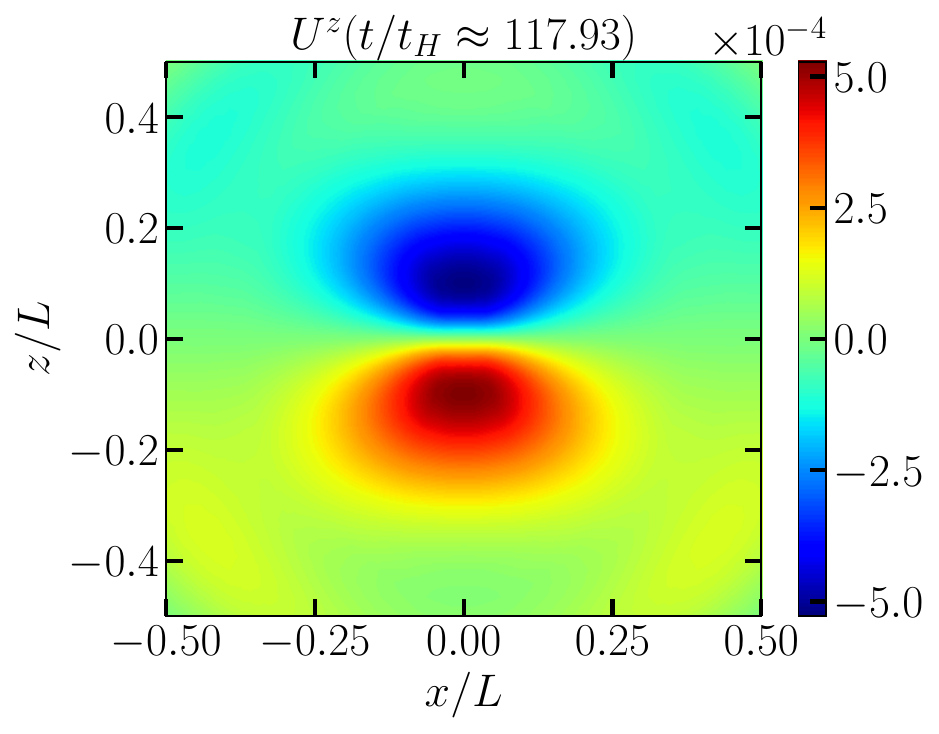}
\hspace*{-0.3cm}
\includegraphics[width=1.5 in]{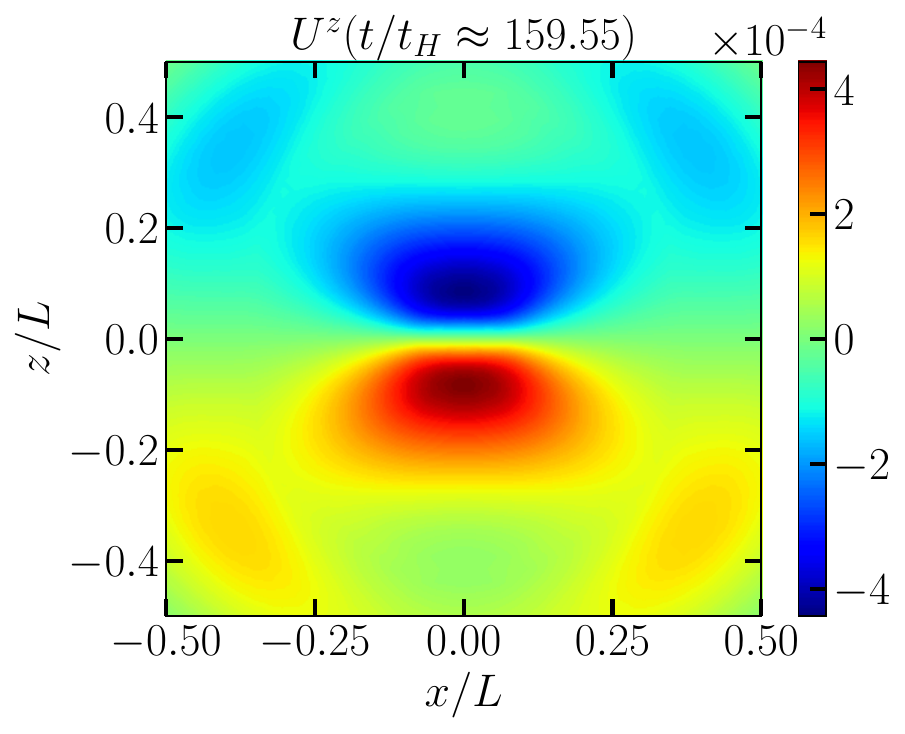}
\hspace*{-0.3cm}
\includegraphics[width=1.5 in]{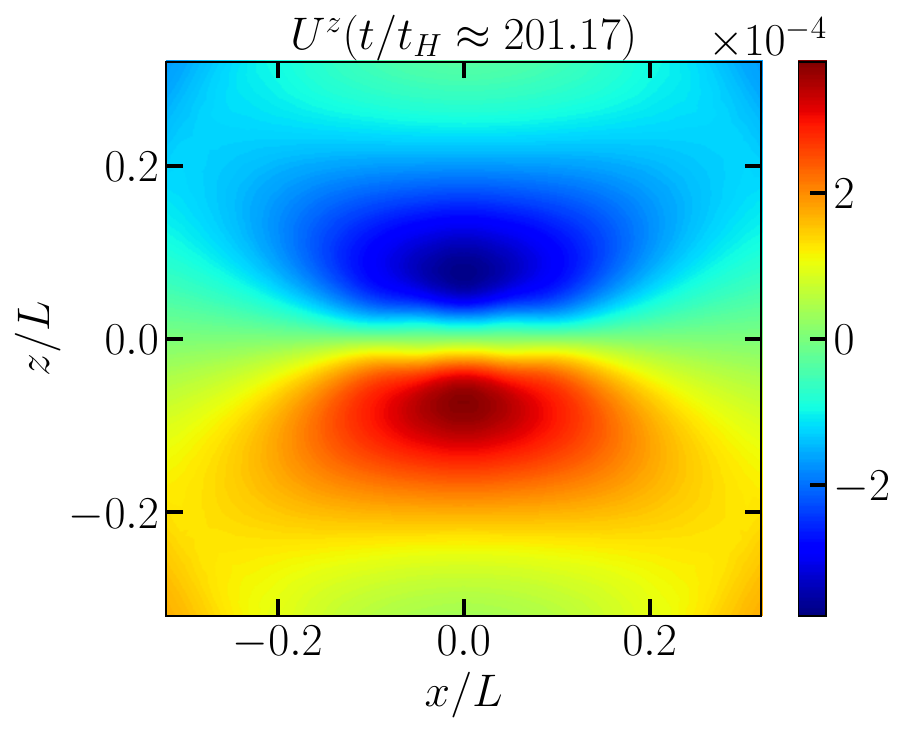}
\hspace*{-0.3cm}
\includegraphics[width=1.5 in]{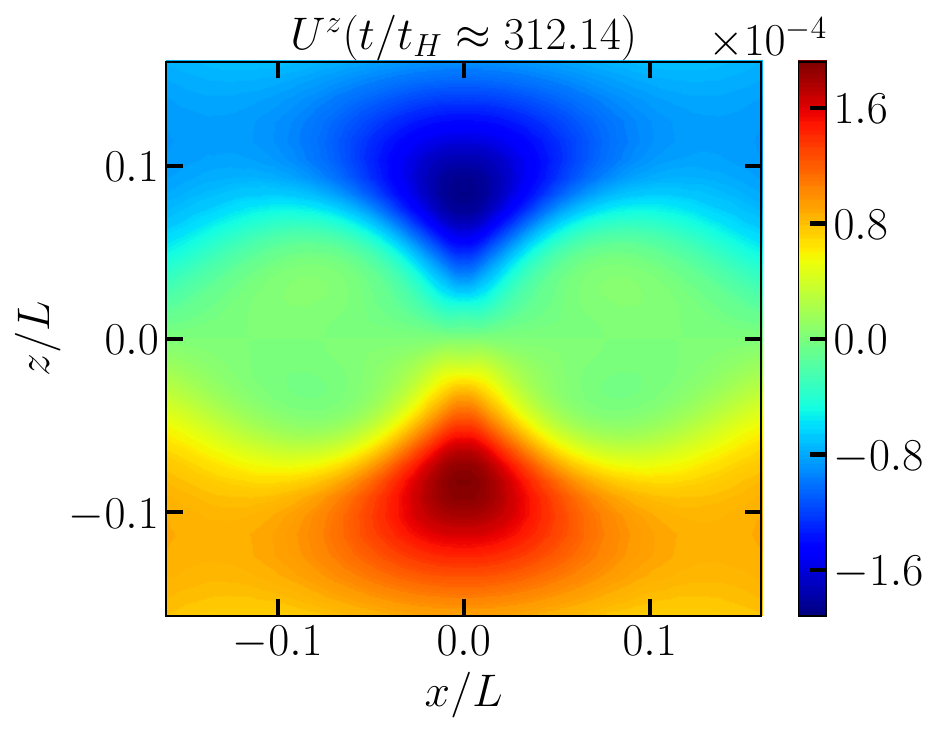}
\hspace*{-0.3cm}
\includegraphics[width=1.5 in]{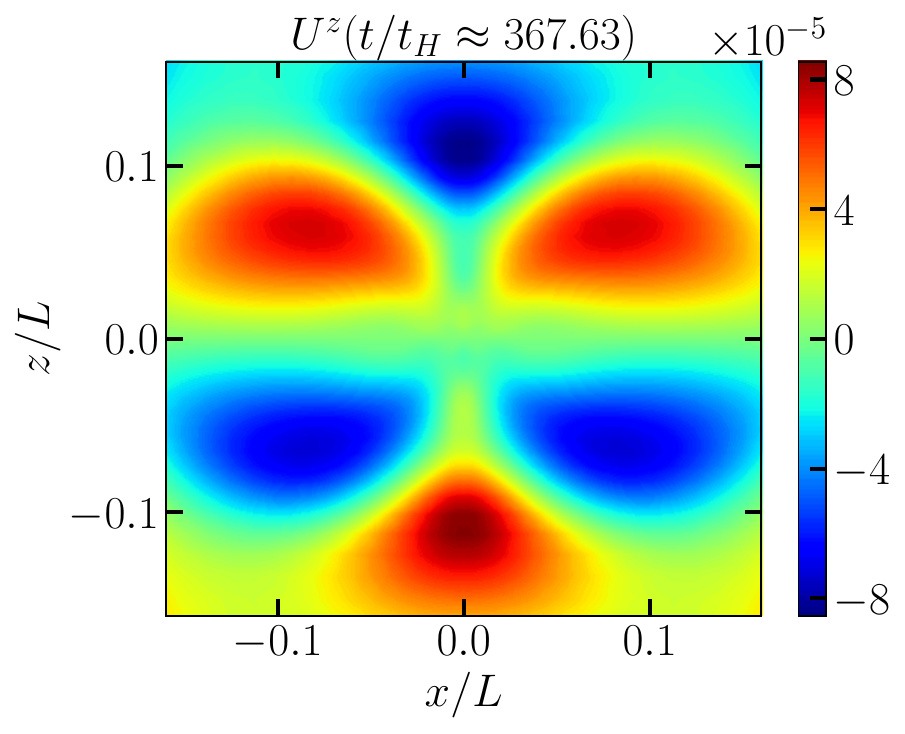}
\hspace*{-0.3cm}
\includegraphics[width=1.5 in]{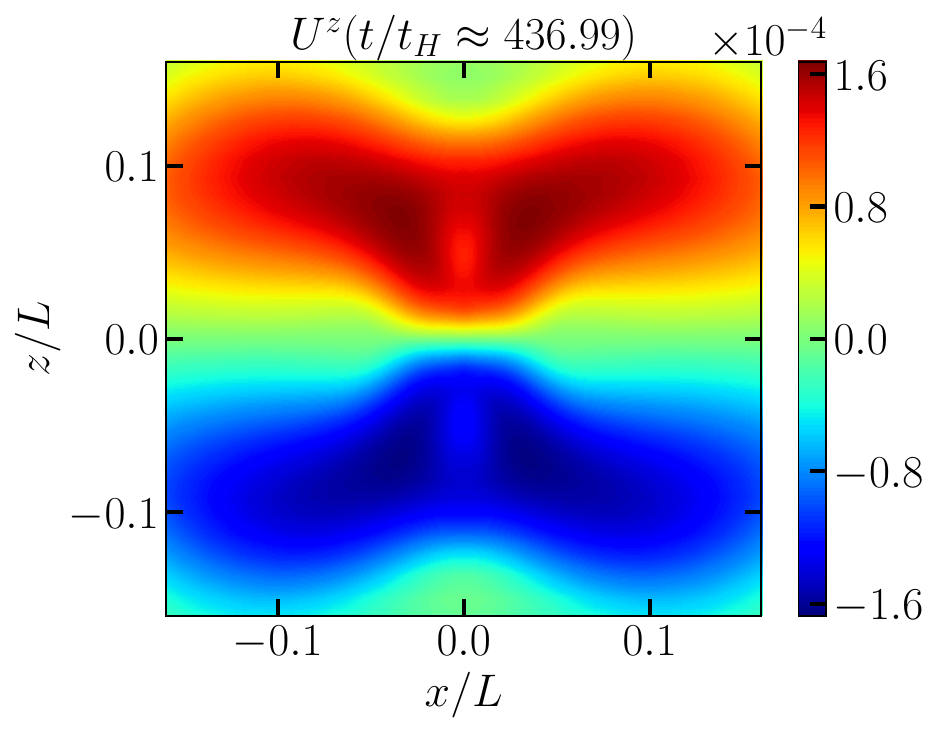}
\caption{Snapshots of the evolution of the Eulerian velocity $U^{z}$ in the plane $y=0$
for
$e=0, p=-0.175$ with $w=1/10$.}
\label{fig:vel_dipersion_soft_z}
\end{figure}

\begin{figure}[!htbp]
\centering
\includegraphics[width=1.9 in]{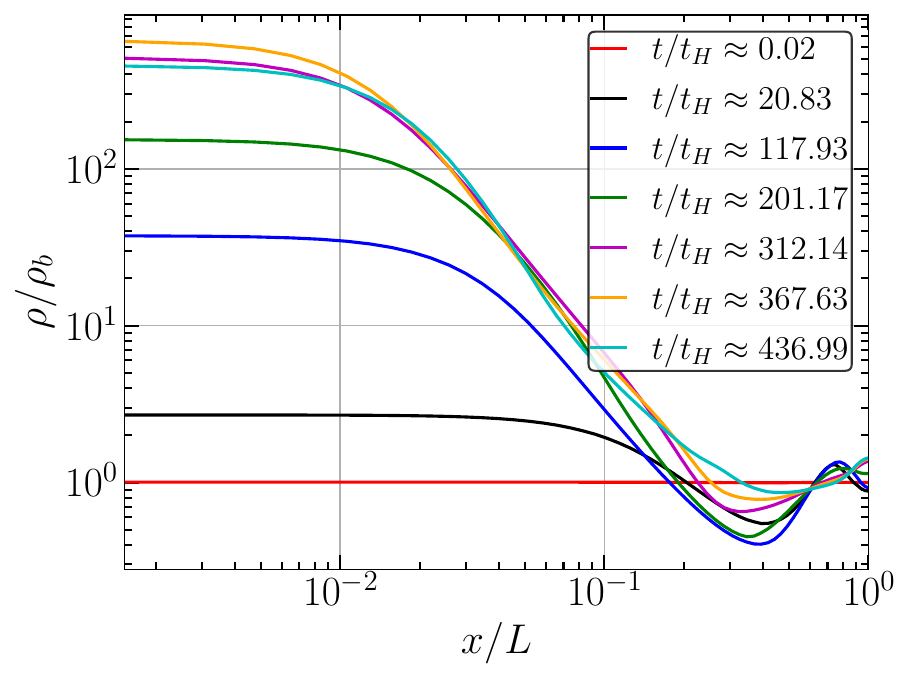}
\includegraphics[width=1.9 in]{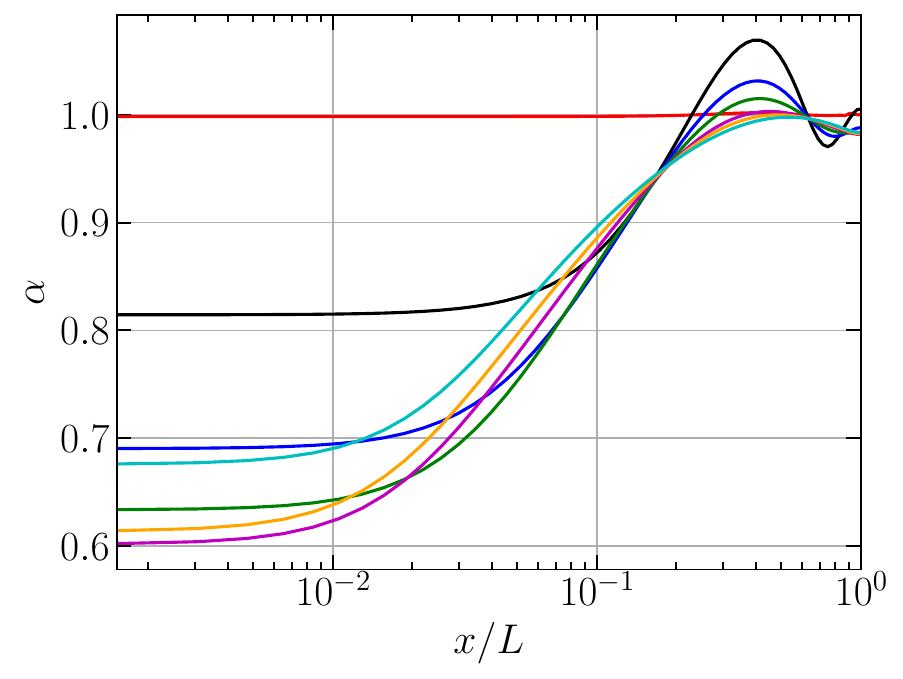}
\includegraphics[width=2.0 in]{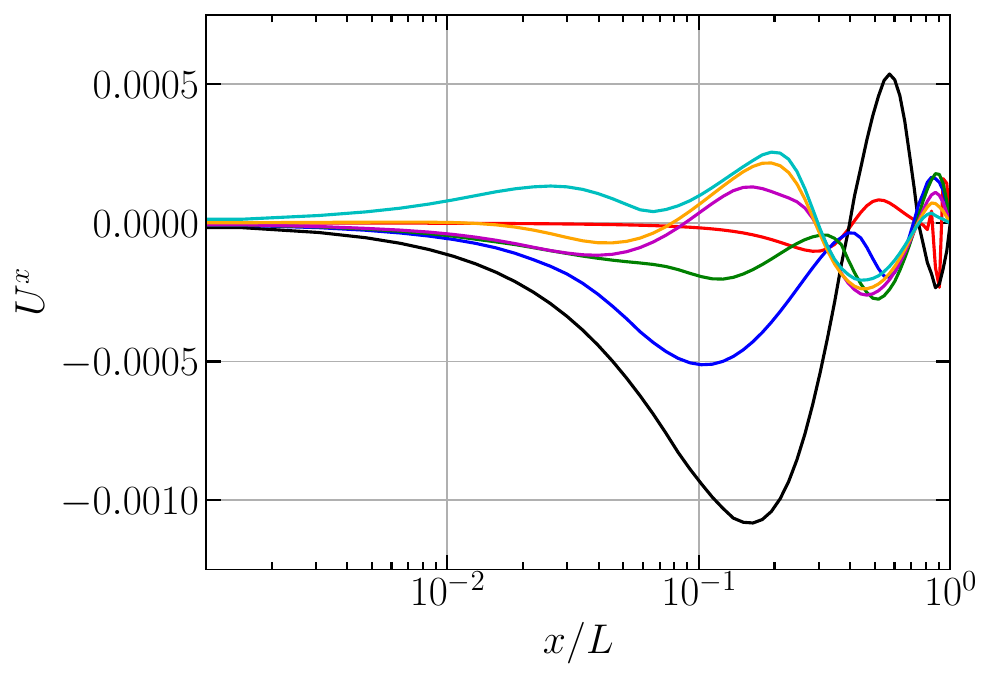}
\caption{
Snapshots of the energy density $\rho/\rho_b$ (left-panel), lapse function $\alpha$ (middle-panel) and Eulerian velocity $U^{x}$ (right-panel)
on the $x$ axis ($y=z=0$) for 
$e=0$ and $p=-0.175$ with $w=1/10$.
}
\label{fig:projection_variables_dispersion_soft}
\end{figure}

In this Appendix, we present additional figures illustrating the gravitational collapse for the case of $w=1/10$. Overall, the qualitative behaviour is similar to the cases discussed in Sec.~\ref{sec:dynamics_collapse}. 
We find that the most notable qualitative difference is that the collapse takes longer than in the $w=1/3$ case, as previously mentioned. Figs.~\ref{fig:energy_density_ratio_soft_collapse}-\ref{fig:projection_variables_collapse_soft} show a scenario where the fluctuation collapses to form a black hole. 
The initial shape of the fluctuation in the $z-x$ plane is nearly spherical, with eigenvalues 
$\lambda_1/( \nu \sigma_2) = 1.175$, $\lambda_2/( \nu \sigma_2) = 1.1$, and $\lambda_3/( \nu \sigma_2) = 0.825$, where $\lambda_1 \approx \lambda_2$. 
In contrast, Figs.~\ref{fig:energy_density_ratio_soft_dispersion}-\ref{fig:projection_variables_dispersion_soft} depict a case with an initial prolate spheroidal geometry, where the eigenvalues are 
$\lambda_1/( \nu \sigma_2) = 0.825$, $\lambda_2/( \nu \sigma_2) = 1.35$, and $\lambda_3/( \nu \sigma_2) = 0.825$, with $\lambda_1 = \lambda_3 < \lambda_2$. 
Although the 
oscillating
behaviour is similar to what was observed previously, it eventually transitions to an oblate geometry, approaching an almost spherical shape in the final stages.

\section{Convergence of the numerical simulations}
\label{sec:appendix_convergence}

\begin{figure}[!htbp]
\centering
\includegraphics[width=3.0 in]{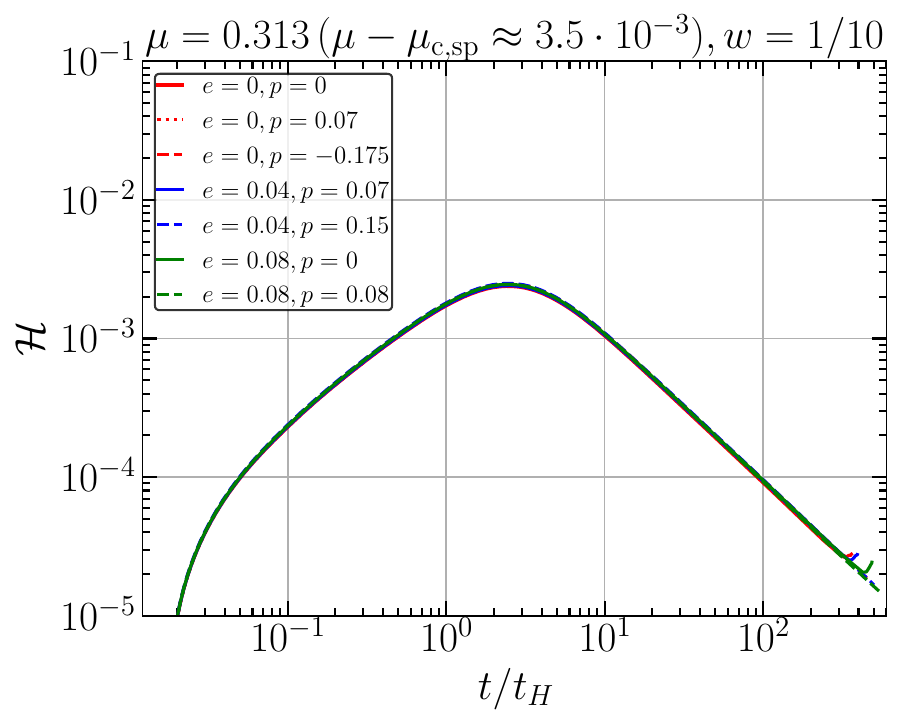}
\includegraphics[width=3.0 in]{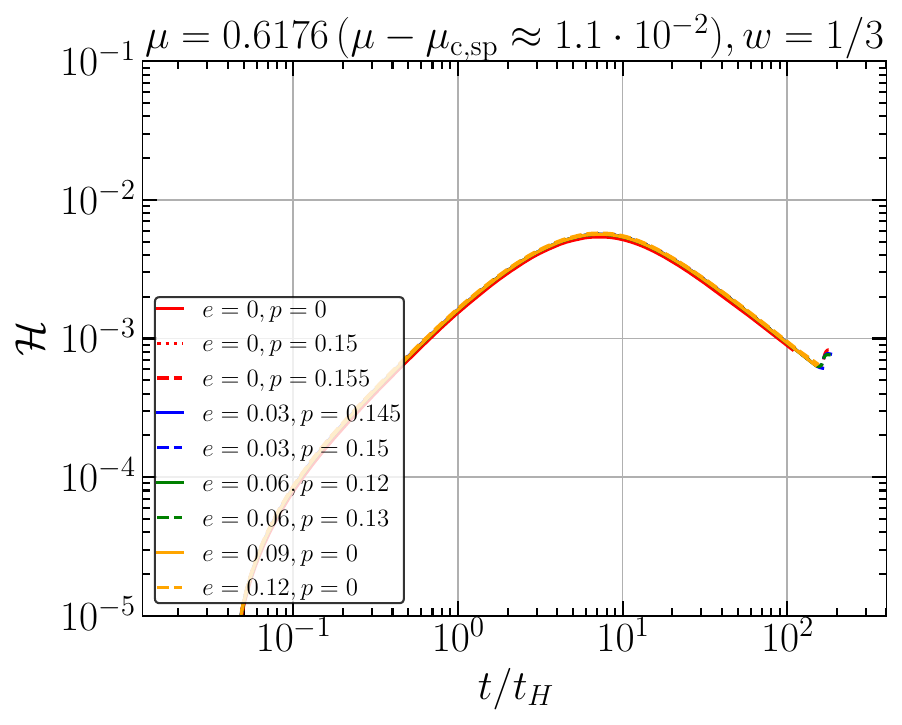}
\includegraphics[width=3.0 in]{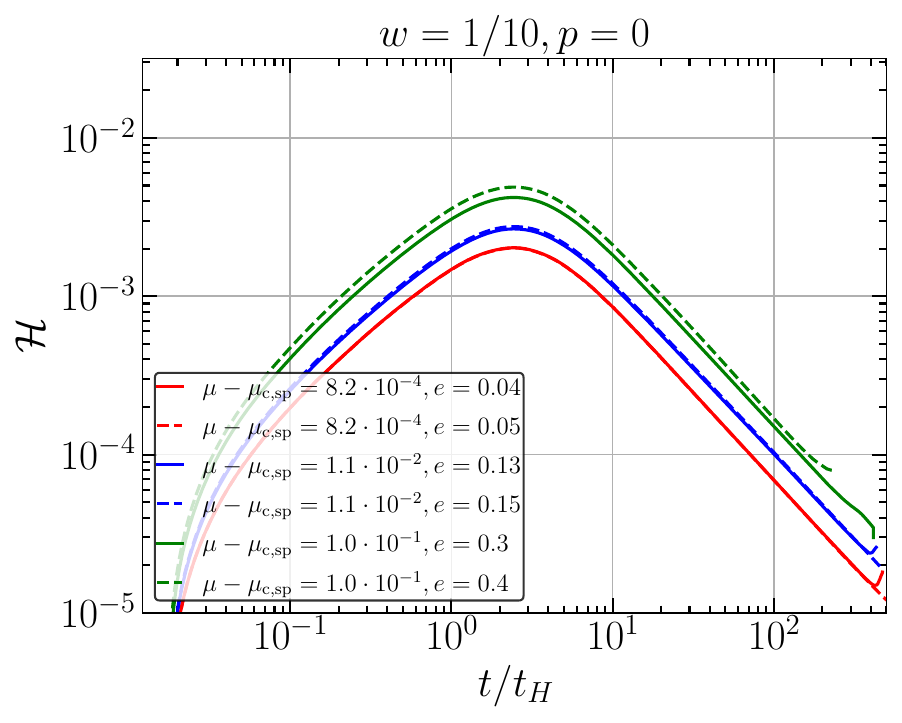}
\includegraphics[width=3.0 in]{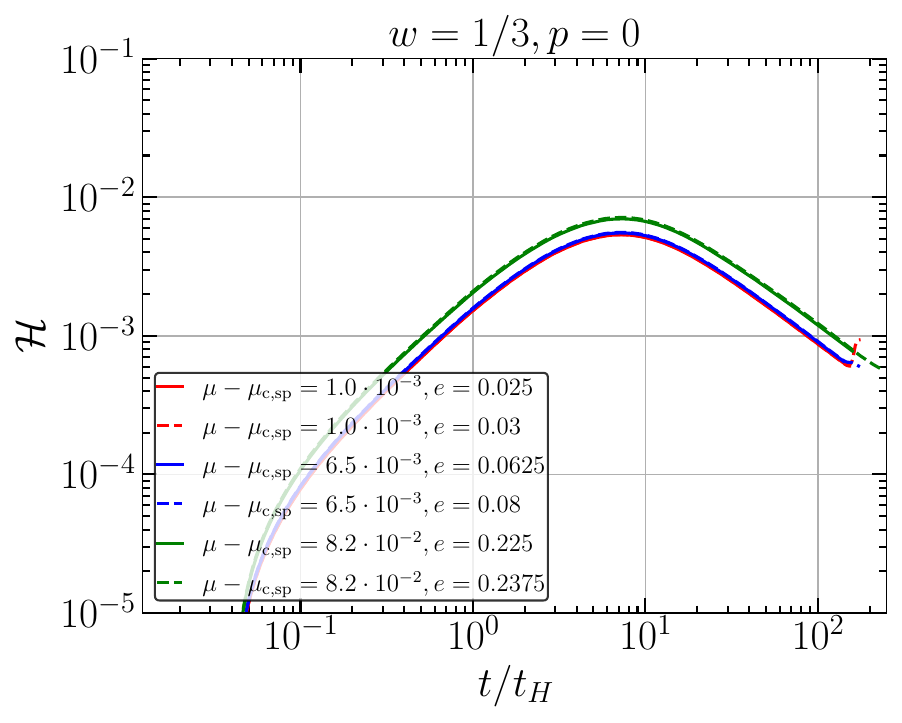}
\includegraphics[width=3.0 in]{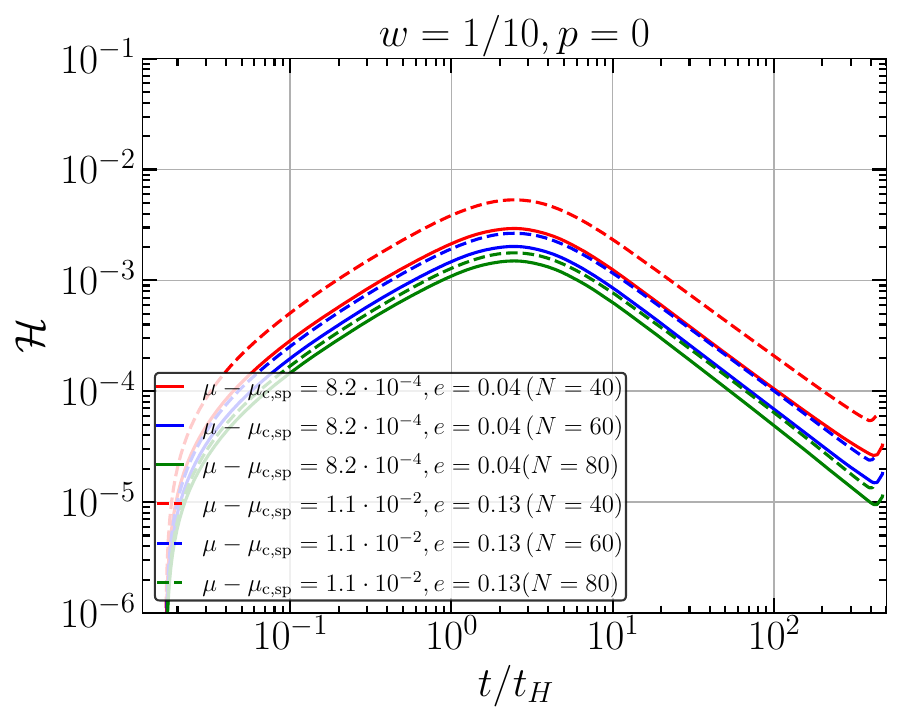}
\includegraphics[width=3.0 in]{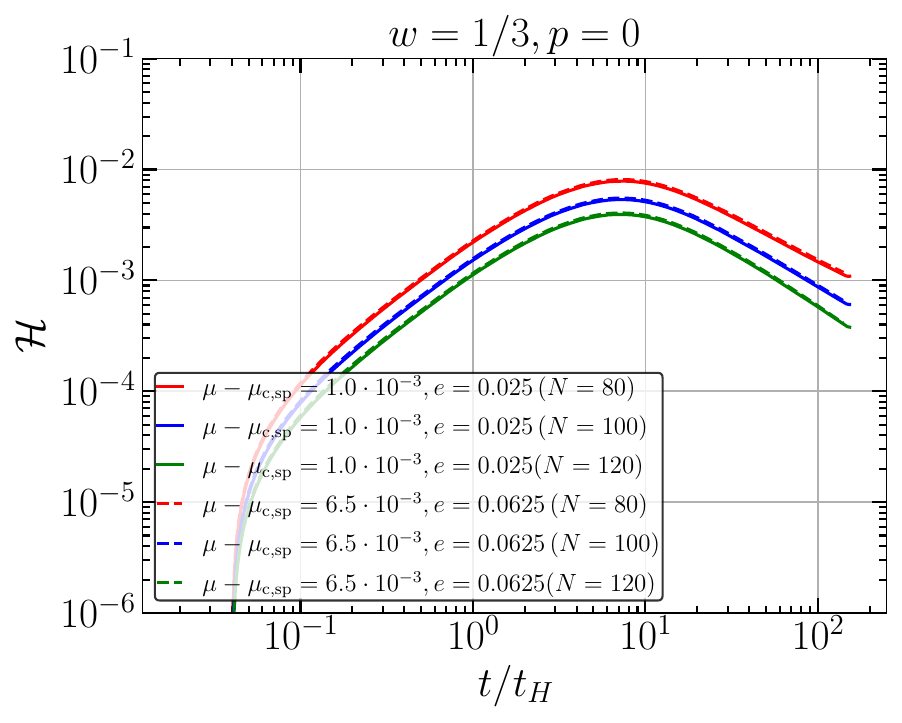}
\includegraphics[width=3.0 in]{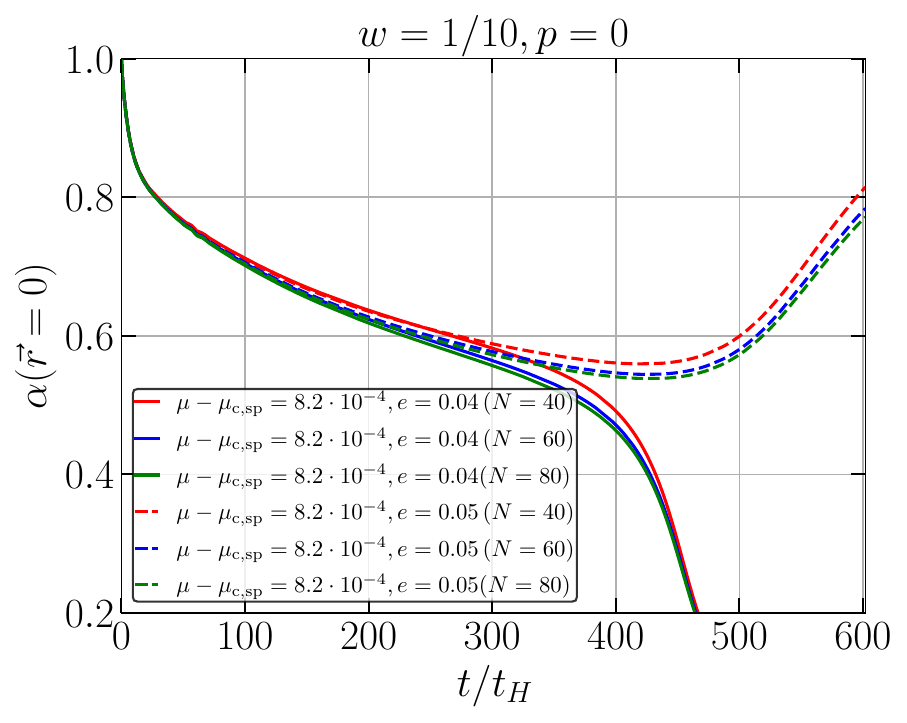}
\includegraphics[width=3.0 in]{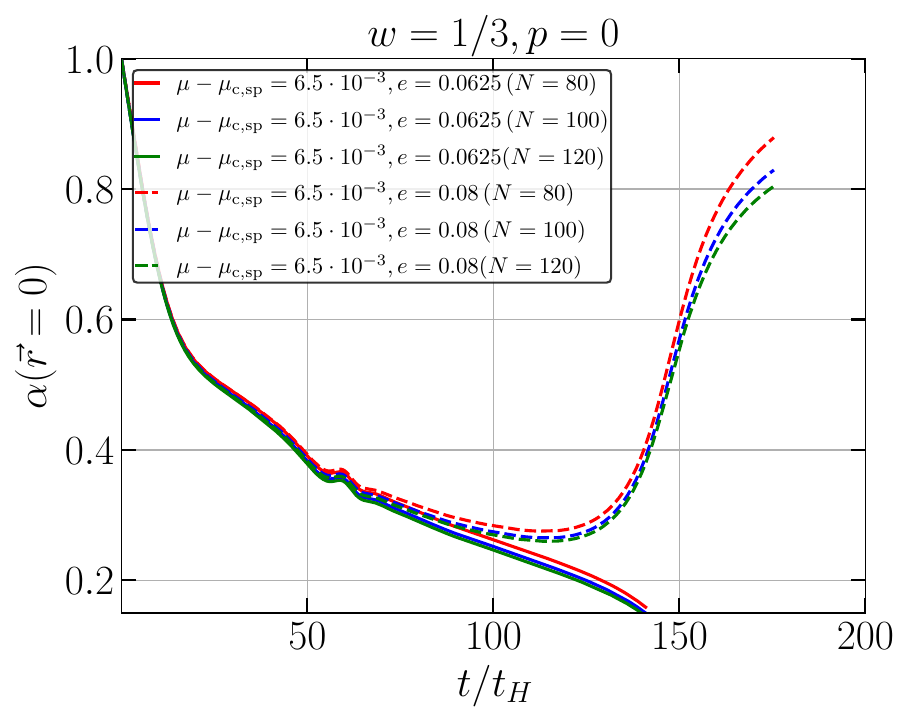}

\caption{Evolution of the averaged Hamiltonian constraint and its convergence in time for the cases shown in Fig.\ref{fig:lapse_evolution_tipical_amplitude} and the convergence of the lapse function at the center $\vec{r}=0$ for some cases.}
\label{fig:H_averaged}
\end{figure}

In this Appendix, we present several figures illustrating the evolution of the Hamiltonian constraint (see \cite{Harada:2015yda} for the equations) and its convergence. 
We compute the averaged Hamiltonian constraint, depicted in Fig.\ref{fig:H_averaged}. Our results indicate that the Hamiltonian constraint is well satisfied until late times, when the simulation breaks down in some cases, particularly in the radiation-dominated scenarios. In some cases for $w=1/10$,  
we terminated the computation 
once the numerical evolution provided sufficient evidence for either black hole formation or not. 
Notably, when the Hamiltonian constraint starts to be violated, a bouncing behaviour of the lapse function at the centre can already be observed. 
This allows us to robustly determine the threshold for black hole formation using a bisection method with different iterations. The convergence of the Hamiltonian constraint reduction as the number of grid points increases is consistent with the second-order spatial grid differentiation in the code. We also demonstrate the convergence of the lapse function at the center, which is a local quantity used to infer the threshold for PBH formation.

\section{Summary of the evolution scheme used}
\label{sec:numerical_scheme}

In this Appendix, we briefly summarize the time-evolution scheme for the fluid
used in the COSMOS code, which employs the MUSCL scheme \cite{KURGANOV2000241,Shibata:2005jv}. We refer the reader to \cite{Gourgoulhon:2007ue,2008LRR....11....7F,2016nure.book.....S} for further details.

\subsection{Fluid quantities and dynamical equations in 3+1 form}
\label{sec:quantities}
We consider the following line element
\begin{equation}
ds^2 = -\alpha^2 dt^2 + \gamma_{ij}( dx^{i}+\beta^{i} dt )( dx^{j}+\beta^{j} dt ),
\label{eq:line_element55}
\end{equation}
where $\alpha$ is the lapse function, $\beta^{i}$ is the shift vector,  $\gamma_{ij}$ is the spatial metric. 
The energy-momentum tensor of a perfect fluid
is given by 
\begin{equation}
T_{\mu\nu}=(\rho + P)u_\mu u_\nu +P g_{\mu\nu}, 
\end{equation}
where $u^\mu$ is the fluid four-velocity and the Lorentz factor $\Gamma$ is given by%
\begin{equation}
\Gamma=-u^\mu n_\mu
\end{equation}
with $n_\mu$ being the normal one-form for the time slice. 
We also introduce the velocity relative to the Eulerian observer $U^\mu$ as%
\begin{equation}
u^\mu=\Gamma(n^\mu+U^\mu) 
\end{equation}
with $n^\mu U_\mu=0$ and
\begin{equation}
\Gamma=\left(1-U^iU_i\right)^{-1/2}. 
\label{eq:Gam}
\end{equation} 
The fluid energy density measured by the Eulerian observer is given by 
\begin{equation}
E=T^{\mu\nu}n_\mu n_\nu=\Gamma^2(\rho+P)-P. 
\label{eq:EGam}
\end{equation}
Let us write the proper rest mass energy density as $\rho_0$. 
Then, the relativistic specific enthalpy is defined by%
\begin{equation}
h=\frac{\rho+P}{\rho_0}=\frac{\rho_0(1+\varepsilon)+P}{\rho_0},
\label{eq:ent}
\end{equation}
where $\rho_0\varepsilon$ is the internal energy. 
Let $D$ denote the baryon rest mass density measured by the Eulerian observer as%
\begin{equation}
D=\rho_0\Gamma. 
\label{eq:DGam}
\end{equation}
For later convenience, we also introduce the fluid momentum density measured by the Eulerian observer $p_\mu$ as%
\begin{equation}
p_\mu=(E+P)U_\mu. 
\label{eq:pmu}
\end{equation}
Then, the fluid equations are written as~\cite{Gourgoulhon:2007ue} 
\begin{eqnarray}
\left(\partial_t-\mathcal L_\beta\right)D&+&D_i(\alpha D U^i)-\alpha K D=0,  \\
\left(\partial_t-\mathcal L_\beta\right)E&+&\alpha\left[D_ip^i-(E+P)(K+K_{ij}U^iU^j)\right]
+p^iD_i\alpha=0,  \\
\left(\partial_t-\mathcal L_\beta\right)p_i&+&
\alpha D_j(P\delta^j_i+p_iU^j)+\left[P\delta^j_i+p_iU^j\right]
D_j\alpha
-\alpha K p_i+ED_i\alpha=0,
\end{eqnarray}
where $D_i$ is the covariant derivative respect the metric $\gamma_{ij}$, $K_{ij}$ is the extrinsic curvature and $K= \gamma^{ij}K_{ij}$.
These equations can be rewritten as
\begin{eqnarray}
\partial_t(\sqrt{\gamma}D)&+&\partial_i\left[\alpha\sqrt{\gamma}D\left(U^i-\frac{\beta^i}{\alpha}\right)\right]=0,  \\
\partial_t(\sqrt{\gamma}E)&+&\partial_i\left[\alpha\sqrt{\gamma}\left(p^i-\frac{\beta^i}{\alpha}E\right)\right]
+\sqrt{\gamma}\left(p^i\partial_i\alpha-\alpha S_{ij}K^{ij}\right)=0,  \\
\partial_t(\sqrt{\gamma}p_i)&+&\partial_j \left[\alpha\sqrt{\gamma}\left\{p_i\left(U^j-\frac{\beta^j}{\alpha}\right)+\delta^j_iP\right\}\right]\cr
&&+\sqrt{\gamma}\left(E\partial_i\alpha-p_j\partial_i\beta^j+\frac{1}{2}\alpha S_{jk}\partial_i\gamma^{jk}\right)=0, 
\end{eqnarray}
where
\begin{equation}
S_{ij}=(E+P)U_iU_j+P\gamma_{ij}. 
\end{equation}
Let us define the following variables:
\begin{eqnarray}
\rho_*&=&\sqrt{\gamma}D,\\
S_0&=&\sqrt{\gamma}E,\\
S_i&=&\sqrt{\gamma}p_i. 
\end{eqnarray}
Then, the equations can be rewritten as 
\begin{eqnarray}
\partial_t\rho_*&+&\partial_i\left[\rho_*V^i\right]=0,  
\label{eq:continuity}\\
\partial_tS_0&+&\partial_i\left[S_0V^i+P\sqrt{\gamma}(V^i+\beta^i)\right]
+S^iD_i\alpha-\alpha \sqrt{\gamma}S_{ij}K^{ij}=0,  
\label{eq:energycons}\\
\partial_tS_i&+&\partial_j \left[S_iV^j+\alpha\sqrt{\gamma}\delta^j_iP\right]+S_0D_i\alpha-S_j\partial_i\beta^j+\frac{1}{2}\alpha \sqrt{\gamma}S_{jk}\partial_i\gamma^{jk}=0, 
\label{eq:euler}
\end{eqnarray}
where we have introduced $V^i$ as 
\begin{equation}
V^i=\alpha U^i-\beta^i,
\end{equation}
and used the following relation
\begin{equation}
\alpha p^i-E\beta^i=EV^i+P(V^i+\beta^i). 
\end{equation}
Note that $V^i=u^i/u^0$.%
We solve the above equations for the dynamical variables $\rho_*$, $S_0$ and $S_i$. 
In contrast to dynamical variables, 
$\rho$, $V^i$ and $\varepsilon$ 
are called primitive variables. 
The fluxes are given by 
\begin{eqnarray}
f_{\rho_*}^i&=&\rho_*V^i, \\
f_{S_0}^i&=&S_0V^i+\sqrt{\gamma}P(V^i+\beta^i), \\
f_{S_j}^i&=&S_jV^i+\alpha\sqrt{\gamma}\delta^i_jP. 
\end{eqnarray}
%
We note that the variable $\gamma$ in the fluxes can be evaluated by $\gamma=\tilde{\psi}^{12}\tilde \gamma$ with $\tilde \gamma$ being the determinant of the reference flat metric and $\tilde{\psi}$ the spatial conformal factor (see Eq.\eqref{eq:line_element}).

From the Jacobi matrix of the fluxes, we obtain the following expression for 
the three characteristic speeds for three directions:
\begin{eqnarray}
\lambda_0^i&=&V^i,\\
\lambda_\pm^i&=&
\frac{\alpha}{1-U^2c_{\rm s}^2}\left\{U^i(1-c_{\rm s}^2)
\pm c_{\rm s}\sqrt{(1-U^2)\left[\gamma^{ii}(1-U^2c_{\rm s}^2)-(1-c_{\rm s}^2)U^iU^i\right]}\right\}, 
\end{eqnarray}
where $c_{\rm s}$ is the sound velocity defined by 
\begin{equation}
c_{\rm s}^2=\left(\frac{\partial P}{\partial \rho}\right)_s 
\end{equation}
with fixed specific entropy $s$. 

\subsection{Primitive variables from dynamical variables}
\label{sec:vartrans}
The equations between the primitive variables and the conserved variables are given by 
\begin{eqnarray}
P&=&P(\rho,\varepsilon), 
\label{eq:ptod1}\\
\rho_*&=&\sqrt{\gamma}\Gamma\frac{\rho}{1+\varepsilon}, 
\label{eq:ptod2}\\
S_0&=&\sqrt{\gamma}\left[\Gamma^2(\rho+P)-P\right], 
\label{eq:ptod3}\\
S_i&=&\sqrt{\gamma}\left(E+P\right)U_i
=\frac{1}{\alpha}\left(S_0+\sqrt{\gamma}P\right)\gamma_{ij}\left(V^j+\beta^j\right). 
\label{eq:ptod4}
\end{eqnarray}
We need to invert these equations to obtain the primitive variables from the dynamical variables. 

From Eqs.~\eqref{eq:Gam} and \eqref{eq:pmu}, we obtain 
\begin{equation}
\Gamma^2p^2=(E+P)^2(\Gamma^2-1), 
\label{eq:Gam2p2}
\end{equation}
where $p^2=p^\mu p_\mu$. 
From Eq.~\eqref{eq:EGam}, we find 
\begin{equation}
\Gamma^2=\frac{E+P}{\rho+P}. 
\label{eq:Gam2}
\end{equation}
Substituting Eq.~\eqref{eq:Gam2}, into Eq.~\eqref{eq:Gam2p2}, 
we obtain 
\begin{equation}
p^2-E^2-(P-\rho)E+\rho P=0. 
\label{eq:Prho}
\end{equation}

\subsection{Barotropic equation of state}
\label{sec:baro}
Let us assume $P=P(\rho)$. 
In this case, we need not solve the continuity equation \eqref{eq:continuity} 
unless we are interested in the value of the internal energy. 
Since the values of $E$ and $p_i$ can be calculated from the dynamical variables $S_0$ and $S_i$, 
Eq.~\eqref{eq:Prho} can be regarded as an equation to get the value of $\rho$ with a given 
equation of state $P=P(\rho)$. 
Once the value of $\rho$ is calculated, 
the value of $\Gamma$ is given by Eq.~\eqref{eq:Gam2}.  
From Eq.~\eqref{eq:ptod4}, the value of $V^i$ is given by 
\begin{equation}
V^i=\alpha U^i-\beta^i=\alpha\frac{\gamma^{ij}S_j}{S_0+\sqrt{\gamma}P}-\beta^i. 
\end{equation}
The value of $\varepsilon$ is given by 
\begin{equation}
\varepsilon=\frac{\rho-\rho_0}{\rho_0}=\frac{\sqrt{\gamma}\Gamma\rho-\rho_*}{\rho_*}. 
\label{eq:varep}
\end{equation}

For the case $P=w\rho$, we obtain 
\begin{equation}
\rho=\frac{1}{2w}\left[-(1-w)E+\sqrt{E^2(1-w)^2+4w(E^2-p^2)}\right] 
\end{equation}
from Eq.~\eqref{eq:Prho}.
From Eq.~\eqref{eq:Gam2}, the value of $\Gamma$ is given by 
\begin{equation}
\Gamma^2=\frac{E+w\rho}{(1+w)\rho}. 
\end{equation}

\subsection{Flux calculation scheme}

For the flux calculation, we use the MUSCL scheme.

For simplicity, let us consider the following 1-dimensional equation in a conserved form: 
\begin{equation}
\partial_t u +\partial_x f(u)=0. 
\end{equation}
We evaluate the flux in a finite difference formula through the values at midpoints between two grid points 
as follows:

\begin{equation}
\partial_t u=-\frac{1}{\Delta x}\left(f_{i+1/2}-f_{i-1/2}\right). 
\label{eq:evo}
\end{equation}

In order to evaluate $f_{i\pm1/2}$, we first introduce the following variables for the value of $u$ at the point specified by $i+1/2$: 
\begin{eqnarray}
(u_L)_{i+1/2}&=&u_i+\frac{1}{4}\left((1-\kappa)(\bar \Delta_-)_i+(1+\kappa)(\bar \Delta_+)_i\right),
\label{eq:uL}\\
(u_R)_{i+1/2}&=&u_{i+1}-\frac{1}{4}\left((1-\kappa)(\bar \Delta_+)_{i+1}+(1+\kappa)(\bar \Delta_-)_{i+1}\right), 
\label{eq:uR}
\end{eqnarray}
where $\kappa$ is a parameter to specify the way 
of
interpolation, and 
$(\bar \Delta_\pm)_i$ is defined by

\begin{eqnarray}
(\bar \Delta_+)_i&=&{\rm minmod}((\Delta_+)_i,b(\Delta_-)_i),\\
(\bar \Delta_-)_i&=&{\rm minmod}((\Delta_-)_i,b(\Delta_+)_i)
\end{eqnarray}
with 
\begin{eqnarray}
(\Delta_+)_i&=&u_{i+1}-u_i,\\
(\Delta_-)_i&=&u_i-u_{i-1}.
\end{eqnarray}
The function minmod($a$,$b$) 
returns $0$ if $a$ and $b$ have different signs, and if not, the value of one of the arguments which has a smaller absolute value. 
That is, 
\begin{equation}
{\rm minmod}(a,b)={\rm sign}(a){\rm max}(0,{\rm min}(|a|,{\rm sign}(a) b)). 
\end{equation}
Then, the flux is evaluated as 
\begin{equation}
f=\frac{1}{2}\left(f(u_L)+f(u_R)-a^*(u_R-u_L)\right), 
\end{equation}
where $a^*$ is the value of the fastest characteristic speed given by 
\begin{equation}
a^*={\rm max}\{|\lambda_{0L}|,|\lambda_{+L}|,|\lambda_{-L}|,|\lambda_{0R}|,|\lambda_{+R}|,|\lambda_{-R}|\}. 
\end{equation}

\bibliographystyle{JHEP}
\bibliography{references_solved.bib}

\end{document}